\documentclass[preprint,11pt]{elsarticle}

\usepackage{url}
\usepackage{hyperref}
\usepackage{amsfonts}
\usepackage{amsthm}
\usepackage{amsmath}
\usepackage{bm}
\usepackage{breqn}
\usepackage[ruled,norelsize]{algorithm2e}

\graphicspath{{figs/}}
\usepackage{subcaption}

\newdefinition{remark}{Remark}

\newcommand{\bmf}{\mathbf{f}}

\newcommand{\bmn}{\mathbf{n}}

\newcommand{\bms}{\mathbf{s}}

\newcommand{\bmu}{\mathbf{u}}

\newcommand{\bmw}{\mathbf{w}}
\newcommand{\bmx}{\mathbf{x}}
\newcommand{\bmy}{\mathbf{y}}

\newcommand{\bmI}{\mathbf{I}}

\newcommand{\bmT}{\mathbf{T}}

\newcommand{\bmdelta}{\bm{\delta}}
\newcommand{\bmepsilon}{\bm{\epsilon}}

\newcommand{\bmeta}{\bm{\eta}}

\newcommand{\bmphi}{\bm{\phi}}

\newcommand{\bmsigma}{\bm{\sigma}}

\newcommand{\bmtheta}{\bm{\theta}}

\newcommand{\bmzero}{\bm{0}}
\newcommand{\bmone}{\bm{1}}

\DeclareMathOperator*{\argmin}{arg\,min}

\newcommand{\mc}{\mathcal}
\DeclareMathOperator{\E}{\mathbb{E}}

\newcommand{\norm}[1]{\left\lVert #1 \right\rVert}

\newcommand{\mb}{\mathbb}
\newcommand{\vc}{\boldsymbol}

\journal{Journal of Computational Physics}

\begin{document}
\abovedisplayskip=6.0pt
\belowdisplayskip=6.0pt
\begin{frontmatter}

\title{Bayesian Deep Convolutional Encoder-Decoder Networks for Surrogate Modeling and Uncertainty Quantification}


\author{Yinhao Zhu}
\ead{yzhu10@nd.edu}
\author{Nicholas Zabaras\corref{cor1}}
\ead{nzabaras@gmail.com}
\ead[url]{https://cics.nd.edu/}
\cortext[cor1]{Corresponding author}

\address{Center for Informatics and Computational Science, \\311I Cushing Hall, University of Notre Dame, Notre Dame, IN 46556, U.S.A}

\begin{abstract}
We are interested in the development of surrogate models for uncertainty quantification and propagation in problems governed by stochastic PDEs using a deep convolutional encoder-decoder network in a similar fashion to approaches considered in deep learning for image-to-image regression tasks. Since normal neural networks are data intensive and cannot provide predictive uncertainty, we propose a Bayesian approach to convolutional neural nets. A recently introduced variational gradient descent algorithm based on Stein's method is scaled to deep convolutional networks to perform approximate Bayesian inference on millions of uncertain network parameters. This approach achieves state of the art performance in terms of predictive accuracy and uncertainty quantification in comparison to other approaches in Bayesian neural networks as well as techniques that include Gaussian processes and ensemble methods even when the training data size is relatively small. To evaluate the performance of this approach, we consider standard uncertainty quantification benchmark problems including flow in heterogeneous media defined in terms of limited data-driven permeability realizations. The performance of the surrogate model developed is very good even though there is no underlying structure shared between the input (permeability) and output (flow/pressure) fields as is often the case in the image-to-image regression models used in computer vision problems. Studies are performed with an underlying stochastic input dimensionality up to $4,225$ where most other uncertainty quantification methods fail.  Uncertainty propagation tasks are considered and the predictive output Bayesian statistics are compared to those obtained with Monte Carlo estimates.
\end{abstract}


\end{frontmatter}

\section{Introduction}
\label{sec:intro}
Uncertainty in complex systems arises from model error and model parametrization, unknown/incomplete material properties, boundary conditions or forcing terms, and other. Uncertainty propagation takes place by reformulating the problem of interest as a system of stochastic partial differential equations (SPDEs). Solution of such problems often needs to rely on the solution of the deterministic problem at a finite number of realizations of the random input using Monte Carlo sampling, or collocation methods. 
Considering the computational cost of solving complex multiscale/multiphysics deterministic problems, one often relies on Bayesian surrogate models that are trained with only a small number of deterministic solution runs while at the same time they are capable of capturing the epistemic uncertainty introduced from the limited training data~\cite{bilionis2016bayesian}.

For realistic problems in science and engineering, we only have access to limited number (e.g. $100$ or so) of deterministic simulation runs. Vanilla Monte Carlo for uncertainty propagation is thus hopeless. A dominant solution is to train a surrogate model using the limited simulation-based training data, and then perform prediction and uncertainty propagation tasks using the surrogate instead of solving the actual PDEs.
Unfortunately most existing surrogate models have difficulty scaling to high-dimensional problems, such as the ones based on Gaussian processes (GP)~\cite{bilionis2013multi, bilionis2012multi} or generalized polynomial chaos expansions  (gPC~\cite{xiu2002wiener}).  High dimensionality often arises from the discretization of properties with small correlation lengths (e.g. permeability in heterogeneous media flows), random distributed sources or force input fields with multiple scales~\cite{torquato2013random}.

To alleviate the curse of stochastic input dimensionality, we usually assume that the given input data lie on an embedded non-linear manifold within the higher dimensional space. This intrinsic dimensionality is captured by dimensionality reduction techniques~\cite{van2009dimensionality}, such as the Karhunen-Lo\`eve expansion (KLE),  t-SNE~\cite{maaten2008visualizing}, auto-encoders~\cite{hinton2006reducing}, probabilistic methods like variational auto-encoders~\cite{kingma2013auto}, Gaussian process latent variable models (GP-LVM)~\cite{lawrence2004gaussian}, and many more. Most  dimensionality reduction models are unsupervised learning problems that do not explicitly take the regression task into account. Thus the classical approach to uncertainty quantification is to first reduce the dimensionality of the input to obtain a low-dimensional latent representation of the input field, then to built a regression model from this latent representation to the output. This approach is certainly not efficient is particular when the map from the latent representation to the physical space of the input data is not available. In~\cite{bilionis2013multi}, KLE was used for dimensionality reduction of the permeability field and then GP was performed as independent task for Bayesian regression. In~\cite{grigo2017bayesian}, this approach was taken one step further with the  probabilistic mappings from input to latent space and from latent space to output being modeled by generalized linear models both trained simultaneously end-to-end using stochastic variational inference instead of performing the unsupervised and supervised/regression tasks separately. 

One of the essential upcoming approaches for handling high-dimensional data is to learn the latent input representation automatically by supervision with the output in regression tasks. This is the central idea of deep neural networks~\cite{bengio2009learning}, especially convolutional neural networks (CNNs)~\cite{krizhevsky2012imagenet} which stack (deeper) layers of linear convolutions with nonlinear activations to automatically extract the multi-scale features or concepts from high-dimensional input~\cite{zeiler2014visualizing}, thus alleviating the hand-craft feature engineering, such as searching for the right set of basis functions, or relying on expert knowledge.

However, the general perspective for using deep neural networks~\cite{lecun2015deep, hinton2006fast} in the context of surrogate modeling is that physical problems in uncertainty quantification (UQ) are not big data problems, thus not suitable for addressing them with deep learning approaches. However, we argue otherwise in the sense that each simulation run generates large amount of data which potentially reveal the essential characteristics about the underlying system. In addition, even for a relatively small dataset, deep neural networks show unique generalization property~\cite{zhang2016understanding, dziugaite2017computing}. These are typically over-parameterized models (hundreds and thousands of times more parameters than training data), but they do not overfit, i.e. the test error does not grow as the network parameters increase. Deep learning has been explored as a competitive methodology across fiels such as fluid mechanics~\cite{kutz2017deep,DBLP:journals/corr/abs-1711-04315}, hydrology~\cite{marccais2017prospective}, bioinformatics~\cite{min2017deep}, high energy physics~\cite{baldi2014searching} and other.

This unique generalization behavior makes it possible to use deep neural networks for surrogate modeling. They are capable to capture the complex nonlinear mapping between high-dimensional input and output due to their expressiveness~\cite{raghu2016expressive}, while they only use small number of data from simulation runs.
In addition, there has been  a resurgence of interest in putting deep neural network under a formal Bayesian framework. Bayesian deep learning~\cite{mackay1992practical, neal2012bayesian, gal2016dropout, blundell2015weight, liu2016stein, kingma2015variational, hernandez2015probabilistic, louizos2017multiplicative, louizos2017bayesian} enables the network to express its uncertainty on its predictions when using a small number of training data. The Bayesian neural networks can quantify the predictive uncertainty by treating the network parameters as random variables, and perform Bayesian inference on those uncertain parameters conditioned on limited observations.

In this work, we mainly consider surrogate modeling of physical systems governed by stochastic partial differential equations with high-dimensional stochastic input such as flow in random  porous media~\cite{CHAN2018493}. The spatially discretized stochastic input field and the corresponding output fields are high-dimensional. We adopt an end-to-end image-to-image regression approach for this challenging surrogate modeling problem. More specifically, a fully convolutional encoder-decoder network is designed to capture the complex mapping directly from the high-dimensional input field  to the output fields without using any explicit intermediate dimensionality reduction method. To make the model more parameter efficient and compact, we use DenseNet to build the feature extractor within the encoder and decoder paths~\cite{huang2016densely}. Intuitively, the encoder network extracts the multi-scale features from the input data  which are used by the decoder network to reconstruct the output fields. In similarity with problems in computer vision, we treat the input-output map as an image-to-image map. To account for the  limited training data and endow the network with uncertainty estimates, we further treat the convolutional encoder-decoder network to be Bayesian and scale a recently proposed approximate inference method called Stein Variational Gradient Descent to modern deep convolutional networks. We will show that the methodology can learn a Bayesian surrogate for a problem with an intrinsic dimensionality of $50$, achieving promising results on both predictive accuracy and uncertainty estimates using as few as $32$ training data. More importantly we develop a surrogate for the case of $4225$ dimensionality using $512$ training data. We also show these uncertainty estimates are well-calibrated using a reliability diagram.
To this end, we believe that Bayesian neural networks are strong candidates for surrogate modeling and uncertainty propagation in high-dimensional problems with limited training data.

The remaining of the paper is organized as follows: In Section~\ref{sec:surrogate_image_regression}, we present the problem setup for surrogate modeling with high-dimensional input and the proposed approach in treating it as an image regression problem.  We then introduce the CNNs and the encoder-decoder network used in our model. In Section~\ref{sec:bayesnn}, we present the Bayesian formulation of neural networks and a non-parametric variational method for the underlying challenging approximate inference task. In Section~\ref{sec:implementationAndresults}, we provide implementation details and show the performed experiments on a porous media flow problem. We finally conclude and discuss the various unexplored research directions in Section~\ref{sec:conclusions}.



\section{Methodology}
\label{sec:surrogate_image_regression}

\subsection{Surrogate Modeling as Image-to-Image Regression}
\label{sec:image_regression}
The physical systems considered here are modeled by stochastic PDEs (SPDEs) with solutions $\vc y(\vc s, \vc x(\vc s))$, i.e. the model response $\vc y \in \mb{R}^{d_y}$ at the spatial location $\vc s \in \mc S \subset \mathbb{R}^{d_s}$ ($d_s = 1, 2, 3$), with one realization $\vc x(\vc s)$ of the random field $\{\vc x(\vc s, \omega), \vc s \in \mc S, \omega \in \Omega\}$, where $\mc S$ is the index set and $\Omega$ is the sample space.  The formulation allows for multiple input channels (i.e. $d_x>1$) even though our interest here is on one input property represented as a vector random field. This random field appears in the coefficients of the SPDEs, and is used to model material properties, such as the permeability or porosity fields in geological media flows.
We assume the computer simulation for the physical systems is performed over a given set of spatial grid locations $\mc S = \{\vc s_1, \cdots, \vc s_{n_s} \}$ (e.g. mesh nodes  in finite element methods). In this case, the random field $\vc x$ is discretized over the fixed grids $\mc S$, thus is equivalent to a high-dimensional random vector, denoted as $\bmx$, where $\bmx \in \mc X \subset \mathbb{R}^{d_x n_s}$. The corresponding response $\vc y$ is solved over $\mc S$, thus can be represented as a vector $\bmy \in \mc Y \subset \mb{R}^{d_y n_s}$.

With the discretization described above and assuming for simplicity fixed boundary and initial conditions and source terms as appropriate, we consider the computation simulation as a black-box mapping of the form:
\begin{equation}
\bmeta: \mc X \to \mc Y.
\end{equation}

In order to tackle the limitations of using the deterministic computationally expensive simulator for uncertainty propagation, a {\em surrogate model} $\bmy = \bmf(\bmx, \bmtheta)$ is trained using \textit{limited} simulation data $\mc D = \{\bmx^i, \bmy^i\}_{i=1}^N$, to approximate the `ground-truth' simulation-induced function $\bmy = \bmeta(\bmx)$, where $\bmtheta$ are the model parameters, and $N$ is the number of simulation runs (number of training simulation-based data).


Let us consider that the PDEs governing the physical system described above are solved over 2D regular grids of $H \times W$, where $H$ and $W$ denote the number of grid points in the two axes of the spatial domain (height and width), and $n_s = H \cdot W$. It is very natural to organize the simulation data as an image dataset $\mc D = \{\bmx^i, \bmy^i\}_{i=1}^N$, where $\bmx^i \in \mathbb{R}^{d_x \times H \times W}$ is one input field realization, and $\bmy^i \in \mathbb{R}^{d_y \times H \times W}$ is the simulated steady-state output fields for $\bmx^i$ discretized over the same grids. Here $d_x$, $d_y$ are the number of dimensions for the input $\vc x$ and the output $\vc y$ at one location. These are treated herein as the number of channels in input and output images, similar to RGB channels in natural images. It is easy to generalize to the 3D spatial domain by adding an extra \textit{depth} axis to the images, e.g. $\bmx^i \in \mathbb{R}^{d_x \times D \times H \times W}$, and $\bmy^i \in \mathbb{R}^{d_y \times D \times H \times W}$.

Therefore, we transform the surrogate modeling problem to an image-to-image regression problem, with the regression function as
\begin{equation}
\bmeta: \mathbb{R}^{d_x \times H \times W} \to \mathbb{R}^{d_y \times H \times W}.
\end{equation}

In distinction from an image classification problem which requires image-wise prediction, the image regression problem is concerned with pixel-wise predictions, e.g. predicting the depth of each pixel in an image, or in our physical problem, predicting the output fields at each grid point. Such problems have been intensively studied within the computer vision community by leveraging the rapid recent progress of convolutional neural networks (CNNs), such as AlexNet~\cite{krizhevsky2012imagenet}, VGG~\cite{simonyan2014very}, Inception~\cite{szegedy2016rethinking}, ResNet~\cite{he2016deep}, DenseNet~\cite{huang2016densely}, and many more. 
A common model design pattern for semantic segmentation~\cite{ronneberger2015u} or depth regression~\cite{eigen2014depth} is the \textit{encoder-decoder} architecture. The intuition behind regression between two high-dimensional objects is to go through a coarse-refine process, i.e. to reduce the spatial dimension of the input image to high-level coarse features using an encoder, and then recover the spatial dimension by refining the coarse features through a decoder.
One of the characteristics shared by those vision tasks is that  the input and output images \textit{share the underlying structure}, or they are different renderings of the same underlying structure~\cite{pix2pix2016}. However, for our surrogate modeling tasks, the input and output images appear to be \textit{quite different}, due to the complex physical influences (defined by PDEs) of the random input field,  forcing terms and boundary conditions on the system response. This was after all the reason of pursuing the training of a surrogate model that avoids the repeated solution of the PDEs for different input realizations. Surprisingly, as we will discuss later on in this paper, the encoder-decoder network still works very well.

\remark{The training data for the surrogate model of interest here include the realizations of the random input field and  the corresponding multi-output obtained from simulation. Of interest is to address problems with limited training data sets considering the high-computational cost of each simulation run.	However, note that  key UQ tasks include the ability to predict the system response and our confidence on it using input realizations (testing dataset) consistent with the given training data but also the computation of the statistics of the response induced by the random input. Both of these tasks require the availability of a high-number of input data sets (e.g. $500$ data points for testing, and $10^4$ input data points for Monte Carlo calculation of the output statistics). The problem of generating more input realizations using only the training dataset  is the solution of a \textit{generative model} problem. There has been significant progress in recent years in the topic, such as the generative adversarial networks (GANs)~\cite{goodfellow2014generative} and its ever \textit{exploding} variants, variational auto-encoders (VAEs)~\cite{kingma2013auto}, autoregressive models like PixelCNN~\cite{van2016conditional}, PixelRNN~\cite{oord2016pixel}, and other. However, note that in this work our focus is on the image-to-image mapping and its performance on uncertainty quantification tasks. We will thus assume that enough input samples are provided both for testing and output statistics calculation even though only a small 
dataset will be used for training. In our examples in Section~\ref{sec:implementationAndresults}, synthetic log-permeability datasets are generated by sampling a Gaussian random field with an exponential kernel. The output for each permeability sample is generated using a deterministic simulator.}

\subsection{Dense Convolutional Encoder-Decoder Networks}
\label{sec:conv_enc_dec}

In this subsection, we briefly introduce a state-of-the-art CNN architecture called DenseNet~\cite{huang2016densely} and fully convolutional encoder-decoder networks developed in computer vision, and then present how to utilize these advances to build our baseline network for surrogate modeling in uncertainty quantification.

\subsubsection{Densely Connected Convolutional Networks}
DenseNet~\cite{huang2016densely} is a recently proposed CNN architecture which extends the ideas of ResNet~\cite{he2016deep} and Highway Networks~\cite{srivastava2015training}  to create dense connections between all layers, so as to improve the information (gradient) flow through the network for better parameter efficiency. 

Let $\bmx_l$ be the output of the $l^{th}$ layer. Traditionally CNNs pass the output of one layer only to the input of the next layer, i.e. $\bmx_l = h_l(\bmx_{l-1})$, where $h_l$ denotes the nonlinear function of the $l^{th}$ hidden layer. In current CNNs, $h$ is commonly defined as a composition of Batch Normalization~\cite{ioffe2015batch} (BatchNorm), Rectified Linear Unit~\cite{glorot2011deep} (ReLU) and Convolution (Conv) or transposed convolution (ConvT)~\cite{2016arXiv160502688short}.
ResNets~\cite{he2016deep} create an additional identity mapping that bypasses the nonlinear layer, i.e. $\bmx_l = h_l(\bmx_{l-1}) + \bmx_{l-1}$. In this way, the nonlinear layer only needs to learn a residual function which facilitates the training of \textit{deeper} networks. 

DenseNets~\cite{huang2016densely} introduce connections from any layer to all subsequent layers, i.e. $\bmx_l = h_l([\bmx_{l-1}, \bmx_{l-2}, \cdots, \bmx_0])$. To put it in another way, the input features of one layer are concatenated to the output features of this layer and this serves as the input features to the next layer. Assume that the input image has $K_0$ channels, and each layer outputs $K$ feature maps, then the $l^{th}$ layer would have input with $K_0 + (l-1) \cdot K$ feature maps, i.e. the number of feature maps in DenseNet grows linearly with the depth. $K$ is here referred to as the \textit{growth rate}.

\begin{figure}[ht!]
	\centering
	\begin{subfigure}[t]{0.6\textwidth}
		\label{subfig:dense_blockA}
		\centering
		\includegraphics[width=\textwidth]{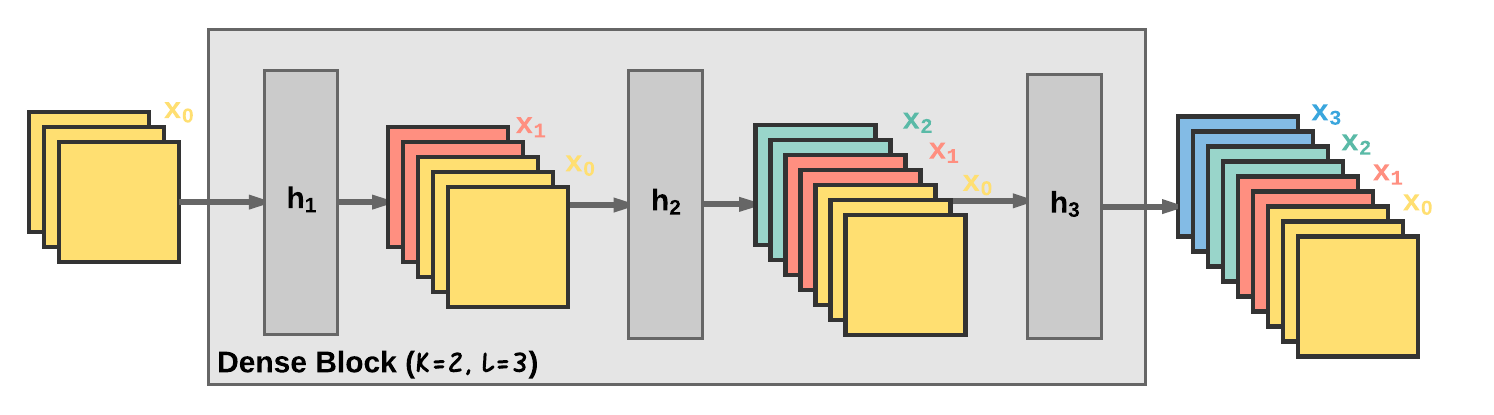}
		\caption{Dense block.}
	\end{subfigure}
	~ 
	\begin{subfigure}[t]{0.38\textwidth}
		\label{subfig:dense_layerA}
		\centering
		\includegraphics[width=0.95\textwidth]{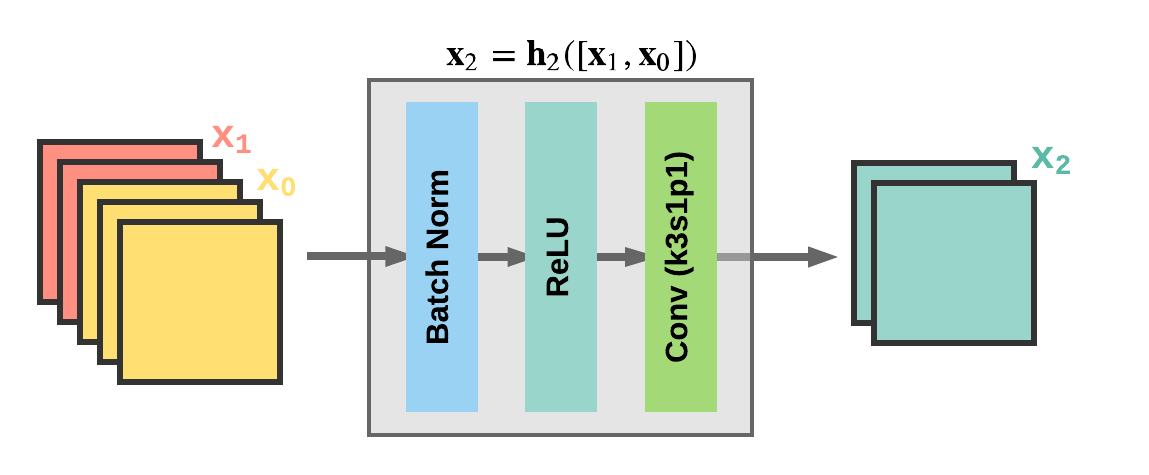}
		\caption{The second layer.}
	\end{subfigure}
	\caption{(a) A dense block contains $L=3$ layers $h_1, h_2, h_3$ with growth rate $K=2$. (b) The second layer $h_2$ of the dense block, where $\bmx_2 = h_2([\bmx_1, \bmx_0])$ is its output feature map. Notice that the input to the third layer is the concatenation of the output and input features of $h_2$, i.e. $[\bmx_2, \bmx_1, \bmx_0]$. As is often the case, each layer is composed of Batch Normalization~\cite{ioffe2015batch} (BatchNorm), Rectified Linear Unit~\cite{glorot2011deep} (ReLU) and Convolution (Conv). The convolution kernel has kernel size $k=3$, stride $s=1$ and zero padding $p=1$, which keep the size of the feature maps the same as the input.}
	\label{fig:dense_block}
\end{figure}

For image regression based on encoder-decoder networks, downsampling and upsampling are required to change the size of feature maps, which makes concatenation of feature maps unfeasible. Dense blocks and transition layers are introduced to solve this problem and modularize the network design. A \textit{dense block} contains multiple densely connected layers whose input and output feature maps are of the same size. It contains two design parameters, namely the number $L$ of layers within and the growth rate $K$ for each layer. An illustration of the dense block is in Fig.~\ref{fig:dense_block}.

\begin{figure}[ht!]
	\centering
	\begin{subfigure}[b]{0.6\textwidth}
		\centering
		\includegraphics[width=0.95\textwidth]{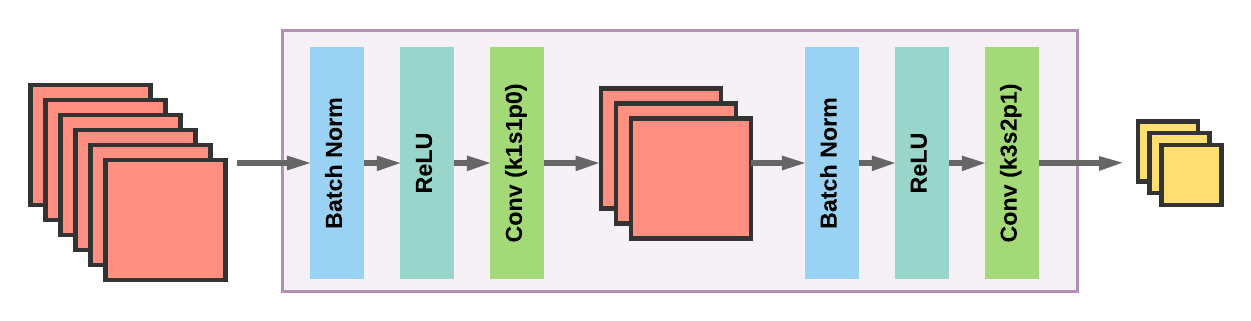}
		\caption{Encoding layer.}
	\end{subfigure}
	~ 
	\begin{subfigure}[b]{0.6\textwidth}
		\centering
		\includegraphics[width=0.95\textwidth]{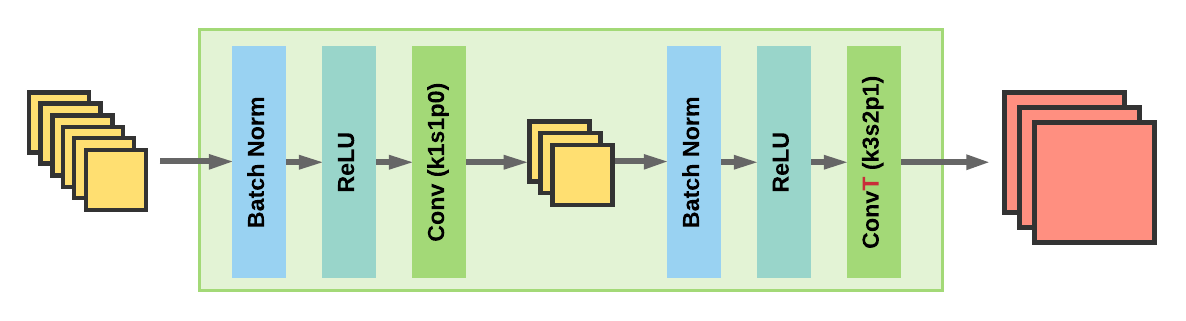}
		\caption{Decoding layer.}
	\end{subfigure}
	\caption{Both (a) encoding layer and (b) decoding layer contain two convolutions. The first convolution  reduces the number of feature maps while keeps  their size the same using a kernel with parameters $k=1, s=1, p=0$; the second convolution changes the size of the feature maps but not their number using a kernel $k=3, s=2, p=1$. The main difference between (a) and (b) is in the type of the second convolution layer, which is Conv and ConvT respectively, for downsampling and upsampling. Note that no pooling is used at transition layers for maintaining the location information. The colored feature maps used here are independent from the feature maps with the same color shown in other figures.}
	\label{fig:transition_layers}
\end{figure}

Transition layers are used to change the size of feature maps and reduce their number between dense blocks. More specifically, the encoding layer typically halfs the size of feature maps, while the decoding layer doubles the feature map size. Both of the two layers reduce the number of feature maps. This is illustrated in Fig.~\ref{fig:transition_layers}.

\subsubsection{Fully Convolutional Networks}

The fully convolutional networks (FCNs)~\cite{long2015fully} are  extensions of CNNs for pixel-wise prediction, e.g. semantic segmentation. FCNs replace the fully connected layers in CNNs with convolution layers, add upsampling layers in the end to recover the input spatial resolution, and introduce the skip connections between feature maps in downsampling and upsampling path to recover finer information lost in the downsampling path. Most of the recent work focuses in improving the upsampling path and increase the connectivity within and between upsampling and downsampling paths. U-nets~\cite{ronneberger2015u} extend the upsampling path as symmetric to the downsampling path and add skip connections between each size of feature maps in the downsampling and upsampling paths. Within SegNets~\cite{badrinarayanan2015segnet}, the decoder uses pooling indices computed in the max-pooling step of the corresponding encoder to perform non-linear upsampling. Fully convolutional DenseNets~\cite{jegou2017one} extend DenseNets to FCNs, which are closest to our network design but with several differences. We keep all the feature maps of a dense block concatenated so far before passing to the transition layers, while they only keep the output feature maps of the last convolution layer within the dense block. The feature maps explosion problem is addressed by the first convolution layer within the transition layer. Besides that we do not use skip connections between encoding and decoding paths because of the weak correspondence between the input and output images. We also do not use max-pooling for encoding layers, instead we use convolution with stride $2$.

\subsection{Network architecture: DenseED}
\begin{figure}[h]
	\centering
	\includegraphics[width=0.8\textwidth]{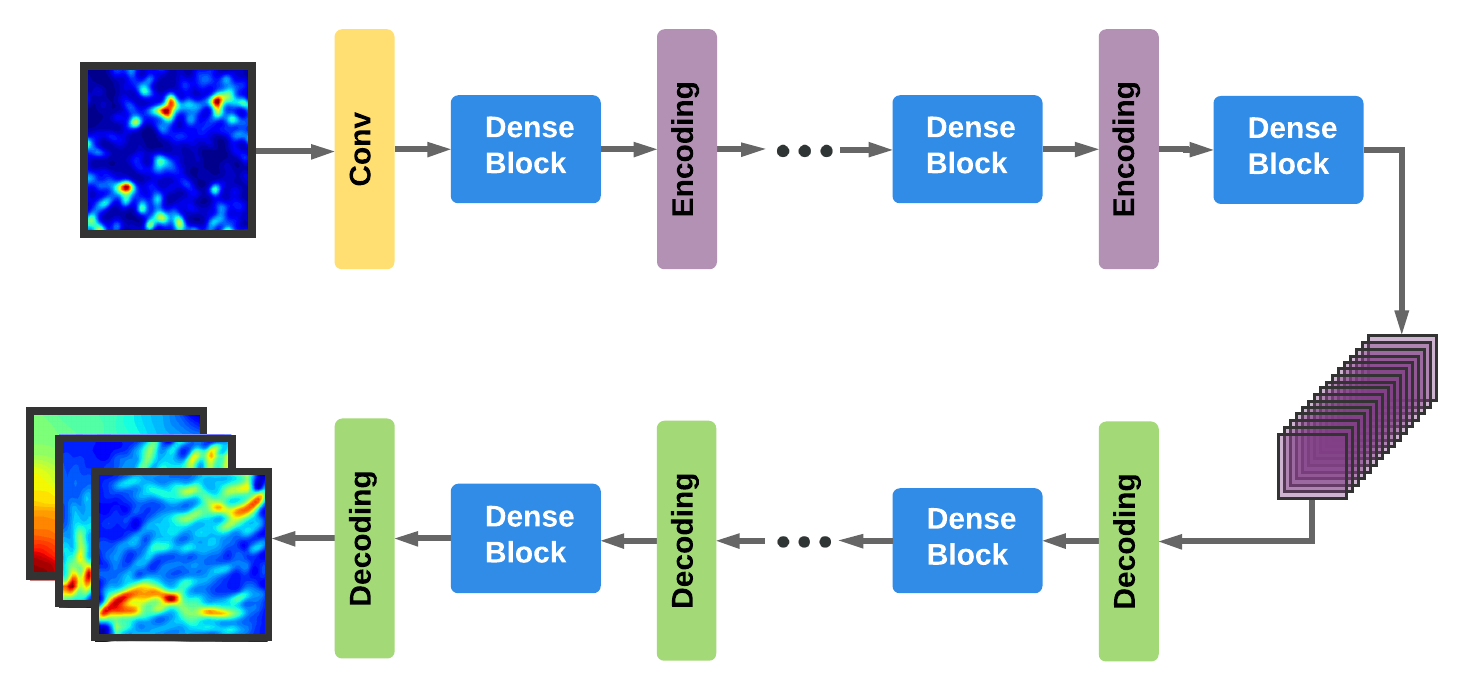}
	\caption{Network architecture: DenseED.}
	\label{fig:dense_ed}
\end{figure}

We follow the fully convolutional networks (FCNs)~\cite{long2015fully} for image segmentation without using any fully connected layers, and encode-decoder architecture similar to U-net~\cite{ronneberger2015u} and SegNet~\cite{badrinarayanan2015segnet} but without the concatenation of feature maps between the encoder paths and decoder paths. Furthermore, we adapt the DenseNet~\cite{huang2016densely} structure into the encoder and decoder networks. After extensive hyperparameter and architecture search, we arrived at a baseline dense convolutional encoder-decoder network, called \texttt{DenseED}, similar to the network proposed in~\cite{jegou2017one} but with noticeable differences as stated above and shown in Fig.~\ref{fig:dense_ed}.

In the encoding path, the input field realizations are fed into the first convolution layer with large kernel size $k=7$, stride $s=2$ and zero padding $p=2$. Then the extracted feature maps are passed through an alternative cascade of dense blocks and encoding layers as introduced in Figs.~\ref{fig:dense_block} and~\ref{fig:transition_layers}. The dense block after the last encoding layer outputs the high-level coarse feature maps extracted from the input, as shown in purple at the right end of the network in Fig.~\ref{fig:dense_ed}, which are subsequently fed into the decoder path. The decoding network consists of an alternation of dense blocks and decoding layers, with the last decoding layer directly leading to the prediction of the output fields.


\subsection{Network architecture engineering and hyperparameter search}
\label{sec:hyper_search}
Network \textit{architecture engineering} and hyperparameter search are among the main challenges and source of innovations in deep learning, mostly problem-specific and empirical. The general network architecture is introduced in Section~\ref{sec:conv_enc_dec}, which is based on the recent development of neural network design for image segmentation. 
For our image regression problem, the main design considerations  include the following:
\begin{itemize}
	\item Downsampling layers: convolution or pooling;
	\item Upsampling layers: bilinear upsampling or transposed convolution;
	\item Smallest spatial dimensions of feature maps: this is determined by the number of downsampling layers; 
	\item Add or not of skip connections between the encoding and decoding paths;
	\item Kernel of convolution layers, kernel size $k$, stride $s$, zero padding $p$;
	\item Number of layers $L$ and growth rate $K$ within each dense block;
	\item Regularizations: weight decay, batch normalization, dropout, etc;
	\item Optimizer: stochastic gradient descent algorithms and their variants, such as Adam, Adagrad, RMSprop, and others;
	\item Training hyperparameters: batch size, learning rate and its scheduler.
\end{itemize}

 The details of architecture search and hyperparameter selection for the particular problem considered are presented in~\ref{sec:DesignTests} where we also report various experiments using DenseED for surrogate modeling with limited training data. No overfitting was observed in our calculations and the obtained results were quite good. This is an intriguing and active research  topic in the deep learning community~\cite{zhang2016understanding}.

In our non-Bayesian calculations, we have considered $\mc L_2$ or $\mc L_1$ regularized MSE training loss function. 
Given an input image $\bmx$, and a target image $\bmy$, the prediction $\bmf(\bmx, \bmw)$, the regularized MSE loss is
\begin{equation}
\label{eq:loss_mse_regularized}
L(\bmf(\bmx, \bmw), \bmy) = \frac{1}{n} \sum_{i=1}^n \Big( \bmf_i - \bmy_i \Big)^2 + \alpha \Omega(\bmw),
\end{equation}
where the penalty function $\Omega(\bmw)=\frac{1}{2} \bmw^\top \bmw$ for $\mc L_2$ regularization, and $\Omega(\bmw) = \norm{\bmw}_1 = \sum_i |w_i|$ for $\mc L_1$ regularization, and $n=C_{out} \cdot H_{out} \cdot W_{out}$ is the number of pixels in all channels of one output image. $\bmw$ denotes all the parameters in the network. For our case, it includes the kernel weights in all the convolution and transposed convolution layers (no bias is used in convolutional kernel), the scale and shift parameters in all the batch normalization layers. See Section~\ref{sec:non_bayes_surrogate} for an example of the fully convolutional encoder-decoder network used for the Darcy flow problem.
Note that $\mc L_2$ regularization is implemented in PyTorch optimizers by specifying \textit{weight decay}, which is $\alpha$ in Eq.~(\ref{eq:loss_mse_regularized}).

The network architecture selected for the non-Bayesian model is the same as that used for our Bayesian model introduced next.

\remark{In the encoder-decoder network, the batch normalization layer used after each convolutional layer can also be considered as an effective regularizer\footnote{\url{https://openreview.net/forum?id=BJlrSmbAZ&noteId=BJlrSmbAZ}}. It is commonly adopted nowadays in deep convolutional networks\footnote{\url{https://github.com/pytorch/vision/tree/master/torchvision/models}} replacing dropout\footnote{\url{https://www.reddit.com/r/MachineLearning/comments/5l3f1c/d\_what_happened_to_dropout/}}. }



\section{Bayesian Neural Networks}
\label{sec:bayesnn}
Consider a deterministic neural net 
$\bmy = \bmf(\bmx, \bmw)$ with input $\bmx$, output $\bmy$,  and all parameters $\bmw$ including the weights and biases.
\footnote{\url{http://pytorch.org/docs/master/nn.html?ht=conv2d##torch.nn.functional.conv2d}}
Bayesian neural networks (BNNs) treat the parameters $\bmw$ as random variables instead of deterministic unknowns to account for epistemic uncertainty induced by lack of training data. Besides that, usually additive noise $\bmn$ is introduced to model the aleatoric uncertainty which can not be reduced by having more observations, also to make the probabilistic model have an explicit likelihood depending on the noise distribution, i.e.
\begin{equation}
	\bmy = \bmf(\bmx, \bmw) + \bmn,
\end{equation}
where $\bmf(\bmx, \bmw)$ is the output of a neural network with the uncertain $\bmw$, and $\bmn$ is the additive noise.

\subsection{Sparsity inducing prior on weights}
Given the large amount of `un-interpretable' parameters $\bmw$ in a deep neural net, there are not many choices of priors. But the demand for compression~\cite{han2015deep,louizos2017bayesian} of the neural net for lower memory and computation cost calls for sparsity promoting priors. 
We assume a fully factorized Gaussian prior with zero mean and Gamma-distributed precision $\alpha$ on parameters $\bmw$
\begin{equation}
	\label{eq:prior_w}
	p(\bmw \mid \alpha) = \mc N(\bmw \mid 0, \alpha^{-1}\bmI), \qquad p(\alpha) = \mathrm{Gamma}(\alpha \mid a_0, b_0).
\end{equation}
This results in a student's t-prior for $\bmw$, which has heavy tails and more mass close to zero.

\subsection{Additive Noise Model}

Additive noise can be considered of the following form:
\begin{itemize}
	\item Output-wise: $\bmn = \sigma \bmepsilon$, same for all output pixels;
	\item Channel-wise: $\bmn = [\sigma_1 \bmepsilon_1, \cdots,  \sigma_{d_y} \bmepsilon_{d_y}]$, same across each of the $d_y$ output channels/fields;
	\item Pixel-wise: $\bmn = \bmsigma \odot \bmepsilon$, distinct for each output pixel.
\end{itemize}
Here, $\sigma$, $\{\sigma_i\}_{i=1}^{d_y}$ are scalars, $\bmsigma$ is a field with the same dimension as the output $\bmy$ and $\odot$ denotes the element-wise product operator. In this work, we have considered both Gaussian noise, $\bmepsilon \sim \mc N (\bmzero, \bmI)$ and Laplacian noise, $\bmepsilon \sim \mathrm{Laplace} (\bmzero, \bmI)$\footnote{\url{http://homepages.inf.ed.ac.uk/rbf/CVonline/LOCAL_COPIES/VELDHUIZEN/node11.html}}. In the numerical results discussed in Section~\ref{sec:implementationAndresults}, we concentrate in the the output-wise and channel-wise cases above. We treat the noise precision $\beta = 1 / \sigma^2$ as a random variable with a conjugate prior $p(\beta) = \mathrm{Gamma}(\beta \mid a_1, b_1)$. For the generated training data, \textit{a priori} we assume that the noise variance (also known as nugget~\cite{gramacy2012cases}) to be very small e.g. $10^{-6}$. Thus the values $a_1=2, b_1=2\cdot 10^{-6}$ provide a good initial guess for the prior hyperparameters.

\remark{
We can also model the noise varying with input (pixel-wise noise model), resulting in a \textit{heteroscedastic} noise model, i.e. $\bmn(\bmx, \bmw) = \sigma(\bmx, \bmw) \bmepsilon$ or $\bmn(\bmx, \bmw) = \bmsigma(\bmx, \bmw) \odot \bmepsilon$. Again $\bmepsilon$ can be Gaussian or Laplacian. The heteroscedastic noise~\cite{nix1994estimating, le2005heteroscedastic,kendall2017uncertainties} can be implemented as extending the output of the neural net as:
\begin{equation}
	\label{eq:expand_noise}
	[\bmf(\bmx, \bmw), \sigma^2(\bmx, \bmw)] ~\mathrm{or} ~ [\bmf(\bmx, \bmw), \bmsigma^2(\bmx, \bmw)].
\end{equation}
The output of the system becomes $\bmy = \bmf(\bmx, \bmw) + \bmn(\bmx, \bmw)$.
The pixel-wise case $\bmsigma^2(\bmx, \bmw)$ may help capture large variations for example near  discontinuous regions of the output. 
In practice, we apply a \textit{softplus} transformation to the second part of the output of the neural net to enforce the positive variance constraint, i.e. $\sigma^2 = \log(1 + \mathrm{exp}(\cdot)) + \mathrm{eps}$, where $\mathrm{eps}=10^{-10}$ for numerical stability.}

\subsection{Stein Variational Gradient Descent (SVGD)}

Approximate inference for Bayesian deep neural network is a daunting task because of the large number of uncertain parameters, e.g. tens or hundreds of millions in modern deep networks. In our surrogate problem, the task is to find a high-dimensional posterior distribution  over millions of random variables using less than hundreds or thousands of training data. As reviewed in Section~\ref{sec:intro}, most of variational inference methods~\cite{blei2017variational} restrict the approximate posterior within certain parametric variational family, while sampling-based methods are slow and difficult to converge. Here we adopt a recently proposed non-parametric variational inference method called stochastic variational gradient descent (SVGD)~\cite{liu2016stein, liu2017stein} that is similar to standard gradient descent while maintaining the efficiency of particle methods.

For a \textit{prescribed} probabilistic model with \textit{likelihood} function $p(\bmy \mid \bmtheta, \bmx)$ and prior $p_0(\bmtheta)$, we are interested in Bayesian inference of the uncertain parameters $\bmtheta$, i.e. to find the posterior distribution $p(\bmtheta \mid \mathcal{D})$, where $\mathcal{D}$ denote the i.i.d. observations (training data), i.e. $\mathcal{D} = \{\bmx^i, \bmy^i\}_{i=1}^N$. For the BNNs with homoescedastic Gaussian noise case, $\bmtheta = \{\bmw, \beta\}$. 
Variational inference aims to approximate the target posterior distribution $p(\bmtheta \mid \mc D)$ with a variational distribution $q^*(\bmtheta)$ which lies in a restricted set of distributions $\mc Q$ by minimizing the KL divergence between the two, i.e.
\begin{equation*}
	q^*(\bmtheta) = \argmin_{q \in \mc Q} \mc{KL}(q(\bmtheta) \parallel p(\bmtheta \mid \mc D)) = \argmin_{q \in \mc Q} \E_q[\log q(\bmtheta) - \log \tilde{p}(\bmtheta \mid \mc D) +\log Z],\end{equation*}

where $\tilde{p}(\bmtheta \mid \mc D) = p(\mathcal{D} \mid \bmtheta) p_0(\bmtheta) = \prod_{i=1}^N p(\bmy^i \mid \bmtheta, \bmx^i) p_0(\bmtheta)$ is the unnormalized posterior, and $Z = \int \tilde{p}(\bmtheta) d\bmtheta$ is the normalization constant or model evidence, which is usually computationally intractable, but can be ignored when we optimize the KL divergence.

The variational family considered here is a set of distributions obtained by smooth transforms from an initial tractable distribution (e.g. the prior) represented in terms of particles. 
The transforms applied to each particle take the following form:
\begin{equation}
	\label{eq:transform}
	\bmT(\bmtheta) = \bmtheta + \epsilon \bmphi(\bmtheta),
\end{equation}
where $\epsilon$ is the step size, $\bmphi(\bmtheta) \in \mc F$ is the perturbation direction within a function space $\mc F$. When $\epsilon$ is small, $\bmT$ transforms the initial density $q(\bmtheta)$ to 
\begin{equation*}
	q_{[\bmT]}(\bmtheta) = q(\bmT^{-1}(\bmtheta)) |\mathrm{det}(\nabla\bmT^{-1}(\bmtheta))|.
\end{equation*}

Instead of using parametric form for the variational posterior, a particle approximation is used, i.e. a set of particles $\{\bmtheta^i\}_{i=1}^S$ with empirical measure $\mu_S(d\bmtheta) = \frac{1}{S}\sum_{i=1}^S \bmdelta(\bmtheta - \bmtheta^i) d\bmtheta$. We would like to have $\mu$ to weakly converge to the measure of the true posterior $\nu_p(d\bmtheta) = p(\bmtheta) d\bmtheta$. We apply the transform $\bmT$ to those particles, and denote the pushforward measure of $\mu$ as $\bmT \mu$.

The problem is to find out the direction to maximally decrease the KL divergence of the variational approximation and the target distribution, i.e. to solve the following functional optimization problem:
\begin{equation}
\max_{\bmphi \in \mc F} \Big\{ -\frac{d}{d\epsilon} \mc{KL}(\bmT \mu \parallel \nu_p) |_{\epsilon=0} \Big\}.
\end{equation}

It turns out that~\cite{liu2016stein}
\begin{equation}
-\frac{d}{d\epsilon} \mc{KL}(\bmT \mu \parallel \nu_p) |_{\epsilon=0} = \E_\mu [\mc T_p\bmphi],
\end{equation}
where $\mc T_p$ is called the Stein operator associated to the distribution $p$,
\begin{equation*}
\mc T_p\bmphi = \frac{\nabla \cdot (p \bmphi)}{p} = \frac{(\nabla p) \cdot \bmphi + p (\nabla \cdot \bmphi)}{p} =  (\nabla \log p) \cdot \bmphi + \nabla \cdot \bmphi.
\end{equation*}
The expectation $\E_\mu [\mc T_p\bmphi]$ evaluates the difference between $p$ and $\mu$, and its maximum is defined as the Stein discrepancy,
\begin{equation}
\mc S(\mu, p) = \max_{\bmphi\in \mc F} \E_\mu[\mc T_p \bmphi].
\end{equation}

It has been shown~\cite{liu2016kernelized} that when the functional space $\mc F$ is chosen to be the unit ball in a product reproducing kernel Hilbert space $\mc H$ with the positive kernel $k(\bmx, \bmx')$, the maximal direction to perturb (or the Stein discrepancy) has a closed-form solution,
\begin{equation*}
\bmphi^*(\bmtheta) \propto \E_{\bmtheta'\sim \mu} [\mc T_p^{\bmtheta'} k(\bmtheta, \bmtheta')] = \E_{\bmtheta'\sim \mu} [\nabla_{\bmtheta'} \log p(\bmtheta') k(\bmtheta, \bmtheta') + \nabla_{\bmtheta'} k(\bmtheta, \bmtheta') ].
\end{equation*}

Thus we have the following algorithm to transform an initial distribution $\mu_0$ to the target posterior $\nu_p$. 

\begin{algorithm}[h]
	\SetKwInOut{Input}{Input}
	\SetAlgoLined
	\Input{A set of initial particles $\{\bmtheta_0^i\}_{i=1}^S$, score function  $\nabla \log p(\bmtheta)$, kernel $k(\bmtheta, \bmtheta')$, step-size scheme $\{\epsilon_t\}$}
	\KwResult{A set of particles $\bmtheta^i$ that approximate the target posterior}
	\For{iteration $t$}{
		$\bmtheta^i_{t+1} \leftarrow  \bmtheta^i_t + \epsilon_t \bmphi(\bmtheta^i_t)$ \;
		$\bmphi(\bmtheta^i_t) = \frac{1}{S} \sum_{i=1}^S \Big[k(\bmtheta^j_t, \bmtheta^i_t) \nabla_{\bmtheta^j_t} \log p(\bmtheta^j_t) + \nabla_{\bmtheta^j_t} k(\bmtheta^j_t, \bmtheta^i_t)\Big]$
	}
	\caption{Bayesian inference by Stein Variational Gradient Descent.}
	\label{algo:svgd}
\end{algorithm}

This is an one-line algorithm, where the gradient $\bmphi(\bmtheta)$ pushes the particles towards the high posterior region by kernel smoothed gradient term $k(\cdot, \cdot) \nabla \log p$, while maintaining a degree of diversity by repulsive force term $\nabla k(\bmtheta, \bmtheta')$. When the number of particles becomes $1$, then the algorithm reduces to the MAP estimate of the posterior. 
This algorithm is implemented in PyTorch with GPU acceleration and scaled to our deep convolutional encoder-decoder network \texttt{DenseED}. 

Here we use one toy example to illustrate the idea of SVGD. We start with the $20$ particles from the Normal distribution $\mc N(-10, 1)$, and tranport the particles iteratively to the target Gaussian mixture distribution $0.8 \mc N(-2, 1) + 0.2 \mc N(2, 1)$ with Algorithm~\ref{algo:svgd}.
\begin{figure}[!h]
	\centering
	\begin{subfigure}{0.17\textwidth}
		\centering
		\includegraphics[width=\textwidth]{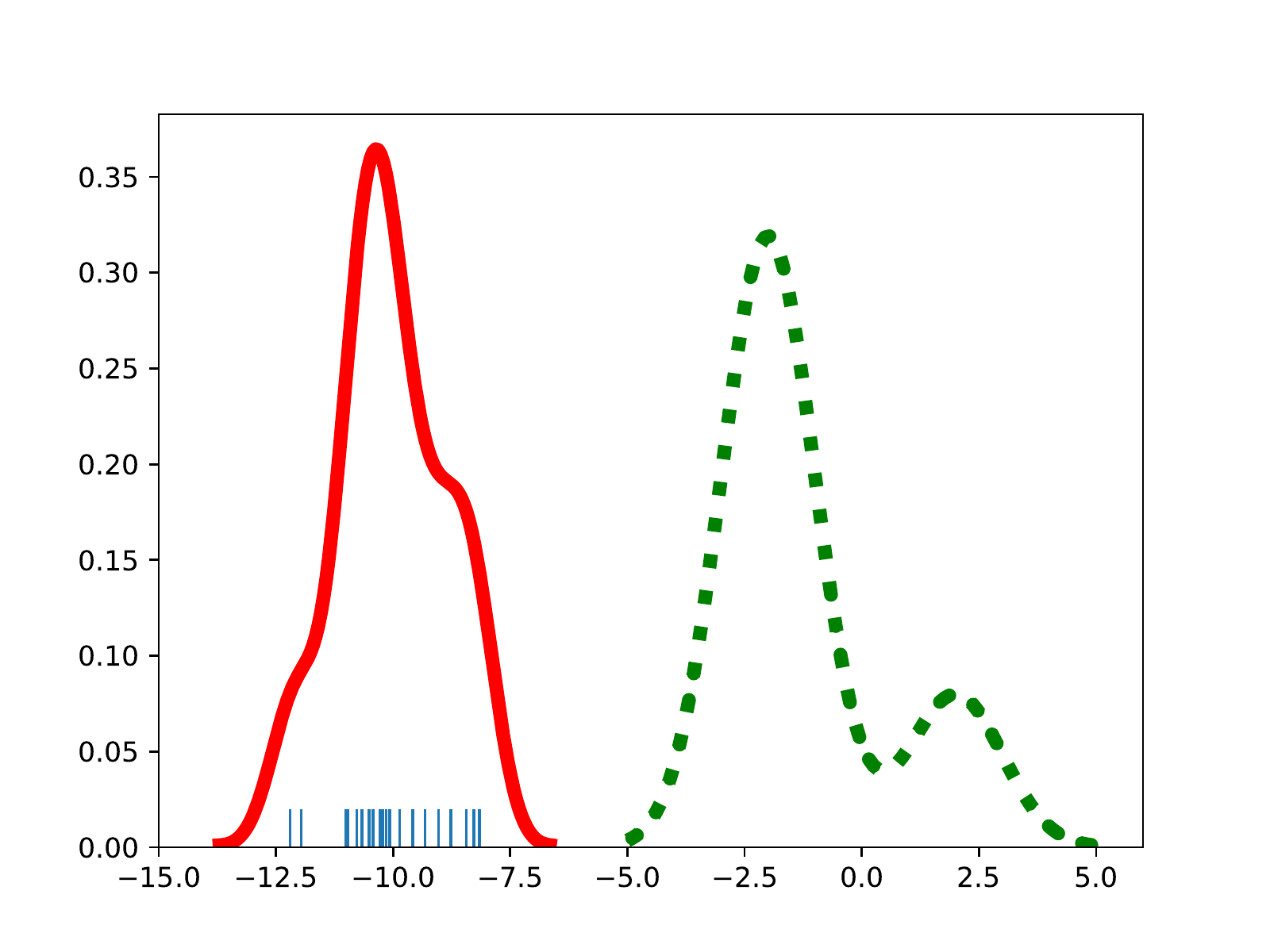}
		\caption{Iter $0$}
		
	\end{subfigure}
	~
	\begin{subfigure}{0.17\textwidth}
		\centering
		\includegraphics[width=\textwidth]{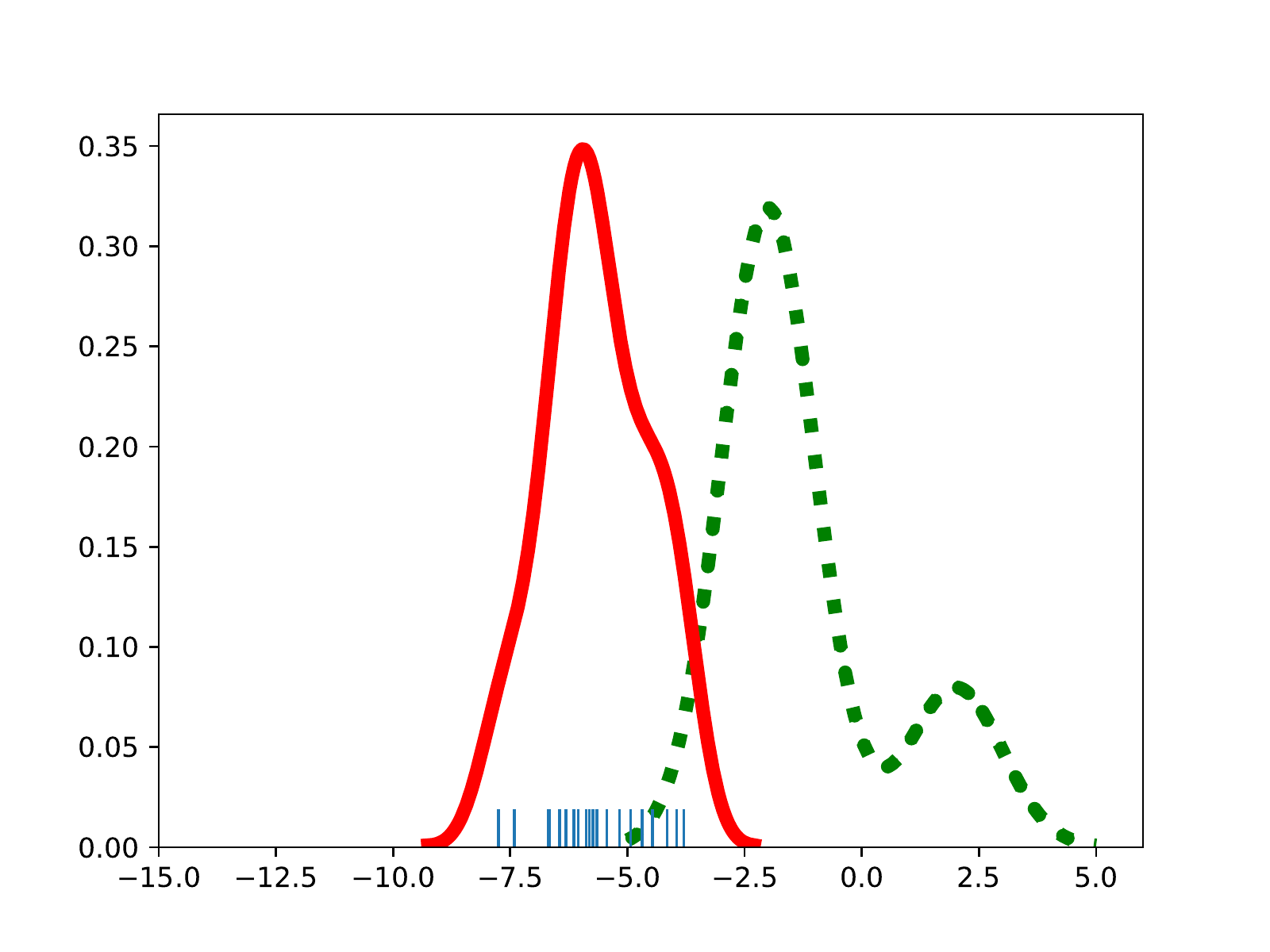}
		\caption{Iter $50$}
		
	\end{subfigure}
	~
	\begin{subfigure}{0.17\textwidth}
		\centering
		\includegraphics[width=\textwidth]{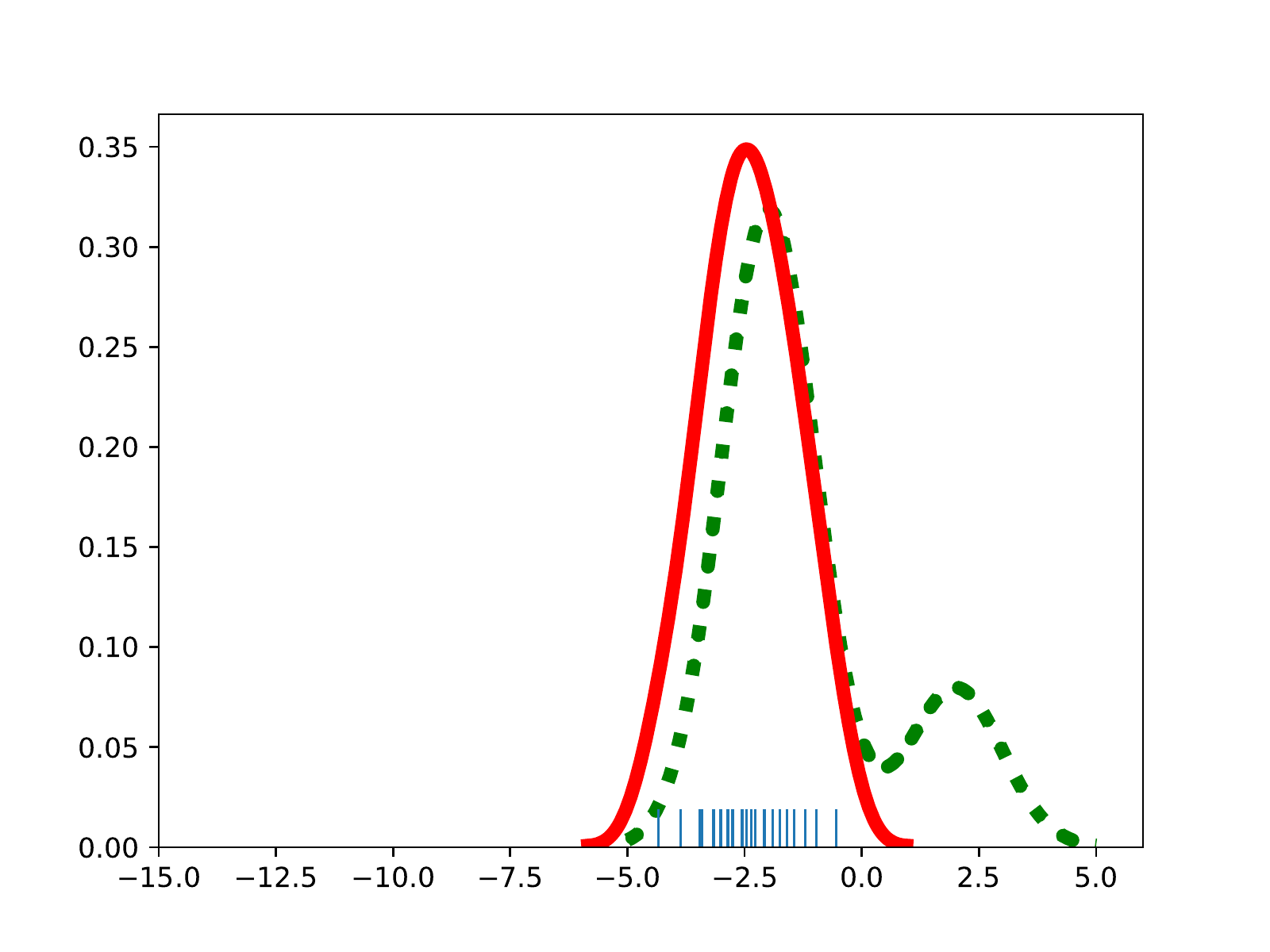}
		\caption{Iter $100$}
		
	\end{subfigure}
	~
	\begin{subfigure}{0.17\textwidth}
		\centering
		\includegraphics[width=\textwidth]{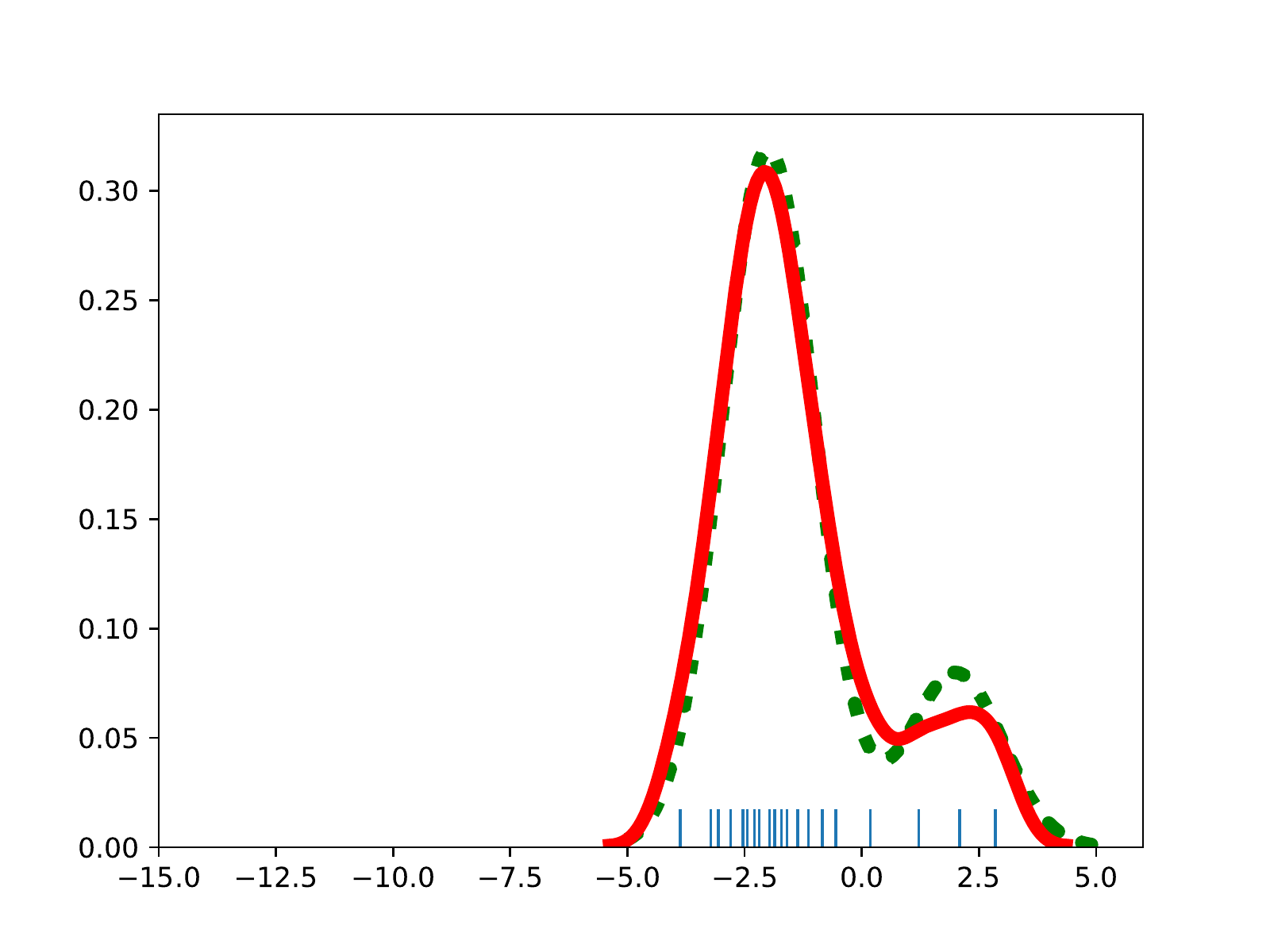}
		\caption{Iter $200$}
		
	\end{subfigure}
	~
	\begin{subfigure}{0.17\textwidth}
		\centering
		\includegraphics[width=\textwidth]{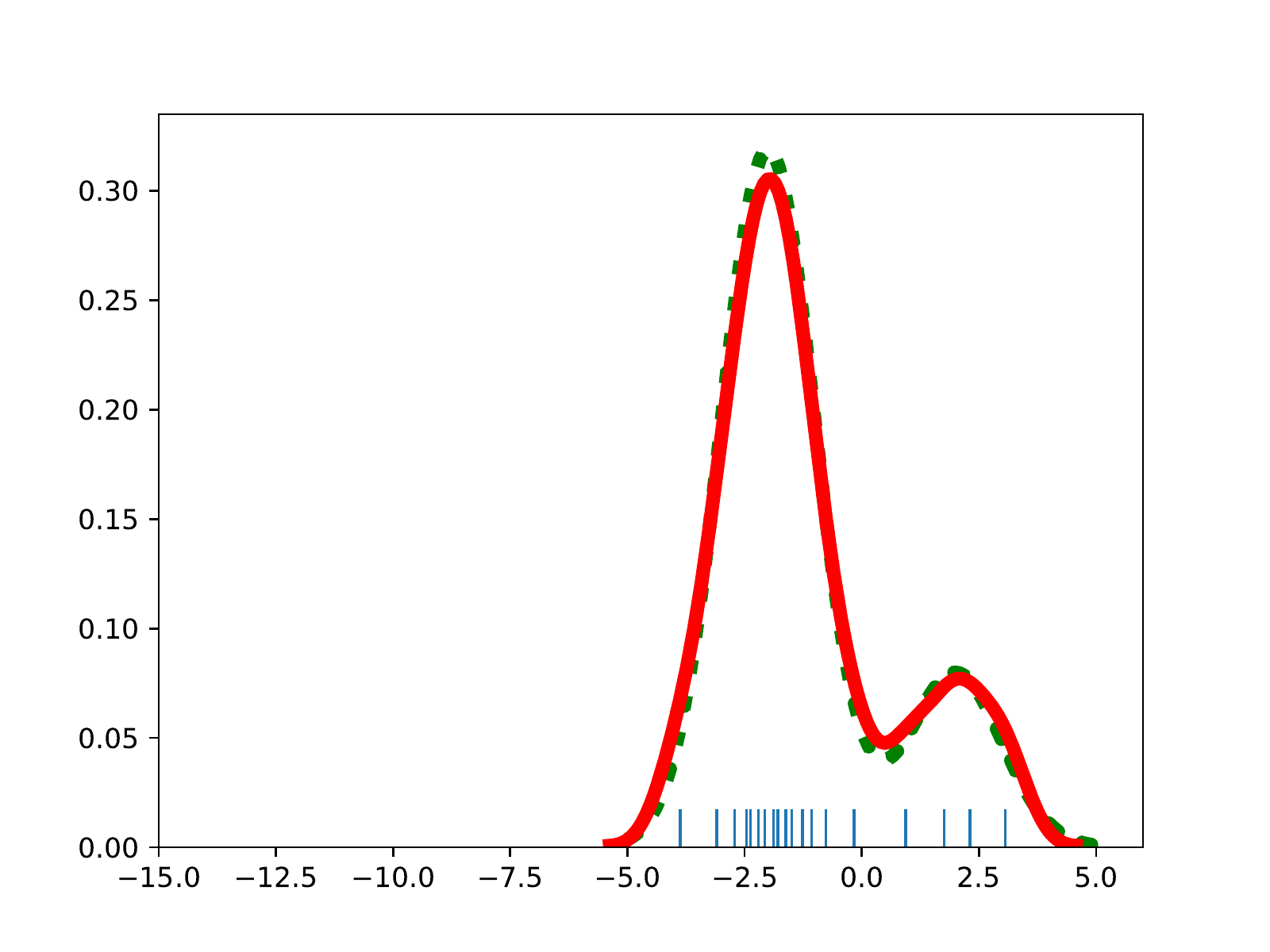}
		\caption{Iter $450$}
		
	\end{subfigure}
	\caption{Example of transporting $20$ particles from  $\mc N(-10, 1)$ to Gaussian mixture $0.8 \mc N(-2, 1) + 0.2 \mc N(2, 1)$ with SVGD using $20$ particles. The green dash represents the target Gaussian mixture.}
\end{figure}

\subsubsection{Implementation and Training}
\begin{itemize}
	
	\item We use $S$ samples of $\bmtheta$ to approximate the empirical measure of the posterior. They are initialized and stored in $20$ deterministic DenseED neural networks.
	
	\item At each step $t$, for each model $j$, the gradient of its joint likelihood (or unnormalized posterior) $\nabla_{\bmtheta^j_t} \log p(\bmtheta^j_t)$ is computed by the automatic differentiation tool in PyTorch. We first compute the joint likelihood $\log p(\bmtheta^j_t) = \prod_{i=1}^N p(\bmy_i \mid \bmtheta^j_t, \bmx_i) p(\bmtheta^j_t)$ by feeding forward the data $\{\bmx^i, \bmy^i\}_{i=1}^N$, then back propagate to compute its gradient $\nabla_{\bmtheta^j_t} \log p(\bmtheta^j_t)$. Note that this gradient is stored in PyTorch module associated to the weights $\bmtheta_t^j$ in each network.
	
	\item Then we can proceed to compute the kernel matrix $\Big[k(\bmtheta^j_t, \bmtheta^i_t)\Big]_{i,j \in \{1, \cdots, S\}}$, and its gradient $\nabla_{\bmtheta^j_t} k(\bmtheta^j_t, \bmtheta^i_t)$, as well as the kernel weighted gradient of the joint likelihood $k(\bmtheta^j_t, \bmtheta^i_t) \nabla_{\bmtheta^j_t} \log p(\bmtheta^j_t)$. For this to happen, we need to vectorize (extract out) the parameters $\bmtheta_t^j$ and the computed gradient $\nabla_{\bmtheta^j_t} \log p(\bmtheta^j_t)$ from each neural network. The optimal perturbation direction $\bmphi$ is further computed by the sum of the two terms as shown in the SVGD algorithm.
	
	\item After $\{\bmphi(\bmtheta_t^j)\}_{j=1}^S$ is computed, we send  $\bmphi(\bmtheta_t^j)$ back to each neural network  the gradient w.r.t. its parameters $\bmtheta_t^j$ (overwriting the previously computed $\nabla_{\bmtheta^j_t} \log p(\bmtheta^j_t)$), and updating locally in each neural network using PyTorch's optimization library such as Adam or SGD to compute $\bmtheta_{t+1}^j$.
	
	\item With one iteration complete, the algorithm repeats the above steps until convergence in the parameters is achieved.
\end{itemize}

\subsection{Uncertainty Quantification}
\label{sec:stats}
Of interest to the classical UQ problem is the computation of 
the posterior predictive distribution (predict the system response for a test input) as well as the computation of the output response averaged over the input probability distribution. In particular, we are interested in 
computing the following:
\begin{itemize}
	\item Predictive uncertainty at $\bmx^*$: $p(\bmy^* \mid \bmx^*, \mc D)$, and in particular the moments $\E [\bmy^* \mid \bmx^*, \mc D], \mathrm{Var}(\bmy^* \mid \bmx^*, \mc D)$.
	\item Propagated uncertainty to the system response by integrating over $p(\bmx)$: $p(\bmy \mid \bmtheta), \bmtheta \sim p(\bmtheta \mid \mc D)$, and in particular $\E [\bmy \mid \bmtheta]$, $\mathrm{Var}(\bmy \mid \bmtheta)$. One can use these moments to compute the statistics of conditional output  statistics, e.g. $\E_\bmtheta \Big[\E[\bmy \mid \bmtheta] \Big]$, $\mathrm{Var}_\bmtheta \Big(\E[\bmy \mid \bmtheta] \Big)$ and $\E_\bmtheta \Big(\mathrm{Var}(\bmy \mid \bmtheta) \Big)$, $\mathrm{Var}_\bmtheta \Big(\mathrm{Var}(\bmy \mid \bmtheta) \Big)$.
\end{itemize}

We can use Monte Carlo to approximate the moments of the predictive distribution 
\begin{equation}
	\label{eq:pred_dist}
	p(\bmy^* \mid \bmx^*, \mc D) = \int p(\bmy^* \mid \bmx^*, \bmw, \beta) p(\bmw, \beta \mid \mc D) d\bmw d\beta,
\end{equation}
with mean (by the law of total expectation)
\begin{eqnarray}
\label{eq:MonteCarloEstimate}
		\E [\bmy^* \mid \bmx^*, \mc D] &=& \E_{p(\bmw, \beta \mid \mc D)} \Big[ \E [\bmy^* \mid \bmx^*, \bmw, \beta] \Big] \nonumber \\
		&=& \E_{p(\bmw \mid \mc D)} [\bmf(\bmx^*, \bmw)] \nonumber \\
		&\approx & \frac{1}{S} \sum_{i=1}^S \bmf(\bmx^*, \bmw^i), \qquad \bmw^i \sim p(\bmw \mid \mc D).
\end{eqnarray}

Note that the SVGD algorithm provides  a sample representation  of the joint posterior of all parameters $p(\bmw, \beta \mid \mc D)$. To obtain 
the samples of the marginal posterior $p(\bmw \mid \mc D)$ as 
needed above, one simply needs to use the samples corresponding to $\bmw$. 

The predictive covariance can also be easily calculated using the law of total variance. The variance and expectation below are w.r.t. to the posterior of the parameters. We can show the following:
\begin{gather}
	\begin{aligned}
		\label{eq:pred_cov}
		\mathrm{Cov} (\bmy^* \mid \bmx^*, \mc D) &= \E_{\bmw, \beta} \Big[ \mathrm{Cov} (\bmy^* \mid \bmw, \beta, \bmx^*)\Big] + \mathrm{Cov}_{\bmw, \beta}\Big(\E [\bmy^* \mid \bmw, \beta, \bmx^*]\Big) \\
		&= \E_{\bmw, \beta}[\beta^{-1} \bmI] + \mathrm{Cov}_{\bmw, \beta}(\bmf(\bmx^*, \bmw))  \\
		&= \E_{\beta}[\beta^{-1} \bmI] + \E_{\bmw} [\bmf(\bmx^*, \bmw) \bmf^\top(\bmx^*, \bmw)]  - \E_\bmw[\bmf(\bmx^*, \bmw)] \E_\bmw^\top[\bmf(\bmx^*, \bmw)] \\
		&\approx \frac{1}{S} \sum_{i=1}^S \Big( (\beta^i)^{-1} \bmI + \bmf(\bmx^*, \bmw^i) \bmf^\top(\bmx^*, \bmw^i) \Big) \\
		&\quad- \Big(\frac{1}{S} \sum_{i=1}^S \bmf(\bmx^*, \bmw^i) \Big) \Big(\frac{1}{S} \sum_{i=1}^S \bmf(\bmx^*, \bmw^i) \Big)^\top,
	\end{aligned}
\end{gather}
where $\beta^i \sim p(\beta \mid \mc D), \bmw^i \sim p(\bmw \mid \mc D)$.
The predictive variance is the diagonal of the predictive covariance:
\begin{eqnarray}
	\label{eq:pred_var}
	\mathrm{Var}(\bmy^* \mid \bmx^*, \mc D) &=& \mathrm{diag} \, \mathrm{Cov} (\bmy^* \mid \bmx^*, \mc D) \nonumber \\
	&=&  \frac{1}{S} \sum_{i=1}^S \Big( (\beta^i)^{-1} \bmone + \bmf^2(\bmx^*, \bmw^i) \Big) - \Big(\frac{1}{S} \sum_{i=1}^S \bmf(\bmx^*, \bmw^i) \Big)^2,
\end{eqnarray}
where $\bmone$ is a vector of ones with the same dimension as $\bmf$, and the square $(\cdot)^2$ is applied element-wise to the vectors.

The above computation is the prediction at a specific input $\bmx^*$. We would also like to compute the average prediction over the distribution of the uncertain input.
We first compute the output statistics given the realizations of the uncertain parameters $\bmtheta=\{\bmw, \beta\}$, where $\bmtheta \sim p(\bmtheta \mid \mc D)$. The conditional predictive mean is
\begin{equation}
	\label{eq:cond_pred_mean}
	\E [\bmy \mid \bmtheta] = \E_\bmx \E [\bmy \mid \bmx, \bmtheta] = \E_\bmx[\bmf(\bmx, \bmw)] \approx \frac{1}{M} \sum_{j=1}^M \bmf(\bmx^j, \bmw), \quad \bmx^j \sim p(\bmx),
\end{equation}
and the conditional predictive covariance is 
\begin{gather}
	\begin{aligned}
		\label{eq:cond_pred_cov}
		\mathrm{Cov} (\bmy \mid \bmtheta) &= \E_\bmx [\mathrm{Cov}(\bmy \mid \bmx, \bmtheta)] + \mathrm{Cov}_\bmx (\E[\bmy \mid \bmx, \theta]) \\
		&= \E_\bmx [(\beta)^{-1}\bmI] + \mathrm{Cov}_\bmx(\bmf(\bmx, \bmw))\\
		&\approx \beta^{-1}\bmI + \frac{1}{M} \sum_{j=1}^M \bmf(\bmx^j, \bmw) \bmf^\top (\bmx^j, \bmw)  -\Big(\frac{1}{M} \sum_{j=1}^M \bmf(\bmx^j, \bmw) \Big) \Big(\frac{1}{M} \sum_{j=1}^M \bmf(\bmx^j, \bmw) \Big)^\top.
	\end{aligned}
\end{gather}
Also the conditional predictive variance (at each spatial location) is
\begin{equation}
	\label{eq:cond_pred_var}
	\mathrm{Var}(\bmy \mid \bmtheta) = \mathrm{diag} \, \mathrm{Cov} (\bmy \mid \bmtheta) = \beta^{-1}\bmone + \frac{1}{M} \sum_{j=1}^M \bmf^2(\bmx^j, \bmw) - \Big(\frac{1}{M} \sum_{j=1}^M \bmf(\bmx^j, \bmw) \Big)^2,
\end{equation}
where $\bmone$ is a vector of ones with the same dimension as $\bmy$, and the square operator is here applied element-wise to the vectors.
Then we can further compute the statistics of the above conditional statistics due to the uncertainty in the surrogate, i.e. $\bmtheta$, such as $\E_\bmtheta \Big[\E[\bmy \mid \bmtheta] \Big]$, $\mathrm{Var}_\bmtheta \Big(\E[\bmy \mid \bmtheta] \Big)$ and $\E_\bmtheta \Big(\mathrm{Var}(\bmy \mid \bmtheta) \Big)$, $\mathrm{Var}_\bmtheta \Big(\mathrm{Var}(\bmy \mid \bmtheta) \Big)$, which are the sample means and sample variances in each output dimension of of the conditional predictive mean and variance.


\section{Numerical Implementation and Results}
\label{sec:implementationAndresults}

We study the two-dimensional, single phase, steady-state flow through a random permeability field following the case study in Section $3.2$ in~\cite{bilionis2013multi}. Consider the random permeability field $K$ on a unit square spatial domain $\mc S = [0, 1]^2$, the pressure field $p$ and velocity field $\bmu$ of the fluid through the porous media are governed by Darcy's law:
\begin{gather}
\begin{aligned}
\label{eq:darcy}
\bmu(\bms) &= -K(\bms) \nabla p(\bms), \qquad \bms \in \mc S, \\
\nabla \cdot \bmu(\bms) &= f(\bms), \qquad \bms \in \mc S, \\
\bmu(\bms) \cdot \hat{\bmn}(\bms) &= 0, \qquad \bms \in \partial \mc S, \\
\int_{\mc S} p(\bms) d\bms &= 0,
\end{aligned}
\end{gather}
where $\hat{\bmn}$ denotes the unit normal vector to the boundary and the source term $f$ is used to model an injection well on the left-bottom corner of $\mc S$ and a production well on the right-top corner. We also enforce no-flux boundary condition, and an integral constraint to ensure the uniqueness of the solution as in~\cite{bilionis2013multi}.
More specifically, 
\begin{equation}
f(\bms) = 
\begin{cases}
r, & \quad \mathrm{if}\, |\bms_i - \frac{1}{2} w| \le \frac{1}{2} w, \mathrm{for}\, i=1,2,\\
-r,  & \quad \mathrm{if}\, |\bms_i - 1 + \frac{1}{2} w| \le \frac{1}{2} w, \text{for}\, i=1,2, \\
0,  & \quad \mathrm{otherwise},
\end{cases}
\end{equation}
where $r$ is the rate of the wells and $w$ is their size.
The input log-permeability field is restricted in this work to be a Gaussian random field, i.e.
\begin{equation}
\label{eq:permeability_Gaussian}
K(\bms) = \mathrm{exp}(G(\bms)), \quad G(\cdot) \sim \mc N (m, k(\cdot, \cdot)),
\end{equation}
where  $m$ is the constant mean and covariance function $k$ is specified in the following form using the $L_2$ norm in the exponent instead of the $L_1$ norm in~\cite{bilionis2013multi}, i.e. 
\begin{equation}
k(\bms, \bms') = \mathrm{exp}(-\norm{\bms - \bms'}_2 / l). 
\label{eq:exp_kernel}
\end{equation}

\subsection{Datasets}
\label{sec:dataset}

The Gaussian random field~\cite{rasmussen2006gaussian} with exponential kernel for the one-dimensional case corresponds to the Ornstein-Uhlenbeck process which is mean-square continuous but not mean-square differentiable. Thus when we discretize the field over a grid, the field value \textit{jumps (varies highly)} as we move from pixel to pixel. The field does not become smoother when we use a finer grid over a fixed spatial domain.
This high variability creates a significant challenge for data-driven models to capture, i.e. the intrinsic dimensionality of the discretized random field is the total number of pixels, e.g. $4,225$ for $65 \times 65$ grids (which will be our reference grid for our calculations). However, a common assumption for natural images is that the underlying dimensionality is actually small (few hundreds) despite their complex appearance. To evaluate the generality and effectiveness of the methodology, we use KLE to control the intrinsic dimensionality of the permeability dataset. We evaluated our model using datasets produced with increasing dimensionality of $50$, $500$, $4225$ (called KLE$50$, KLE$500$, and KLE$4225$, respectively). Notice that when the number of KLE terms is $4225$, the permeability field is directly sampled from the exponential Gaussian field without any dimensionality reduction.
The intrinsic dimensionality of dataset is hidden from our model, i.e. our model do not built a map from the KLE terms to the system output. Instead, it models an end-to-end mapping from input fields to output fields. In fact, we will show one specific network architecture that works well for all three datasets obtained from the 
different intrinsic input data dimensions.

\begin{figure}[h]
	\centering
	\includegraphics[width=0.45\textwidth]{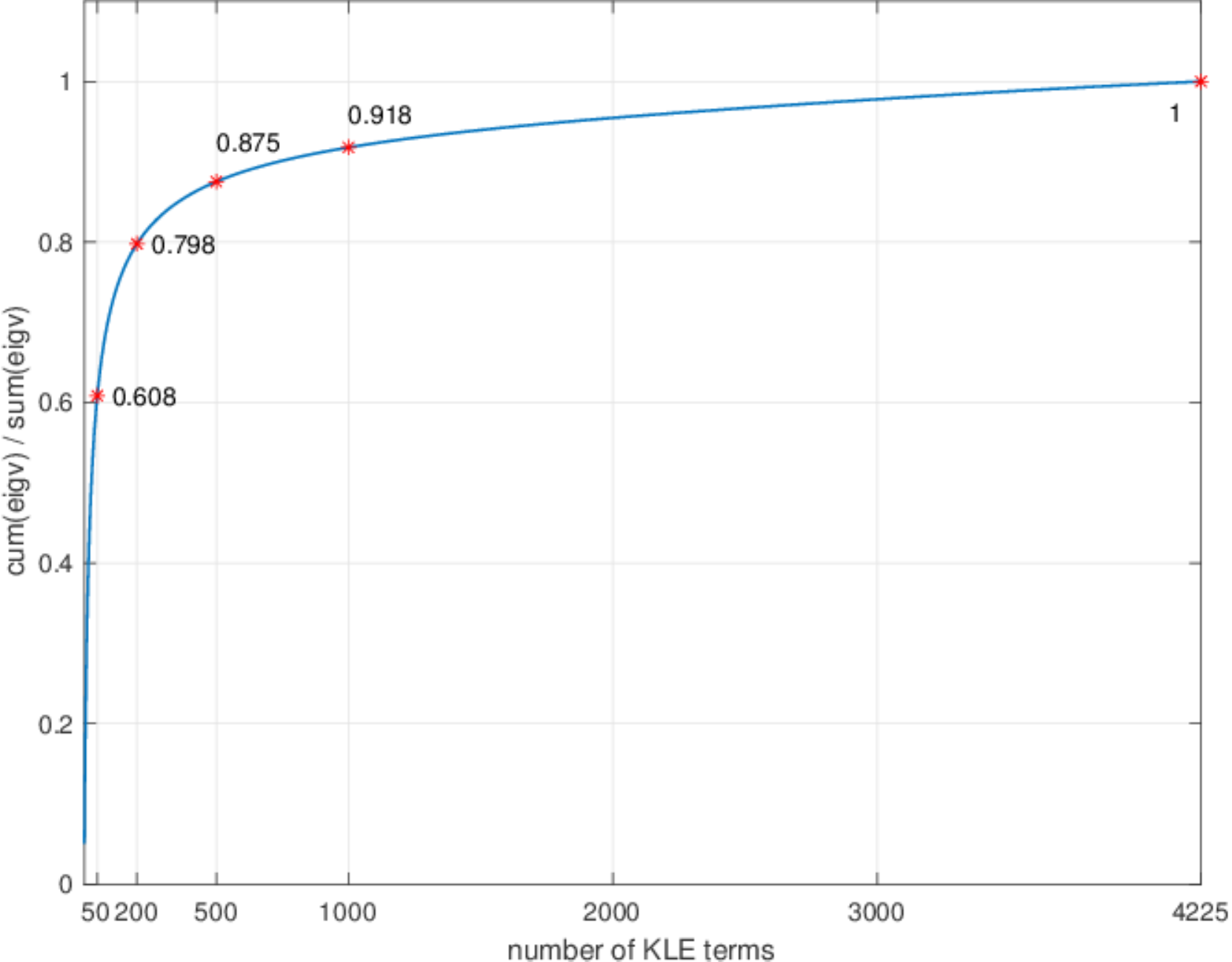}
	\caption{KLE profile.}
	\label{fig:kle_profile}
\end{figure}

\begin{figure}[!h]
	\centering
	\begin{subfigure}{0.45\textwidth}
		\centering
		\includegraphics[width=0.95\textwidth]{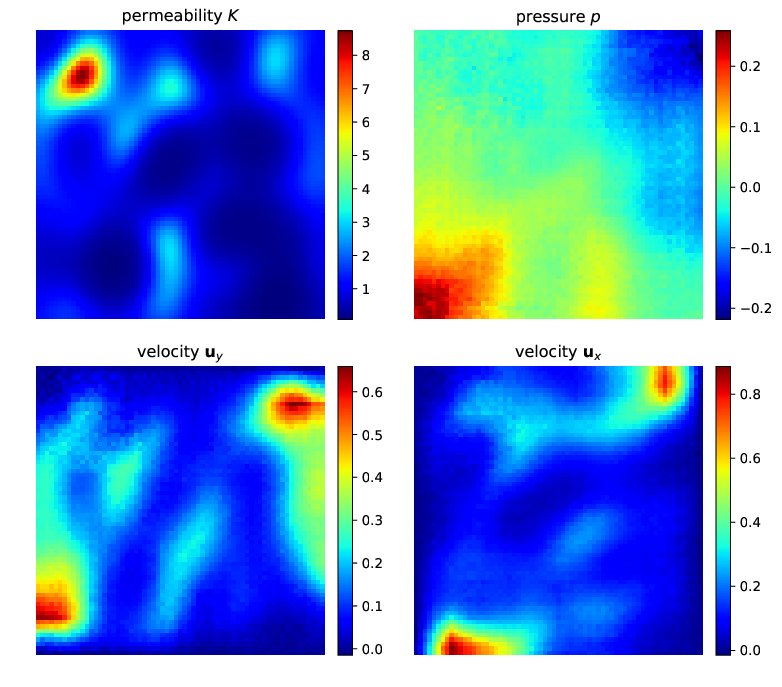}
		\caption{KLE$50$}
		\label{fig:sample_kle50}
	\end{subfigure}
	~
	\begin{subfigure}{0.45\textwidth}
		\centering
		\includegraphics[width=0.95\textwidth]{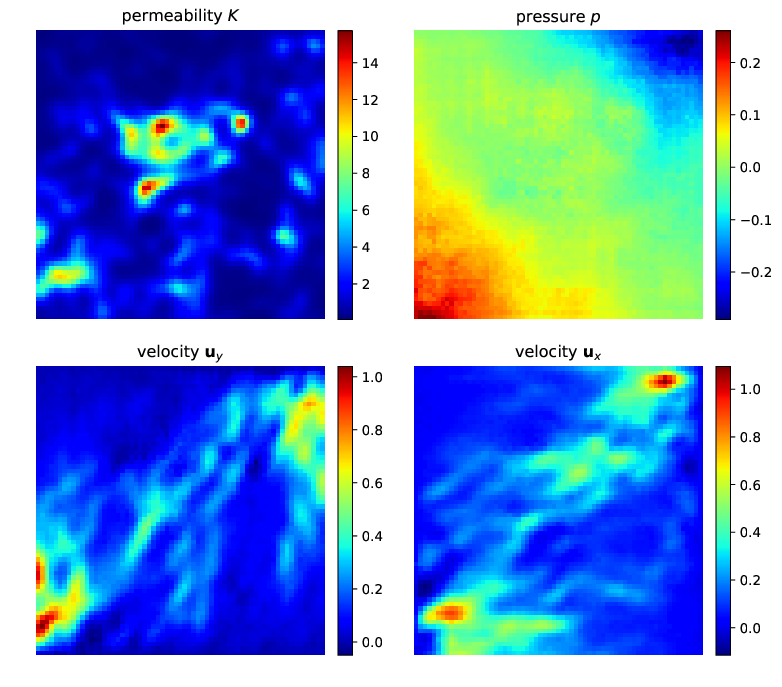}
		\caption{KLE$500$}
		\label{fig:sample_kle500}
	\end{subfigure}
	\\
	\begin{subfigure}{0.6\textwidth}
		\centering
		\includegraphics[width=0.7\textwidth]{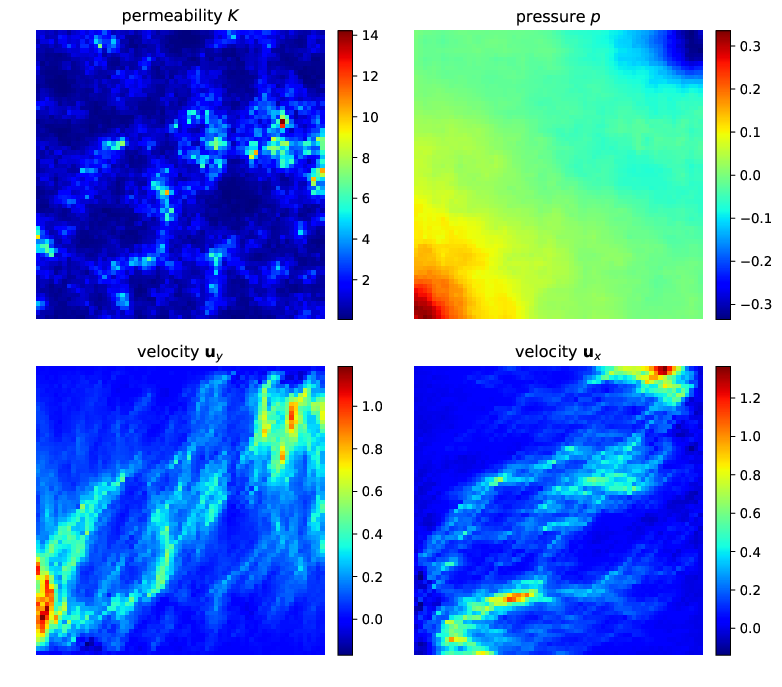}
		\caption{KLE$4225$}
		\label{fig:sample_kle4225}
	\end{subfigure}
\caption{Sample permeability $K$ obtained  from the (a) KLE$50$ dataset, (b) KLE$500$ dataset, and (c) KLE$4225$ dataset (no dimensionality reduction) and the corresponding velocity  components $\bmu_x$, $\bmu_y$ and pressure $p$ obtained from the simulator. All figures are shown using pixels (\texttt{imshow}) to reveal the high variability of the input and output fields.}
	\label{fig:samples_dataset}
\end{figure}

We consider solving the Darcy flow Eq.~(\ref{eq:darcy}) over a unit squared domain $\mc S = [0, 1]^2$ with fixed $65 \times 65$ grid, and length scale $l=0.1$, kernel mean $m=0$, rate of source $r=10$, size of source $w=0.125$. The ratio of the cumulative sum of eigenvalues (in decreasing order) over the total sum of them is shown in Fig.~\ref{fig:kle_profile}.
The Darcy flow equation is solved using mixed finite element formulation implemented in FEniCS~\cite{AlnaesBlechta2015a} with third-order Raviart-–Thomas elements for the velocity, and fourth-order discontinuous elements for the pressure. The sample input permeability field and computed output pressure and velocity fields for three datasets are shown in Fig.~\ref{fig:samples_dataset}.

When the available data is limited, it is common practice to use cross-validation to evaluate the model. Since our dataset is synthetically generated, we have access to any number of training and test data up to computing constraints to solve the Darcy flow equations. Our current data includes four sets: the training set, validation set, test set, and 
uncertainty propagation set.
The training set is sampled using the simplest design of experiment method, Latin hypercube sampling. More specifically, the KLE for the log-permeability field is 
\begin{equation}
\label{eq:kle}
G(\bms) = m + \sum_{k=1}^{q} \sqrt{\lambda_k} z_k \phi_k(\bms),
\end{equation}
where $\lambda_k$ and $\phi_k(\bms)$ are the eigenvalues and eigenfunctions of the exponential covariance function of the Gaussian field specified in Eqs.~(\ref{eq:permeability_Gaussian}) and~(\ref{eq:exp_kernel}), $z_k$'s are i.i.d. standard Normal, and $q$ is the number of KLE coefficients maintained in the expansion. The maximum number that can be used is finite and equal to the number of grid points used in the discetization of the field over the unit square. We first use Latin hypercube design to sample $\xi_k$ from the hypercube $[0, 1]^q$, then obtain the eigenvalue by $z_k=\Phi^{-1}(\xi_k)$, where $\Phi$ is the cumulative distribution function of the standard normal distribution. The KLE$50$ case contains $32$, $64$, $128$, and $256$ training data; KLE$500$ contains $64$, $128$, $256$, and $512$ training data; and KLE$4225$ contains $128$, $256$, $512$, and $1024$ training data.

The log-permeability fields in the other three sets are reconstructed directly with $z_k$, which are sampled from standard normal.
The validation and test set each contains $500$ input permeability fields, and the dataset for uncertainty propagation contains $10,000$ realizations. All datasets are organized as images as discussed in Section~\ref{sec:image_regression}.

\subsection{Evaluation Metrics}
\label{sec:Metrics}

Several metrics are used to evaluate the trained models on test data $\{\bmx^i, \bmy^i\}_{i=1}^T$. In particular, we consider the following:

\noindent {\em Coefficient of determination ($R^2$-score)}: 
\begin{equation}
R^2 = 1 - \frac{\sum_{i=1}^T \norm{\bmy^i - \hat{\bmy}^i}_2^2}{\sum_{i=1}^T \norm{\bmy^i - \bar{\bmy}}_2^2},
\label{eq:R2_score}
\end{equation}
where $\hat{\bmy}^i$ is the output mean of the Bayesian surrogate, i.e. $\sum_{i=1}^S \bmf(\bmx, \bmw^i) / S$ as in Eq.~(\ref{eq:MonteCarloEstimate}) or predictive output of the non-Bayesian surrogate, i.e. just $\bmf(\bmx)$, $\bmy^i$ is the test target,  $\bar{\bmy}$ is the mean of test target, and $T$ is the total number of test data. This metric enables the comparison between different datasets since the error is normalized, with the score closer to $1$ corresponding to better regression. This is the only metric used for evaluating non-Bayesian surrogate, the following metrics are additional metrics for evaluating the Bayesian surrogate. Note that this metric is also used for tracking the performance of the training process, thus it is evaluated for both the training and test data sets.

\noindent {\em Root Mean Squared Error (RMSE)}: 
$$\sqrt{\frac{1}{T} \sum_{i=1}^T \norm{\hat{\bmy}^i - \bmy^i}_2^2}.$$
This is a common metric for regression that is used in our experiments for monitoring the convergence of training.

\noindent {\em Mean Negative Log-Probability (MNLP)}: 
$$ \mathrm{MNLP} = -\frac{1}{T} \sum_{i=1}^T \log p(\bmy^i \mid \bmx^i, \mc D).$$
This metric evaluates the likelihood of the observed data. It is is  used to assess the quality of the predictive model.

\noindent {\em Predictive uncertainty and Propagated Uncertainty:} These metrics were introduced in Section~\ref{sec:stats}.

\noindent {\em Estimated Distributions}: They include histograms or  kernel density estimates for the output fields at certain locations of the physical domain.

\noindent {\em Reliability Diagram:}  
Given a trained Bayesian surrogate and a test data set, we can compute the $p \%$ predictive interval for each test data point based on the Gaussian quantiles using the predictive mean and variance~\cite{lakshminarayanan2016simple}. We then compute the frequency of the test targets that fall within this predictive interval. For a well-calibrated regression model, the observed frequency should be close to $p \%$. The reliability diagram is the plot of the observed frequency with respect to $p$. Thus a well-calibrated model should have a reliability diagram close to the diagonal.

\subsection{Non-Bayesian Surrogate Model}
\label{sec:non_bayes_surrogate}

\begin{figure}[hbtp]
	\centering
	\includegraphics[width=0.6\textwidth]{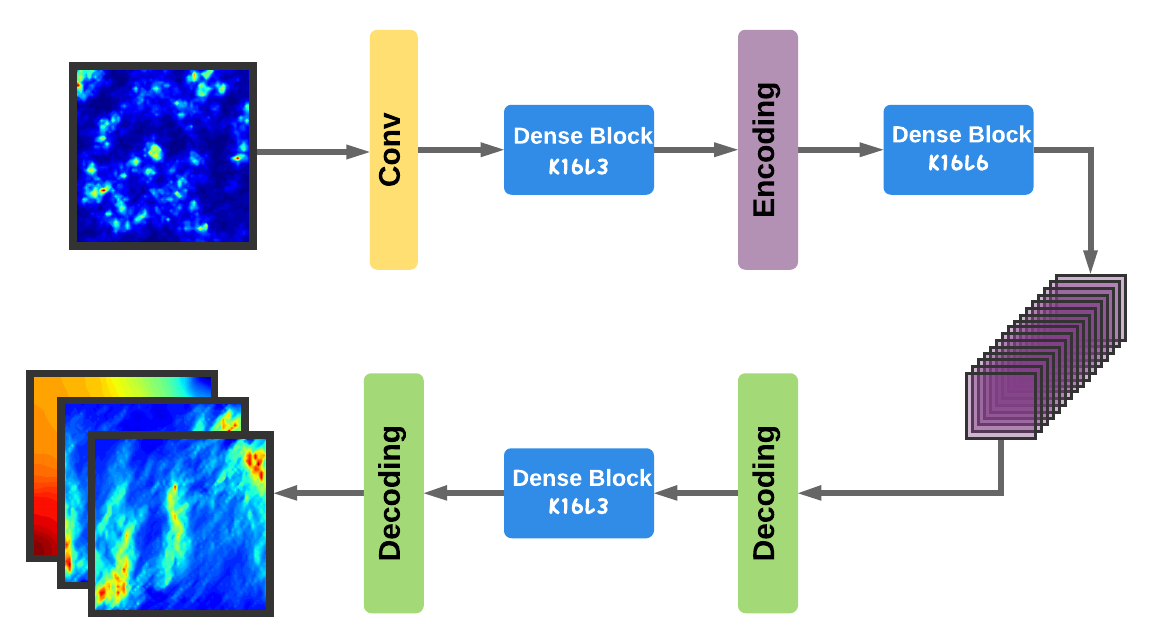}
	\caption{\texttt{DenseED-c16} with blocks $(3, 6, 3)$, growth rate $16$, and $48$ initial feature maps after the first convolution layer (yellow in the figure). There are in total $19$ (conv) layers and $241,164$ parameters in the network. The number of network parameters is optimized based on the generalization error analysis reported in~\ref{sec:Generalization}. It contains two down-sampling layers, thus the smallest spatial dimension (code dimension) of feature maps (purple in the figure) is $16 \times 16$, hence the network is named  \texttt{DenseED-c16}. The first convolution kernel is of $k7s2p2$. The last transposed conv kernel in the decoding layer is of $k5s2p1$. The number of its output feature maps is $3$ (corresponding to $3$ output fields). For the decoding layers, the output padding is set to $1$. The other conv kernels in dense blocks and encoding, decoding layers are described in Section~\ref{sec:conv_enc_dec}.}
	\label{fig:dense_ed_c16}
\end{figure}

The hyperparameters to search include the parameters that determine the network architecture and the ones that specify training process, which both affect model performance. We use Hyperband~\cite{li2016hyperband} algorithm to optimize those hyperparameters with a constraint that the number of model parameters being less than $0.25$ million. The details of these experiments are given in~\ref{sec:DesignTests}.  
The network configuration with the highest $R^2$-score that Hyperband finds is shown in Fig.~\ref{fig:dense_ed_c16} with more details provided in Table~\ref{table:dense_ed_c16}. This configuration is referred to as \texttt{DenseED-c16}. The $2$nd--$4$th columns of Table~\ref{table:dense_ed_c16} show
the number $C_f$ of output feature maps, the spatial resolution $H_f \times W_f$ of output feature maps, the number of parameters of each layer in the network.

\begin{table}[!hbtp]
	\centering
	\caption{\texttt{DenseED-c16} architecture for the Darcy flow dataset}
	\begin{tabular}{ c c c c } 
		\hline
		Layers &  $C_f$ &  Resolution $H_f \times W_f$ & Number of parameters  \\
		\hline
		Input                               & $1$     & $65 \times 65 $ & - \\
		Convolution \texttt{k7s2p2}         & $48$    & $32 \times 32 $ & 2352 \\
		Dense Block (1) \texttt{K16L3}      & $96$    & $32 \times 32 $ & 28032 \\
		Encoding Layer                      & $48$    & $16 \times 16 $ & 25632 \\
		Dense Block (2) \texttt{K16L6}      & $144$   & $16 \times 16 $ & 77088 \\
		Decoding Layer (1)                  & $72$    & $32 \times 32 $ & 57456 \\
		Dense Block (3) \texttt{K16L3}      & $120$   & $32 \times 32 $ & 38544 \\
		Decoding Layer (2)                  & $3$     & $65 \times 65 $ & 12060 \\
		\hline
	\end{tabular}
	\label{table:dense_ed_c16}
\end{table}


The network \texttt{DenseED-c16} is trained with Adam~\cite{kingma2014adam}, a variant of stochastic gradient descent, with the loss function being $L_2$ regularized MSE which is implemented as weight decay in modern neural net frameworks, such as PyTorch and TensorFlow. Other loss functions may achieve better results, such as smoothed $L_1$ loss, or conditional GAN loss~\cite{pix2pix2016}. This requires further investigations to be considered in future publication.  The initial learning rate is $0.015$, weight decay (regularization on weights) is $0.0005$, the batch size is $16$. We also use a learning rate scheduler which drops $10$ times on plateau of the rooted MSE. The model is trained $200$ epochs. We train the model with the dataset introduced in Section~\ref{sec:dataset}.

\begin{figure}[htbp]
	\centering
	\includegraphics[width=0.5\textwidth]{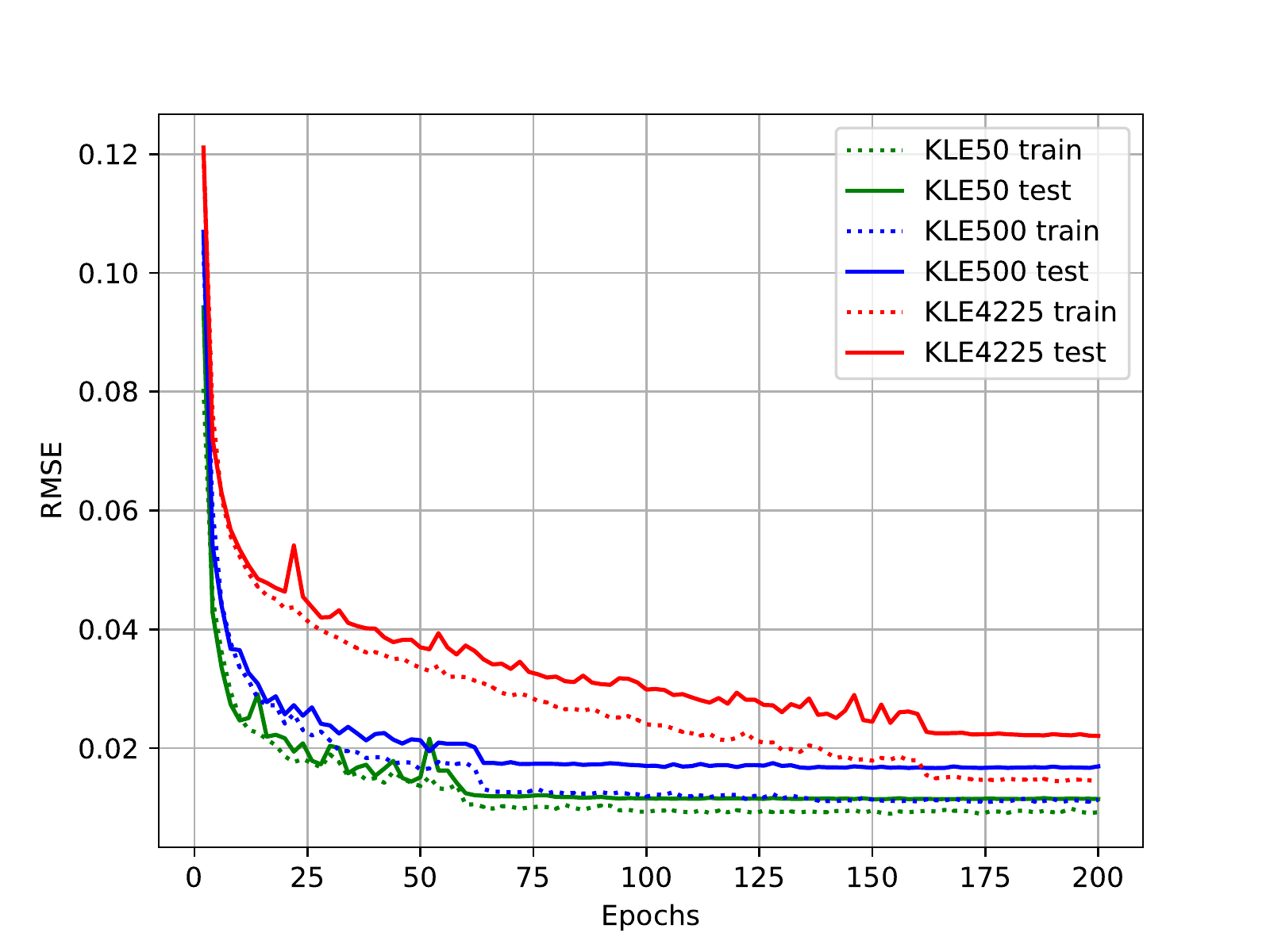}
	\caption{Training process of \texttt{DenseED-c16} with $128$ training data.}
	\label{fig:dense_ed_c16_training}
\end{figure}

Training the deterministic neural networks with $L_2$ regularized MSE is equivalent to finding the maximum \textit{a posterior} of the uncertain parameters in Bayesian neural networks whose prior is independent normal. The typical training process is shown in Fig.~\ref{fig:dense_ed_c16_training}.

\begin{figure}[!h]
	\centering
	\includegraphics[width=0.5\textwidth]{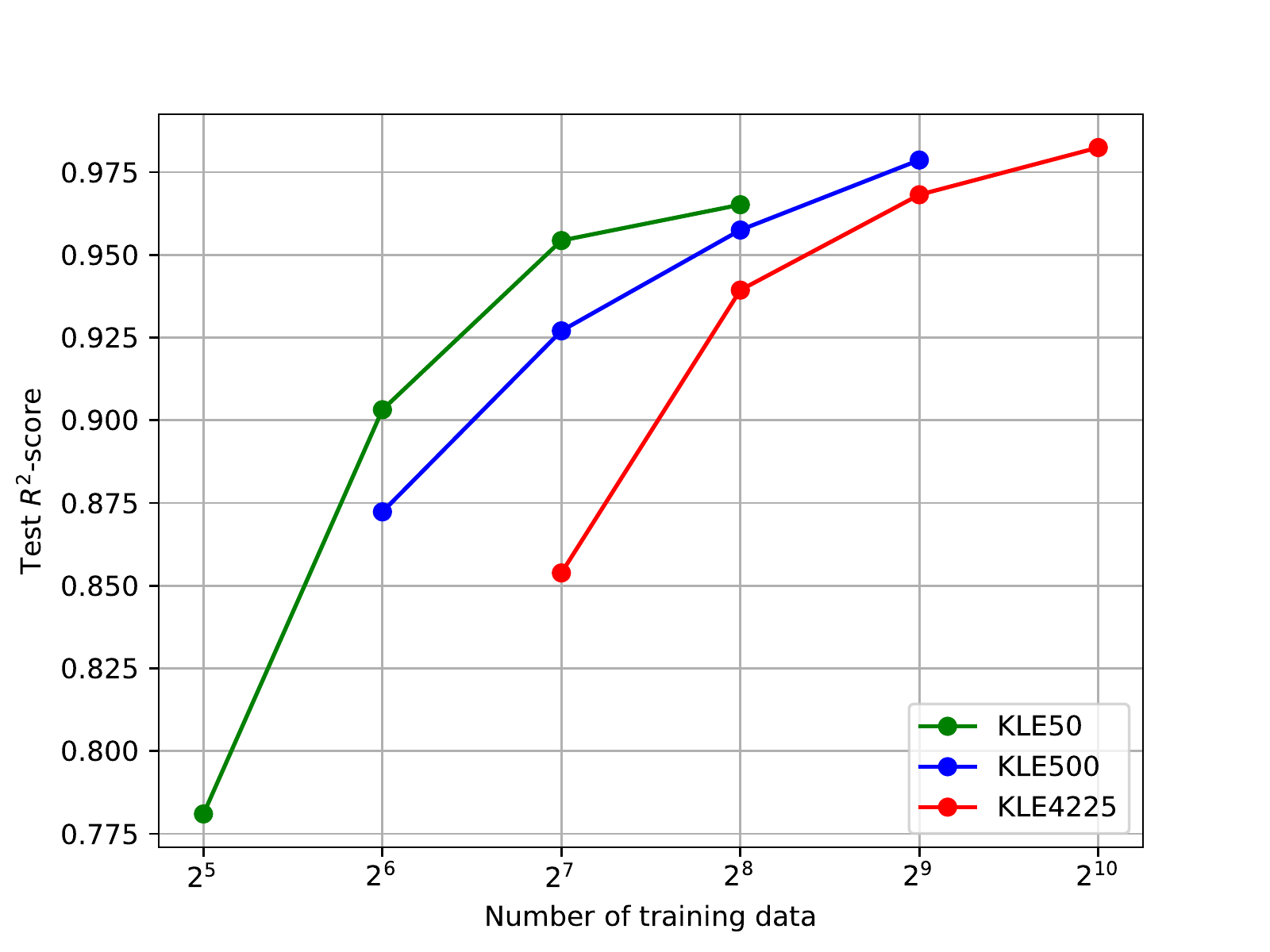}
	\caption{Test $R^2$ scores for the non-Bayesian surrogate.}
	\label{fig:dense_ed_c16_r2}
\end{figure}

We train each network with different number of training data of KLE$50$, KLE$500$, and KLE$4225$. 
The validation $R^2$-score is shown in Fig.~\ref{fig:dense_ed_c16_r2}, which shows that, with the same training data, the $R^2$-score is closer to $1$ when the intrinsic dimensionality is smaller, and the $R^2$-score is higher with more training data of the same dimensionality. Note that the score is more than $0.9$ with reasonably small size training data set for all the three cases which have dimensionality from $50$ to $4225$. This shows the  effectiveness of the network \texttt{DenseED-c16} for both low-dimensional and high-dimensional problems.

\begin{figure}[htbp]
	\centering
	\begin{subfigure}[t]{0.475\textwidth}
		\centering
		\includegraphics[width=0.95\textwidth]{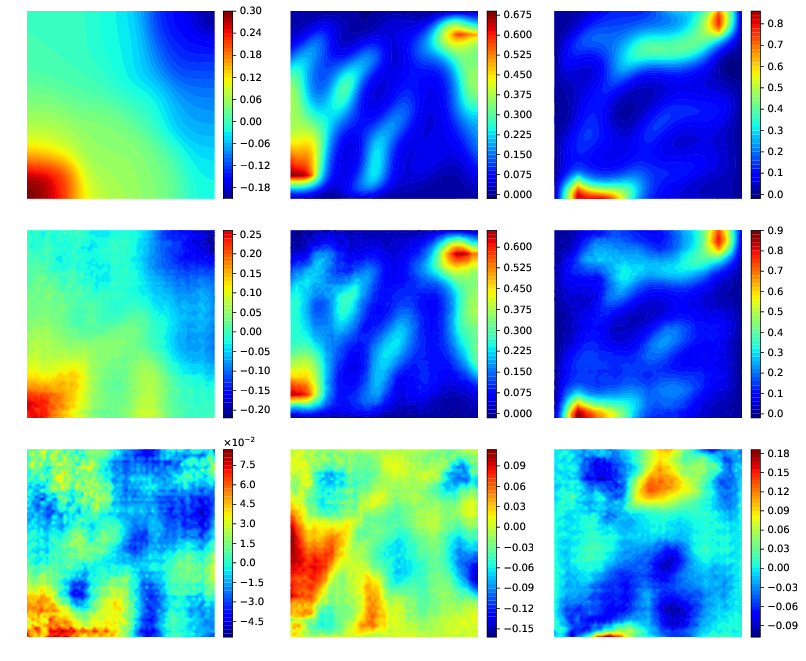}
		\caption{$32$ training data}
	\end{subfigure}
	~ 
	\begin{subfigure}[t]{0.475\textwidth}
		\centering
		\includegraphics[width=0.95\textwidth]{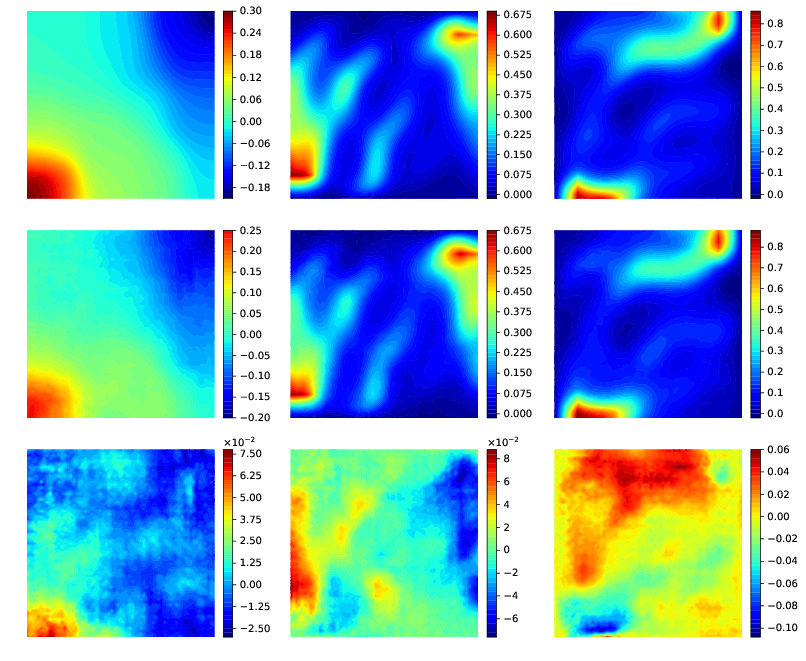}
		\caption{$128$ training data}
	\end{subfigure}
	\caption{Prediction for the input realization shown in Fig.~\ref{fig:sample_kle50} from the KLE$50$ dataset  
		using \texttt{DenseED-c16} which is trained with datasets of sizes (a) $32$ and (b) $128$, respectively. In both subfigures, the first row shows the three test target fields (simulation output), i.e. pressure $p$ and velocity $\bmu_y$, $\bmu_x$, the second row shows the corresponding model predictions, the third row shows the error.}
	\label{fig:kle50_pred_at_47}
\end{figure}

\begin{figure}[htbp]
	\centering
	\begin{subfigure}[t]{0.475\textwidth}
		\centering
		\includegraphics[width=0.95\textwidth]{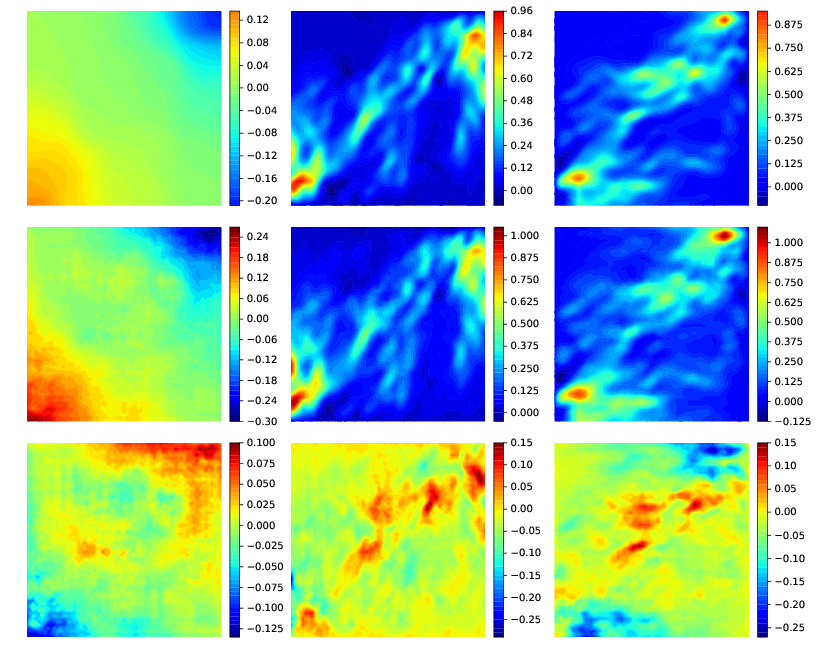}
		\caption{$64$ training data}
	\end{subfigure}
	~ 
	\begin{subfigure}[t]{0.475\textwidth}
		\centering
		\includegraphics[width=0.95\textwidth]{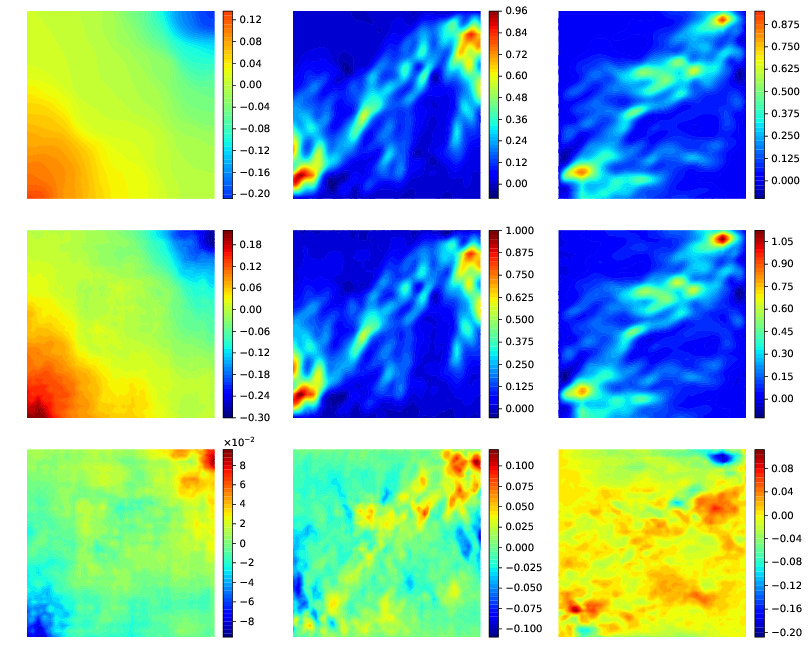}
		\caption{$256$ training data}
	\end{subfigure}
	\caption{Prediction for the input realization as shown in Fig.~\ref{fig:sample_kle500} from KLE$500$ dataset  
		using \texttt{DenseED-c16} which is trained with datasets of sizes (a) $64$ and (b) $256$, respectively. In both subfigures, the first row shows the three test target fields (simulation output), i.e. pressure $p$ and velocity $\bmu_y$, $\bmu_x$, the second row shows the corresponding model predictions, the third row shows the error.}
	\label{fig:kle500_pred_at_293}
\end{figure}

\begin{figure}[htbp]
	\centering
	\begin{subfigure}[t]{0.475\textwidth}
		\centering
		\includegraphics[width=0.95\textwidth]{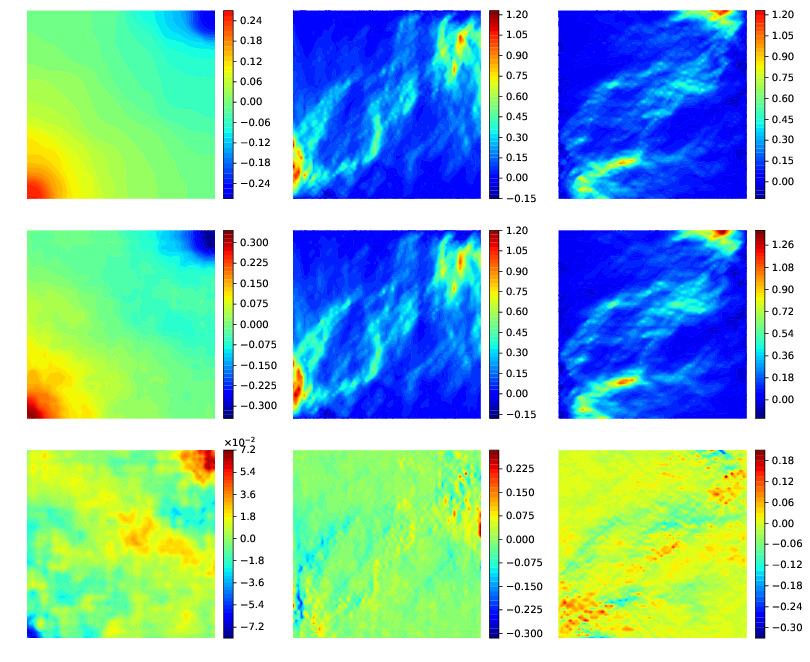}
		\caption{$128$ training data}
	\end{subfigure}
	~ 
	\begin{subfigure}[t]{0.475\textwidth}
		\centering
		\includegraphics[width=0.95\textwidth]{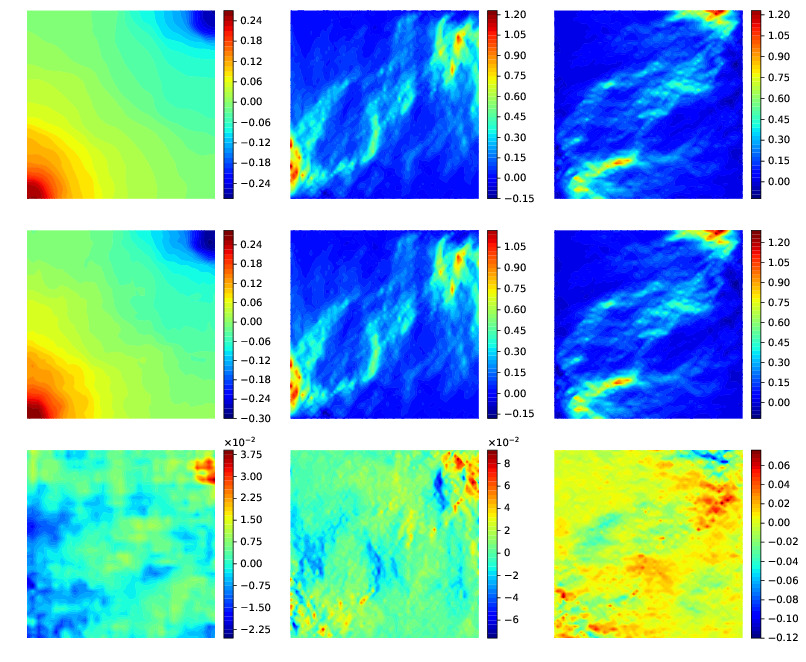}
		\caption{$512$ training data}
	\end{subfigure}
	\caption{Prediction for the input realization as shown in Fig.~\ref{fig:sample_kle4225} from KLE$4225$ dataset  using \texttt{DenseED-c16} 
		which is trained with datasets of sizes (a) $128$ and (b) $512$, respectively. In both subfigures, 
		the first row shows the three test target fields (simulation output), i.e. pressure $p$ and velocity $\bmu_y$, $\bmu_x$, the second row shows the corresponding model predictions, the third row shows the error.}
	\label{fig:kle4225_pred_at_312}
\end{figure}

The prediction of the output fields can be easily obtained in the test time by feeding the test input permeability field $\bmx^*$ into the trained network, i.e. $\hat{\bmy}^* = \bmf(\bmx^*)$.
We show the prediction of the test input shown in Fig.~\ref{fig:samples_dataset} using \texttt{DenseED-c16}, which is trained with three datasets (KLE$50$, KLE$500$, KLE$4225$) in Figs.~\ref{fig:kle50_pred_at_47},~\ref{fig:kle500_pred_at_293}, and~\ref{fig:kle4225_pred_at_312}, respectively. The predictions are quite good even for the KLE$4225$ case, where both the input and output fields vary rapidly in certain regions of the domain.

\subsection{Bayesian Surrogate Model}

For all the experiments we only consider the homoscedastic noise model for Bayesian neural networks, i.e. output-wise Gaussian noise with Gamma prior on its precision $\beta$, and Student's t-prior on $\bmw$.

The set of all uncertain parameters is denoted as $\bmtheta = \{\bmw, \beta\}$.
We apply SVGD to the Bayesian neural network with $S$ samples $\{\bmtheta^i\}_{i=1}^S$ from the posterior $p(\bmtheta \mid \mc D)$, i.e. $S$ set of deterministic model parameters $\{\bmw^i\}_{i=1}^S$ of \texttt{DenseED}'s  and noise precision $\{\beta^i\}_{i=1}^S$. In implementation, this corresponds to $S$ different initializations for the deterministic \texttt{DenseED} and noise precision (a scalar). 
We update the parameters of $S$ \texttt{DenseED}'s and the corresponding noise precision using the SVGD algorithm as in Algorithm~\ref{algo:svgd}. The kernel is chosen to be $k(\bmx, \bmx') = \mathrm{exp}(-\norm{\bmx - \bmx'}_2^2 / h)$, with median heuristic for the choice of the kernel bandwidth $h=H^2/\log S$, where $H$ is the median of the pairwise distances between the current samples $\{\bmtheta^i\}_{i=1}^S$. We typically use $S=20$ samples of $\bmtheta$ to approximate the empirical measure of the posterior. 
For large number of training data, the unnormalized posterior is evaluated using mini-batches of training data, i.e. $\tilde{p}(\bmtheta \mid \mc D) = \prod_{i=1}^N p(\bmy^i \mid \bmtheta, \bmx^i) p_0(\bmtheta) \approx N / B \prod_{i=1}^B p(\bmy^i \mid \bmtheta, \bmx^i) p_0(\bmtheta)$. We observe that even for small training data such as $512$, using smaller batch size (e.g. $16$) helps to get lower training and test errors, but with more time for training.
We use Adam~\cite{kingma2014adam} to update $\bmtheta$ using the gradient $\bmphi$, instead of the vanilla stochastic gradient descent, for $300$ epochs, with learning rate $0.002$ for $\bmw$ and $0.01$ for $\beta$, and a learning rate scheduler that decreases by $10$ times when the training RMSE is on plateau.

The algorithm is implemented in PyTorch and runs on a single NVIDIA GeForce GTX $1080$ Ti X GPU which requires about $2000 - 7000$ seconds for training $300$ epochs, when the training data size varies from $32$ to $512$. The training time depends heavily on the training mini-batch size, which is $16$ for all cases. Potential ways to speed up significantly the training process include increasing the mini-batch size, or implementing the SVGD in  parallel using multi-GPUs. The python source code will become available upon publication at \url{https://github.com/bmmi/bayesnn}.

\begin{figure}[!htb]
	\centering
	\includegraphics[width=0.6\textwidth]{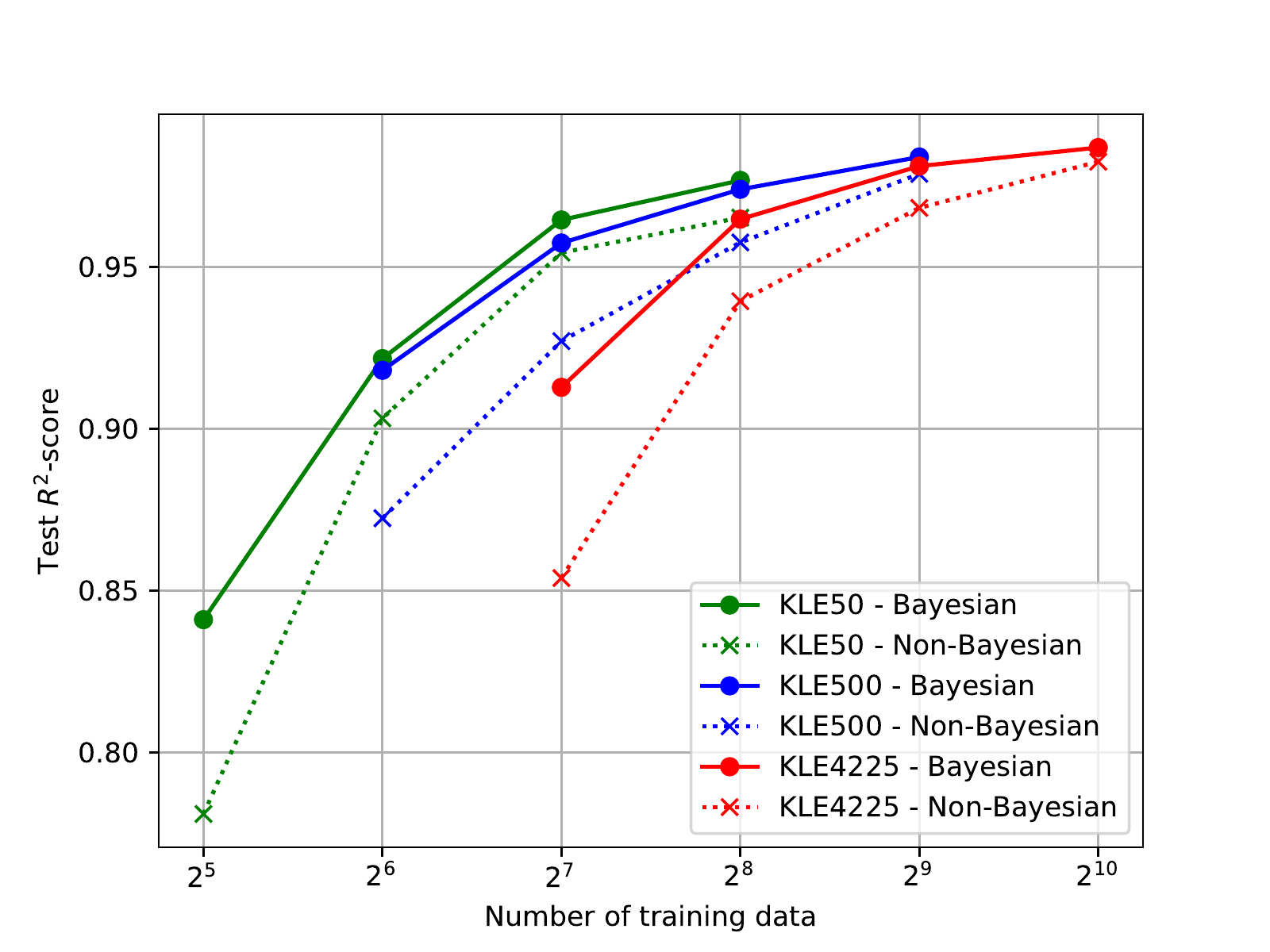}
	\caption{$R^2$--scores comparison for the non-Bayesian and  Bayesian models.}
	\label{fig:bayes_r2}
\end{figure}

We report next the $R^2$-score computed similar to the non-Bayesian case, except the predicted output mean is used to compare with the test target. The scores are shown in Fig.~\ref{fig:bayes_r2}. We can see that the Bayesian surrogate improves the $R^2$-score significantly over the non-Bayesian version.  

\begin{figure}[!htb]
	\centering
	\includegraphics[width=0.5\textwidth]{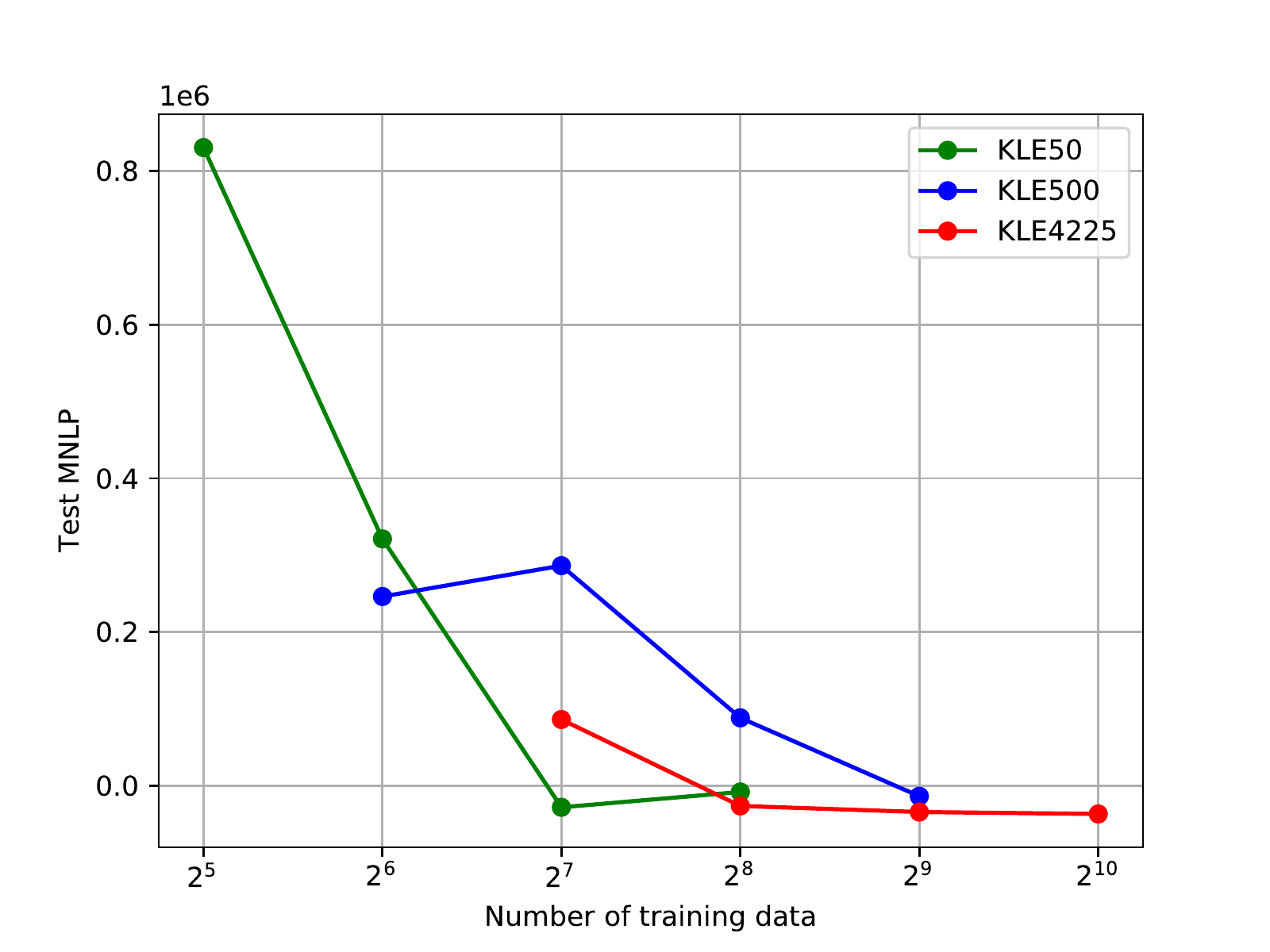}
	\caption{MNLP.}
	\label{fig:mnlp}
\end{figure}

We also report the MNLP for test data in Fig.~\ref{fig:mnlp}, which is a proper scoring rule and usually used to access the quality of predictive uncertainty~\cite{quinonero2006evaluating}. 

\begin{figure}[htbp]
	\centering
	\begin{subfigure}[t]{0.425\textheight}
		\centering
		\includegraphics[width=0.95\textwidth]{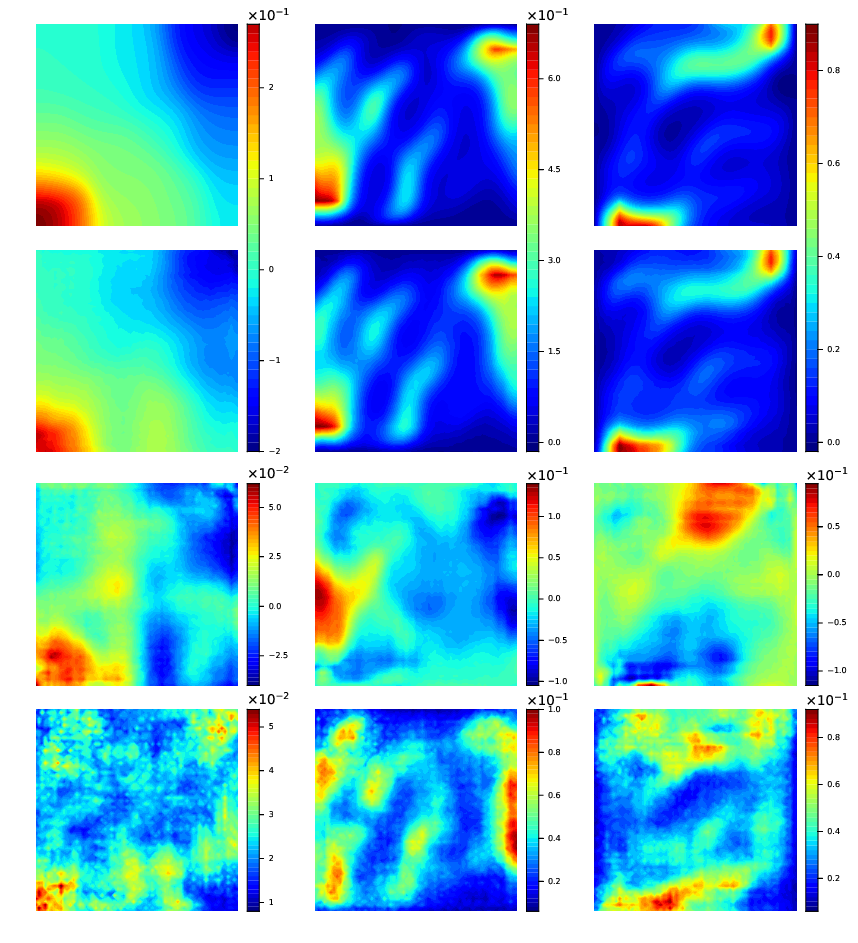}
		\caption{$64$ training data}
	\end{subfigure}
	~ 
	\begin{subfigure}[t]{0.425\textheight}
		\centering
		\includegraphics[width=0.95\textwidth]{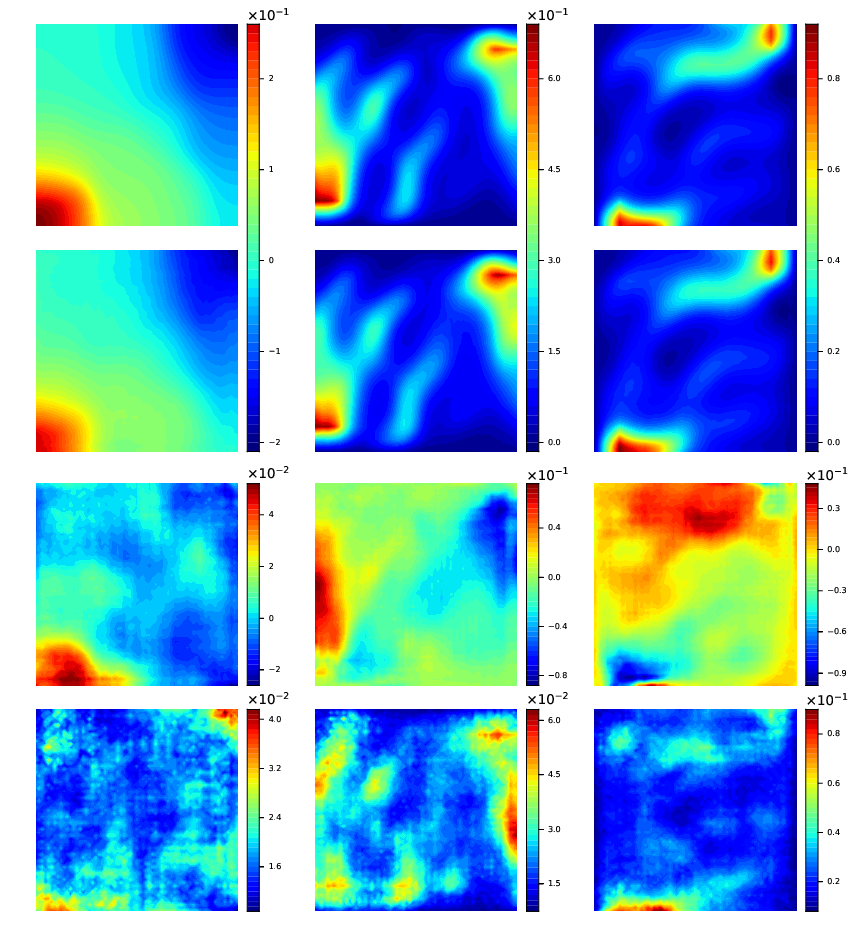}
		\caption{$256$ training data}
	\end{subfigure}
	\caption{Prediction for the input realization shown in Fig.~\ref{fig:sample_kle50} from the KLE$50$ dataset. The rows from top to bottom show the simulation output fields (ground truth), predictive mean $\E [\bmy^* \mid \bmx^*, \mc D]$, the error of the above two, and two standard deviation of predictive output distribution per pixel $\mathrm{Var} (\bmy^* \mid \bmx^*, \mc D)$. The three columns from left to right correspond to pressure field $p$, and two velocity fields $\bmu_y$, $\bmu_x$, respectively.}
	\label{fig:kle50_pred_at_47_bayes}
\end{figure}

\begin{figure}[htbp]
	\centering
	\begin{subfigure}[t]{0.425\textheight}
		\centering
		\includegraphics[width=0.95\textwidth]{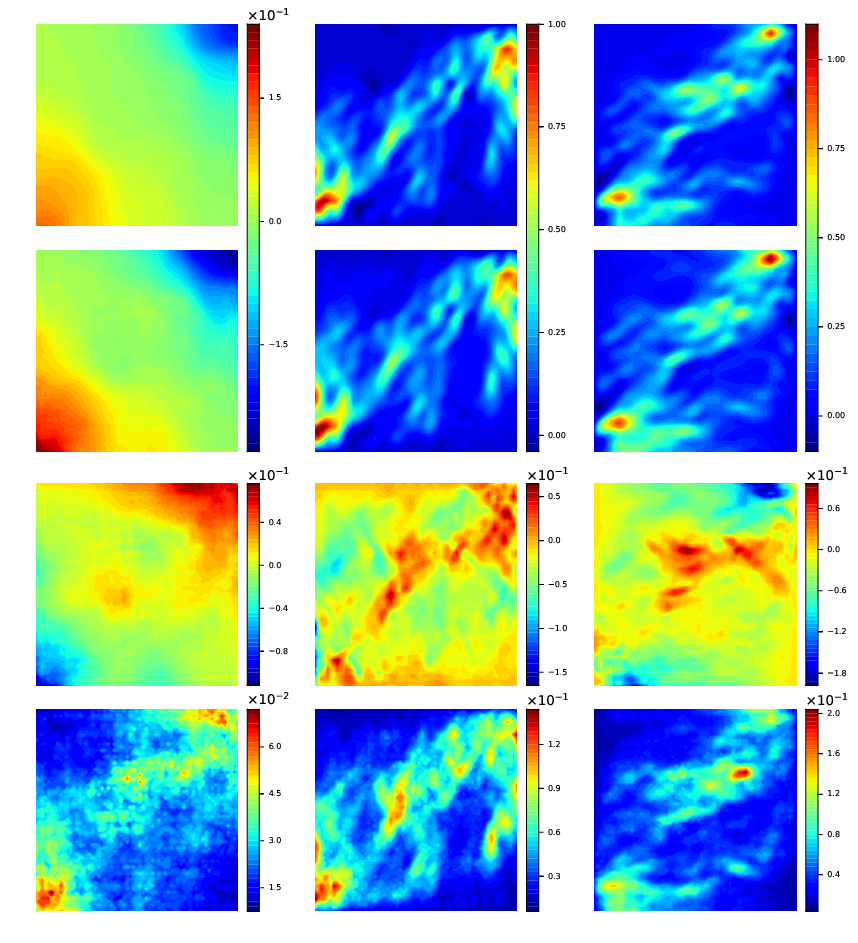}
		\caption{$64$ training data}
	\end{subfigure}
	~ 
	\begin{subfigure}[t]{0.425\textheight}
		\centering
		\includegraphics[width=0.95\textwidth]{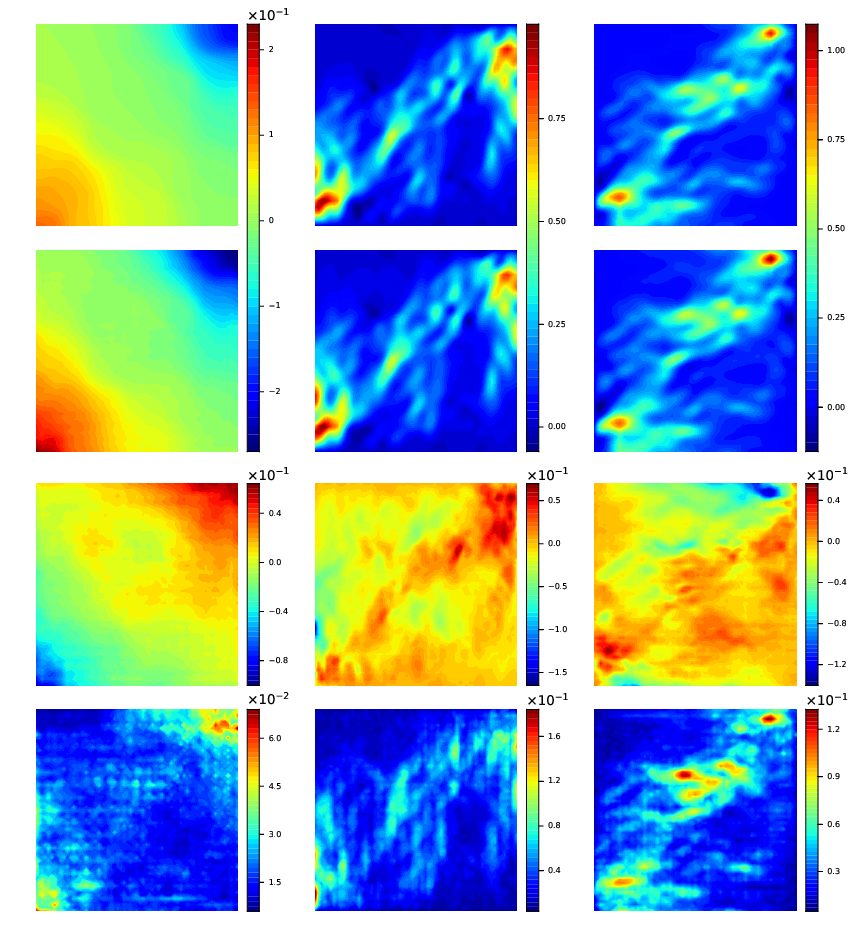}
		\caption{$256$ training data}
	\end{subfigure}
	\caption{Prediction for the input realization  shown in Fig.~\ref{fig:sample_kle500} from the KLE$500$ dataset. The rows from top to bottom show the simulation output fields (ground truth), predictive mean $\E [\bmy^* \mid \bmx^*, \mc D]$, the error of the above two, and two standard deviation of predictive output distribution per pixel $\mathrm{Var} (\bmy^* \mid \bmx^*, \mc D)$. The three columns from left to right correspond to pressure field $p$, and two velocity fields $\bmu_y$, $\bmu_x$, respectively.}
	\label{fig:kle500_pred_at_293_bayes}
\end{figure}

\begin{figure}[htbp]
	\centering
	\begin{subfigure}[t]{0.425\textheight}
		\centering
		\includegraphics[width=0.95\textwidth]{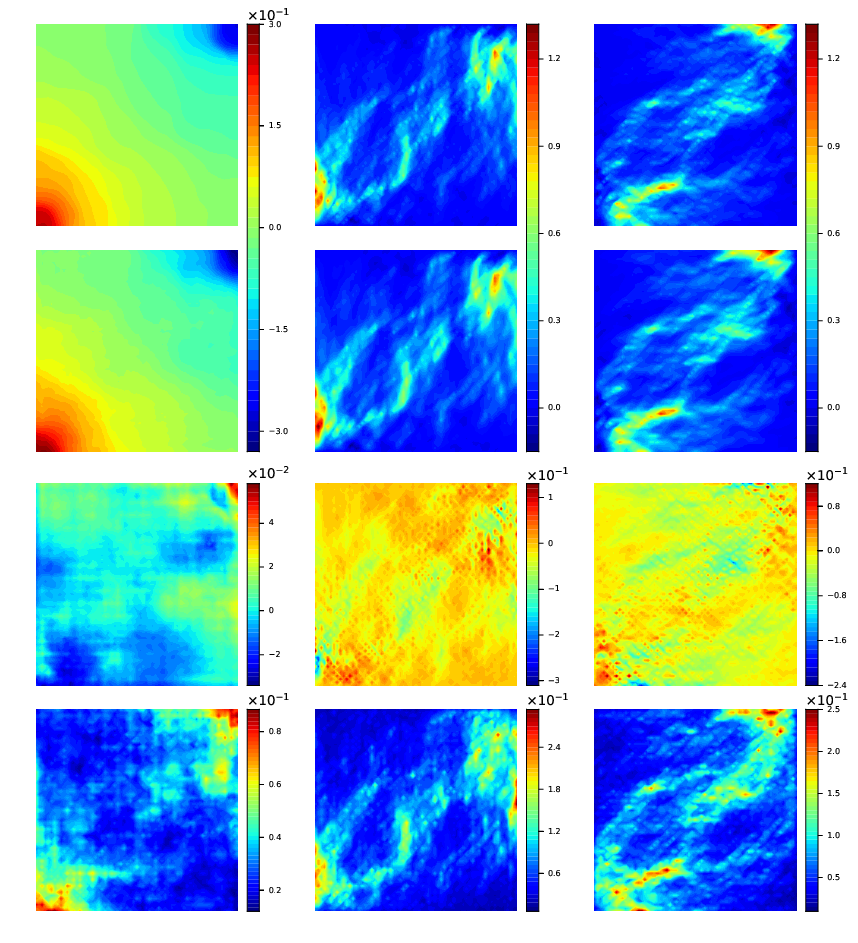}
		\caption{$128$ training data}
	\end{subfigure}
	~ 
	\begin{subfigure}[t]{0.425\textheight}
		\centering
		\includegraphics[width=0.95\textwidth]{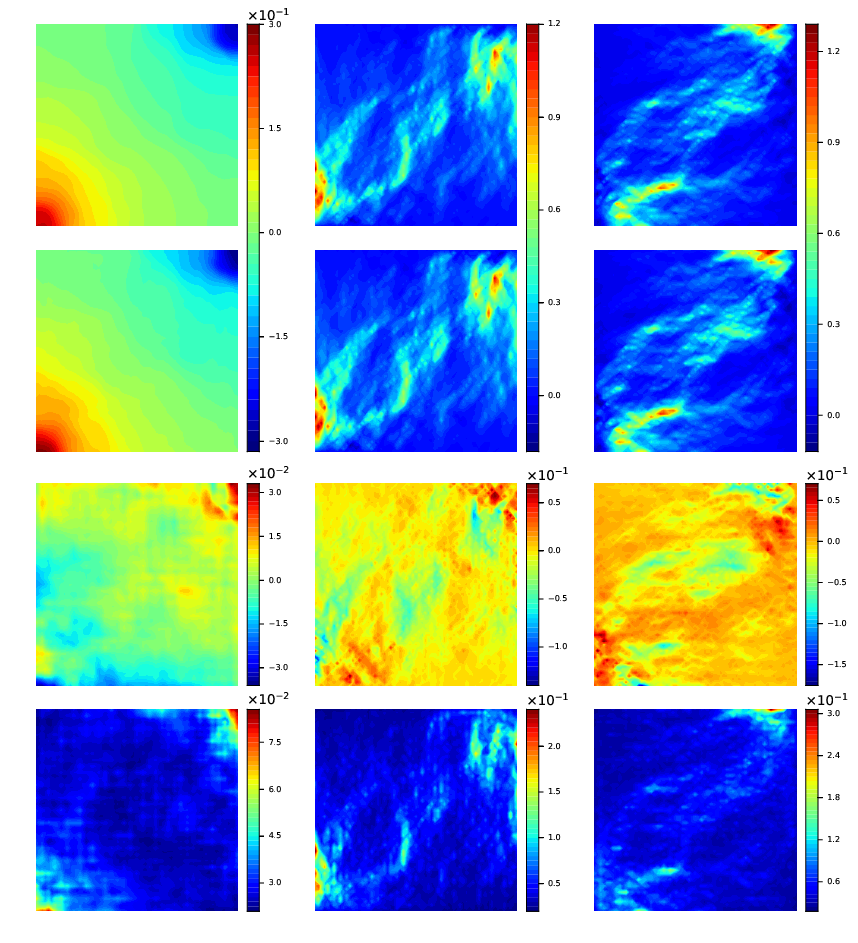}
		\caption{$512$ training data}
	\end{subfigure}
	\caption{Prediction for the input realization shown in Fig.~\ref{fig:sample_kle4225} from the KLE$4225$ dataset. 
		The rows from top to bottom show the simulation output fields (ground truth), predictive mean $\E [\bmy^* \mid \bmx^*, \mc D]$, the error of the above two, and two standard deviation of predictive output distribution per pixel $\mathrm{Var} (\bmy^* \mid \bmx^*, \mc D)$. The three columns from left to right correspond to pressure field $p$, and two velocity fields $\bmu_y$, $\bmu_x$, respectively.}
	\label{fig:kle4225_pred_at_312_bayes}
\end{figure}

The $R^2$-score gives us a general estimate of how well the regression performs.
For a given input permeability field, the Bayesian neural network can predict the mean of corresponding output fields, and also gives uncertainty estimate represented as predictive variance at each spatial location, which is unavailable for deterministic models, and desirable when the training data is small. 
In Figs.~\ref{fig:kle50_pred_at_47_bayes},~\ref{fig:kle500_pred_at_293_bayes}, and~\ref{fig:kle4225_pred_at_312_bayes},  we show predictions for the test input  shown in Fig.~\ref{fig:samples_dataset} with training data from KLE$50$, KLE$500$, and KLE$4225$, respectively. We can see that the predictive accuracy improves as the size of the training dataset increases, while the predictive uncertainty drops.

\begin{figure}[hbtp]
	\centering
	\begin{subfigure}[b]{0.45\textwidth}
		\centering
		\includegraphics[width=\textwidth]{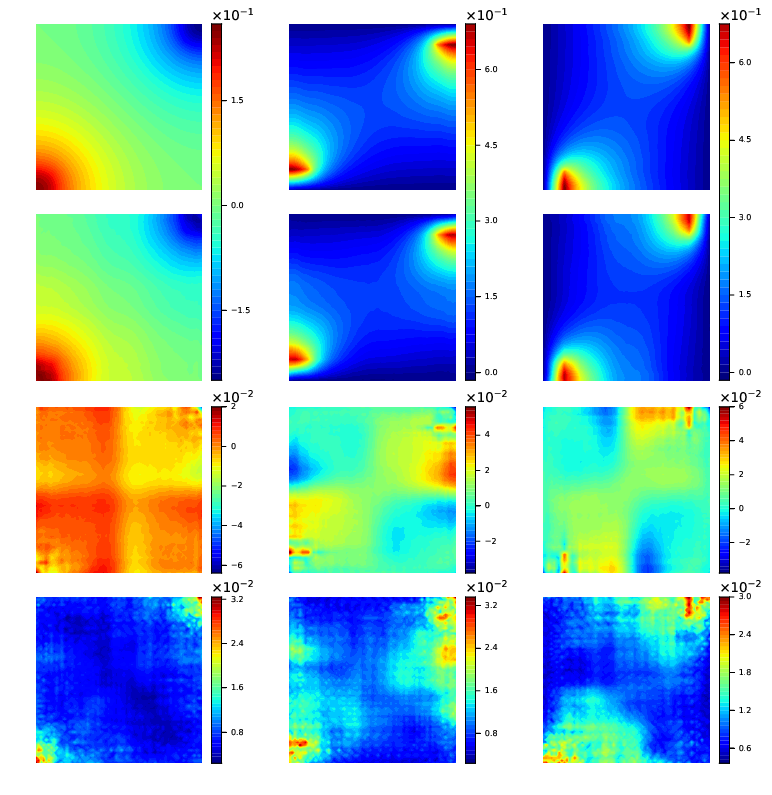}
		\caption{$32$ training data}
	\end{subfigure}
	~
	\begin{subfigure}[b]{0.45\textwidth}
		\centering
		\includegraphics[width=\textwidth]{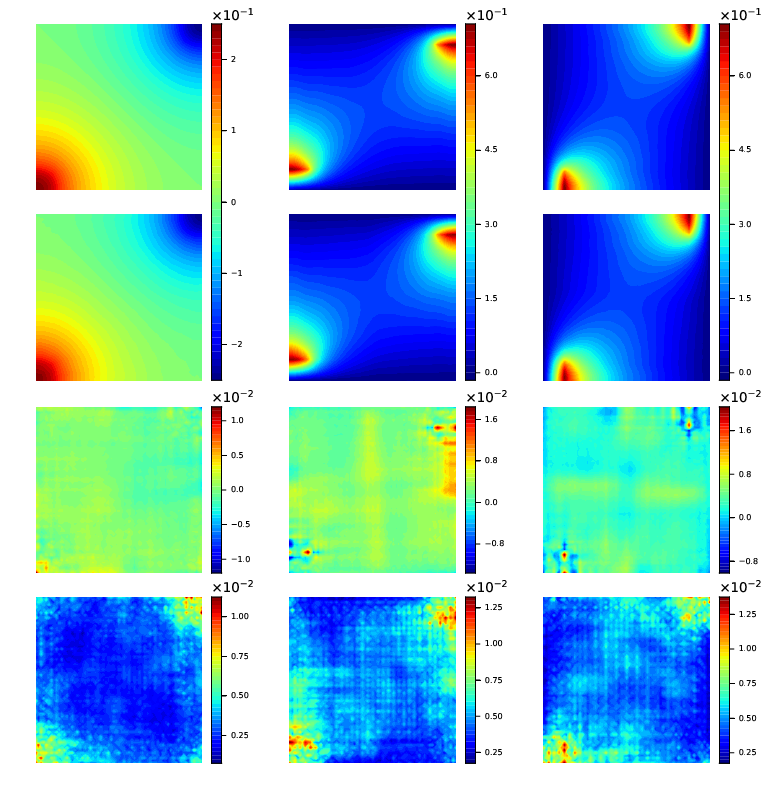}
		\caption{$128$ training data}
	\end{subfigure}
	
	\begin{subfigure}[b]{0.45\textwidth}
		\centering
		\includegraphics[width=\textwidth]{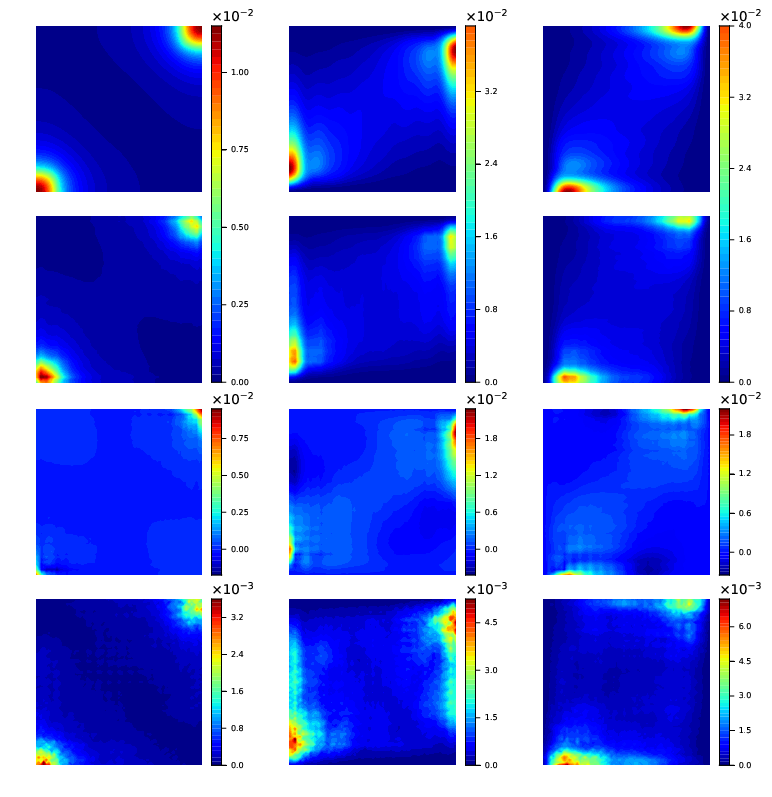}
		\caption{$32$ training data}
	\end{subfigure}
	~
	\begin{subfigure}[b]{0.45\textwidth}
		\centering
		\includegraphics[width=\textwidth]{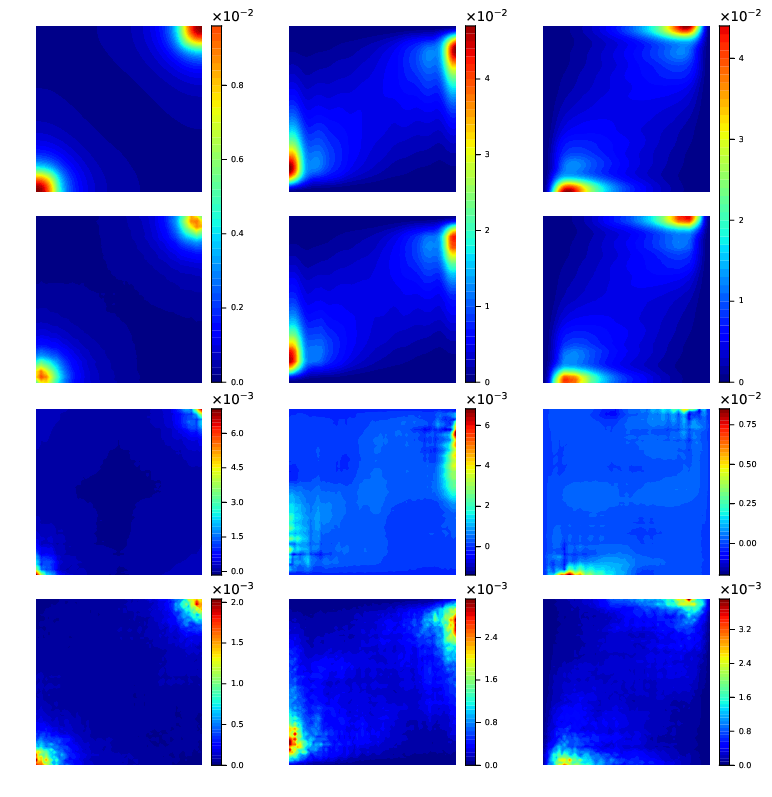}
		\caption{$128$ training data}
	\end{subfigure}
	\caption{Uncertainty propagation for KLE$50$: In (a) and (b) from top to bottom, we show the Monte Carlo output mean, predictive output mean $\E_\bmtheta [\E[\bmy \mid \bmtheta]]$, the error of the above two, and two standard deviation of conditional predictive mean $\mathrm{Var}_\bmtheta (\E[\bmy \mid \bmtheta] )$. The three columns from left to right correspond to pressure field $p$, and two velocity fields $\bmu_y$, $\bmu_x$, respectively. In (c) and (d) from top to bottom, we show the Monte Carlo output variance, predictive output variance $\E_\bmtheta [\mathrm{Var}(\bmy \mid \bmtheta)]$, the error of the above two, and two standard deviation of conditional predictive variance $\mathrm{Var}_\bmtheta (\mathrm{Var}(\bmy \mid \bmtheta) )$. The three columns are the same as (a) and (b).}
	\label{fig:mc_stats_kle50_32_128}
\end{figure}

\begin{figure}[hbtp]
	\centering
	\begin{subfigure}[b]{0.45\textwidth}
		\centering
		\includegraphics[width=\textwidth]{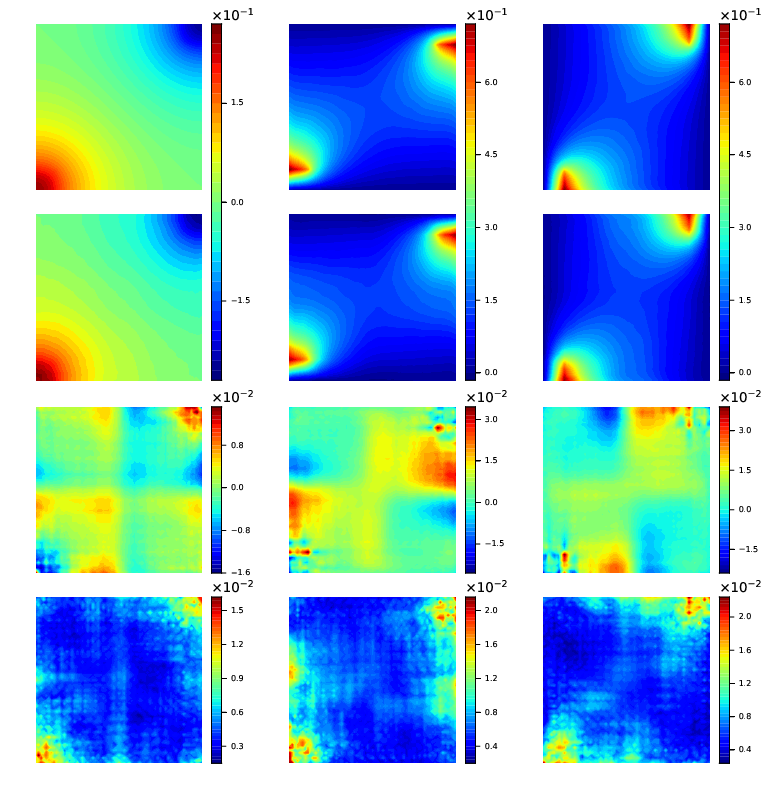}
		\caption{$64$ training data}
	\end{subfigure}
	~
	\begin{subfigure}[b]{0.45\textwidth}
		\centering
		\includegraphics[width=\textwidth]{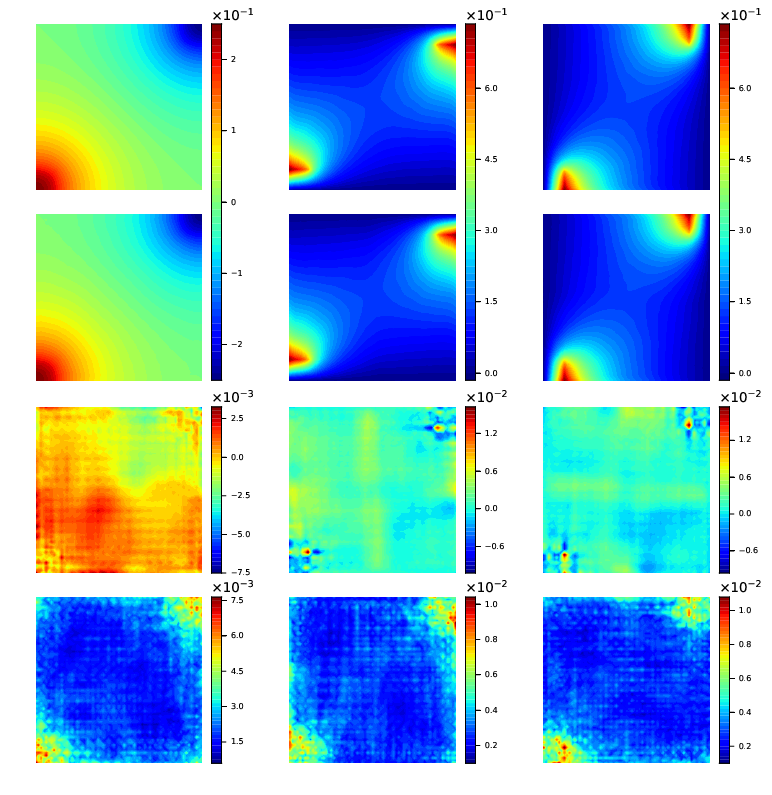}
		\caption{$256$ training data}
	\end{subfigure}
	
	\begin{subfigure}[b]{0.45\textwidth}
		\centering
		\includegraphics[width=\textwidth]{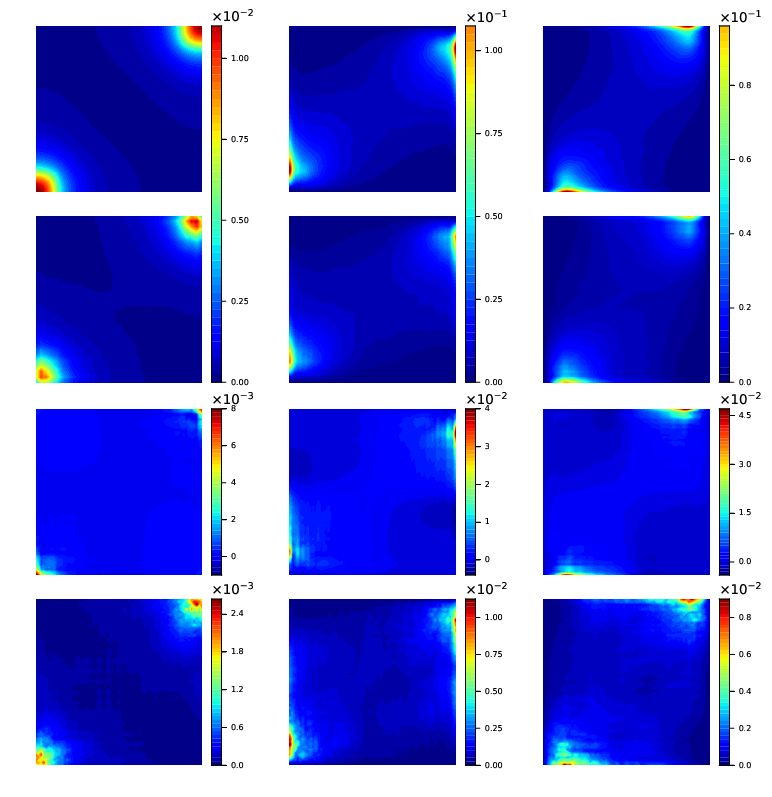}
		\caption{$64$ training data}
	\end{subfigure}
	~
	\begin{subfigure}[b]{0.45\textwidth}
		\centering
		\includegraphics[width=\textwidth]{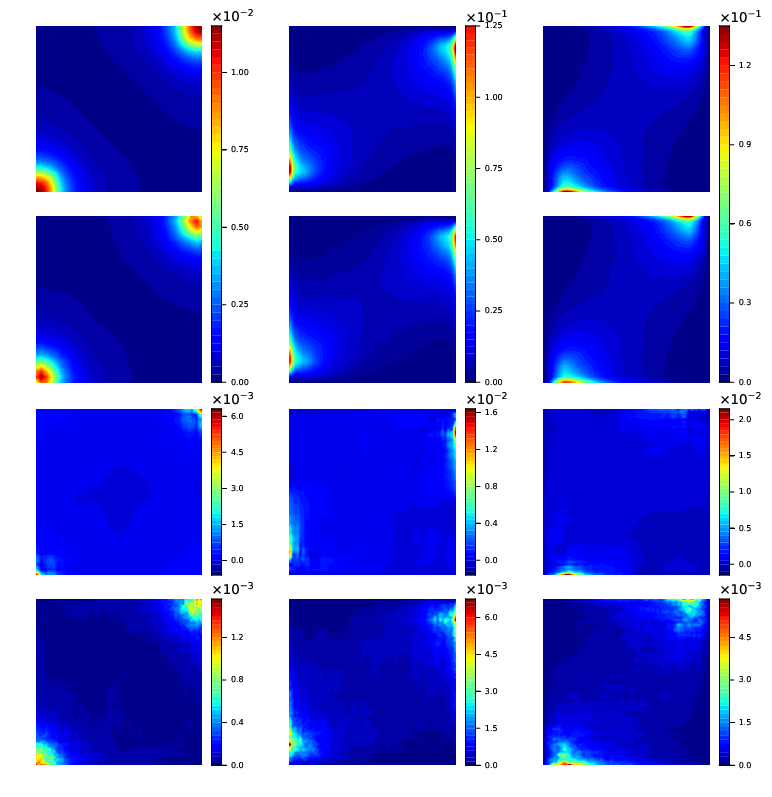}
		\caption{$256$ training data}
	\end{subfigure}
	\caption{Uncertainty propagation for KLE$500$: (a), (b), (c), and (d) refer to Fig.~\ref{fig:mc_stats_kle50_32_128}.}
	\label{fig:mc_stats_kle500_64_256}
\end{figure}

\begin{figure}[hbtp]
	\centering
	\begin{subfigure}[b]{0.45\textwidth}
		\centering
		\includegraphics[width=\textwidth]{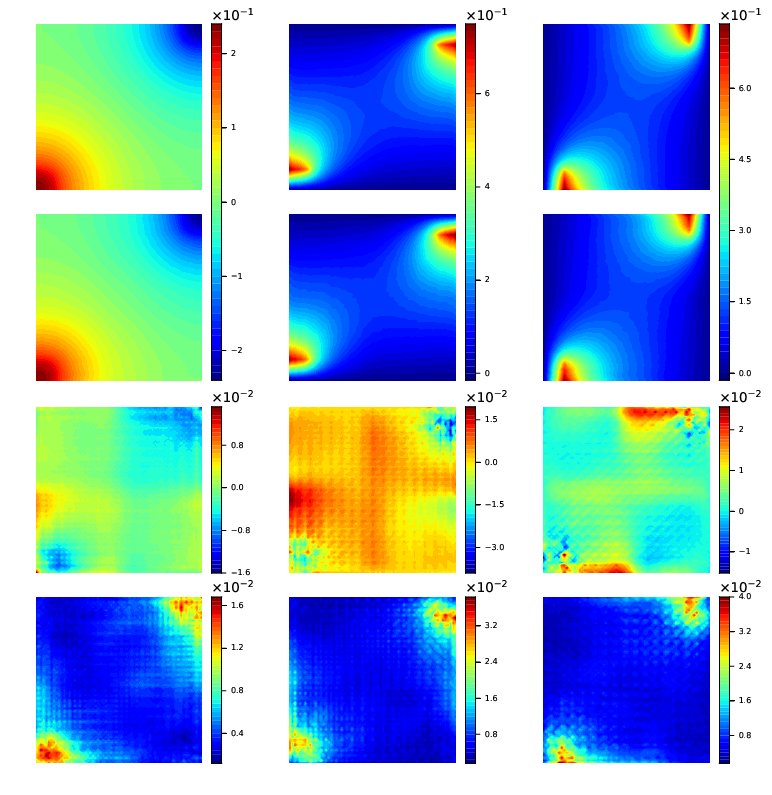}
		\caption{$128$ training data}
	\end{subfigure}
	~
	\begin{subfigure}[b]{0.45\textwidth}
		\centering
		\includegraphics[width=\textwidth]{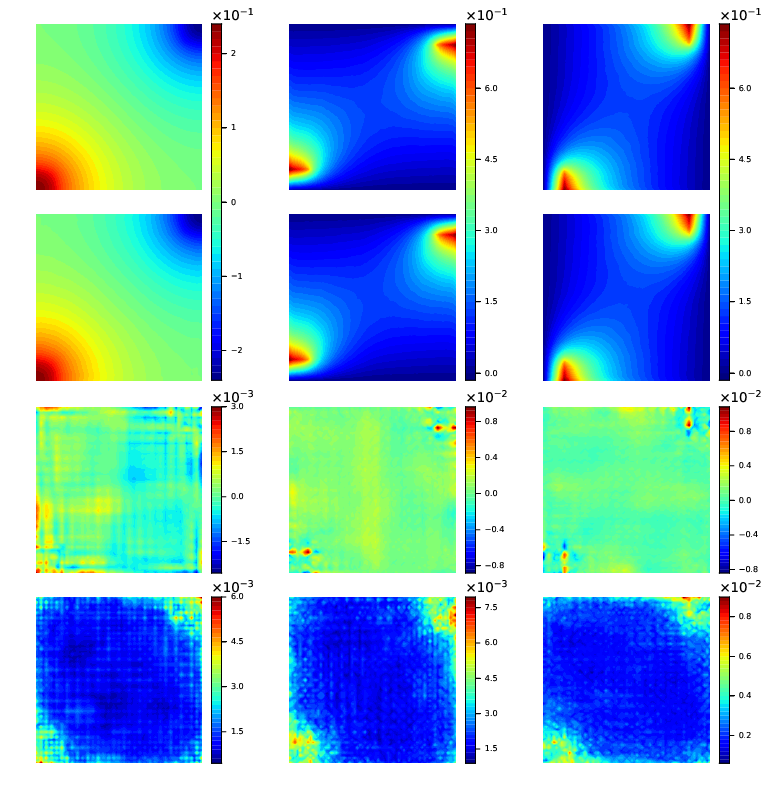}
		\caption{$512$ training data}
	\end{subfigure}
	
	\begin{subfigure}[b]{0.45\textwidth}
		\centering
		\includegraphics[width=\textwidth]{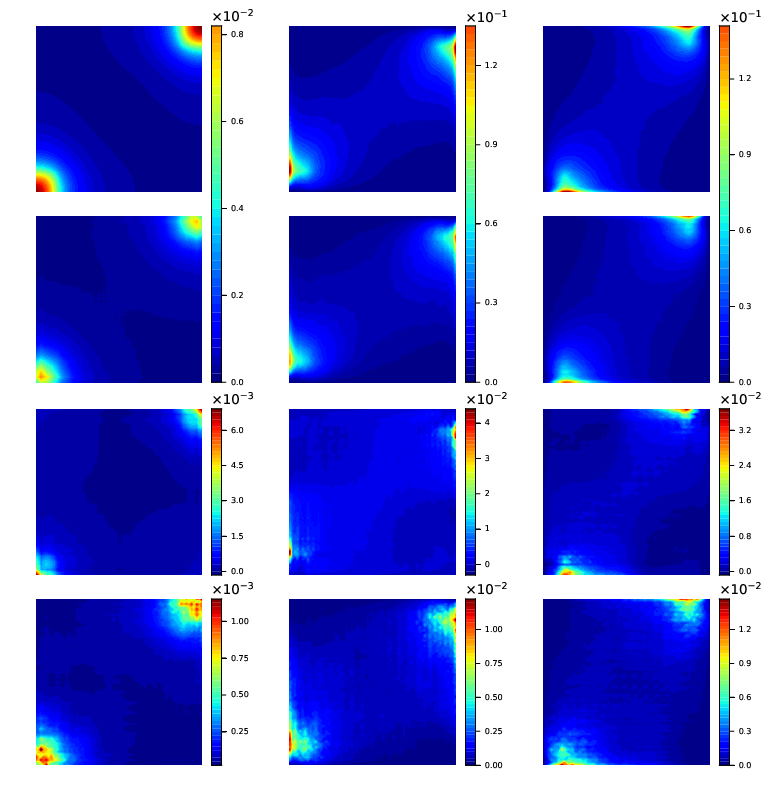}
		\caption{$128$ training data}
	\end{subfigure}
	~
	\begin{subfigure}[b]{0.45\textwidth}
		\centering
		\includegraphics[width=\textwidth]{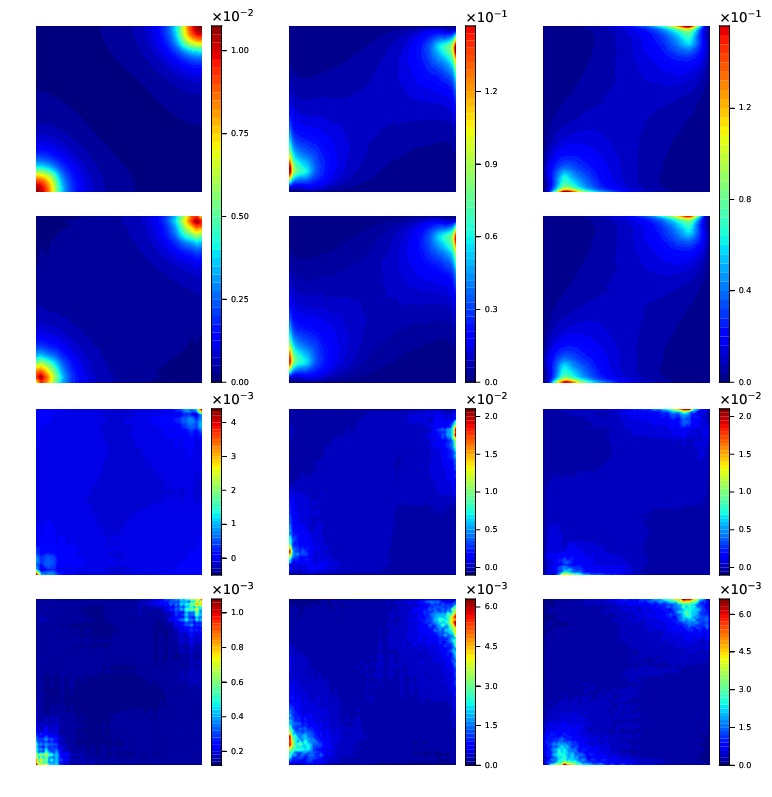}
		\caption{$512$ training data}
	\end{subfigure}
	
	\caption{Uncertainty propagation for KLE$4225$: (a), (b), (c), and (d) refer to Fig.~\ref{fig:mc_stats_kle50_32_128}.}
	\label{fig:mc_stats_kle4225_128_512}
\end{figure}

We also performed uncertainty propagation by feeding the trained Bayesian surrogate with $10,000$ input realizations sampled from the Gaussian field, and calculating the output statistics as in Section~\ref{sec:stats}. In Figs.~\ref{fig:mc_stats_kle50_32_128},~\ref{fig:mc_stats_kle500_64_256} and~\ref{fig:mc_stats_kle4225_128_512}, we show the uncertainty propagation results and compare with  Monte Carlo using the $10,000$ UP data for the Bayesian surrogate trained with the datasets KLE$50$, KLE$500$, and KLE$4225$.

\begin{figure}[hbtp]
	\centering
	\begin{subfigure}[b]{0.75\textwidth}
		\includegraphics[width=1\linewidth]{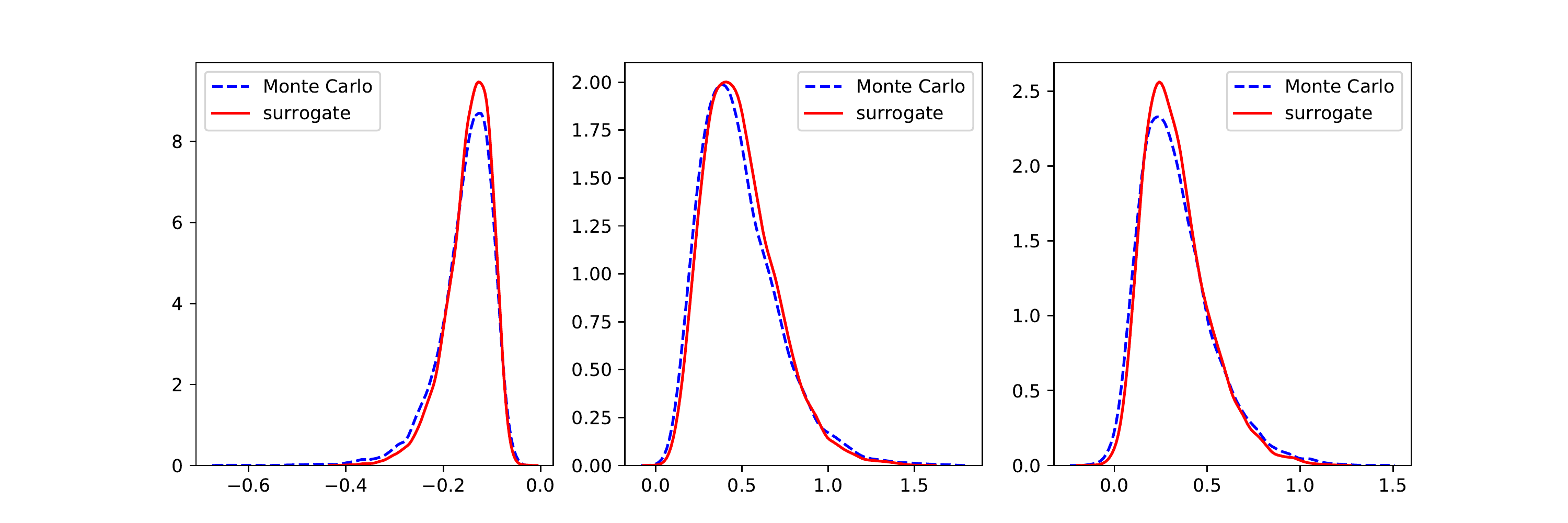}
		\caption{Location $(0.85, 0.88)$ $128$ training data}
		\label{fig:pdfs_kle4225_n128_loc_85_88} 
	\end{subfigure}
	
	\begin{subfigure}[b]{0.75\textwidth}
		\includegraphics[width=1\linewidth]{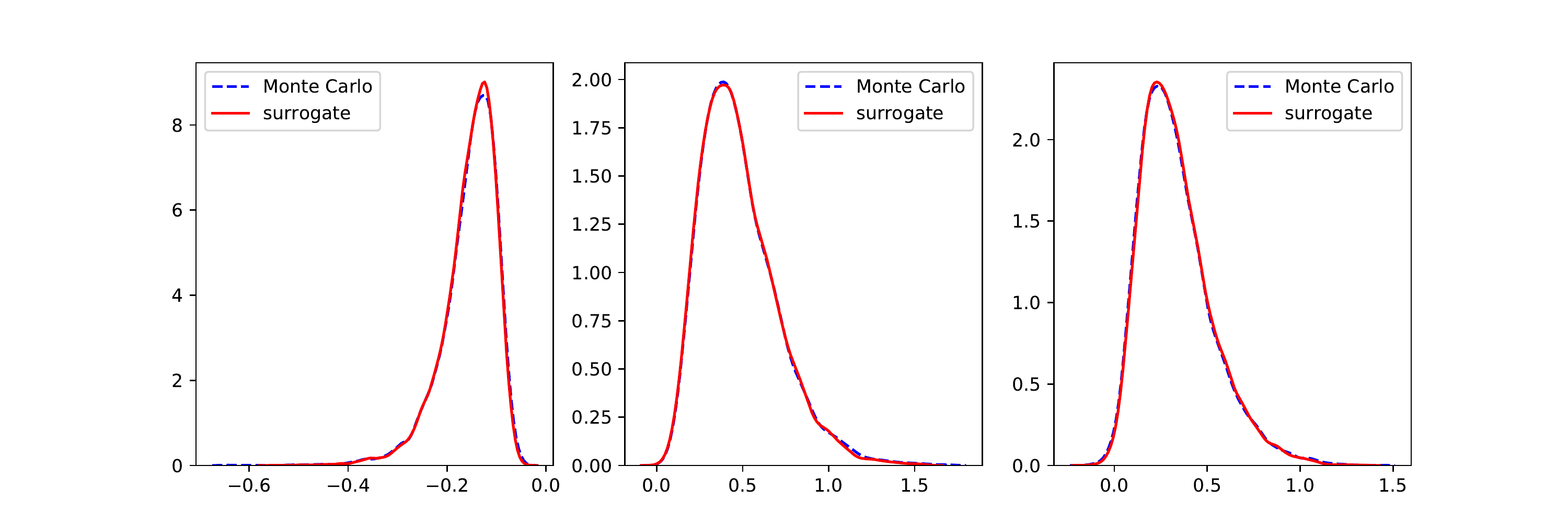}
		\caption{Location $(0.85, 0.88)$ $512$ training data}
		\label{fig:pdfs_kle4225_n256_loc_85_88}
	\end{subfigure}
	\caption{Distribution estimate for the pressure, and the two velocity components (from left to right) at location $(0.85, 0.88)$ 
		with Bayesian surrogate trained with KLE$4225$ dataset.}
	\label{fig:pdfs_kle4225_loc_85_88}
\end{figure}
\begin{figure}[hbtp]
	\centering
	\begin{subfigure}[b]{0.75\textwidth}
		\includegraphics[width=1\linewidth]{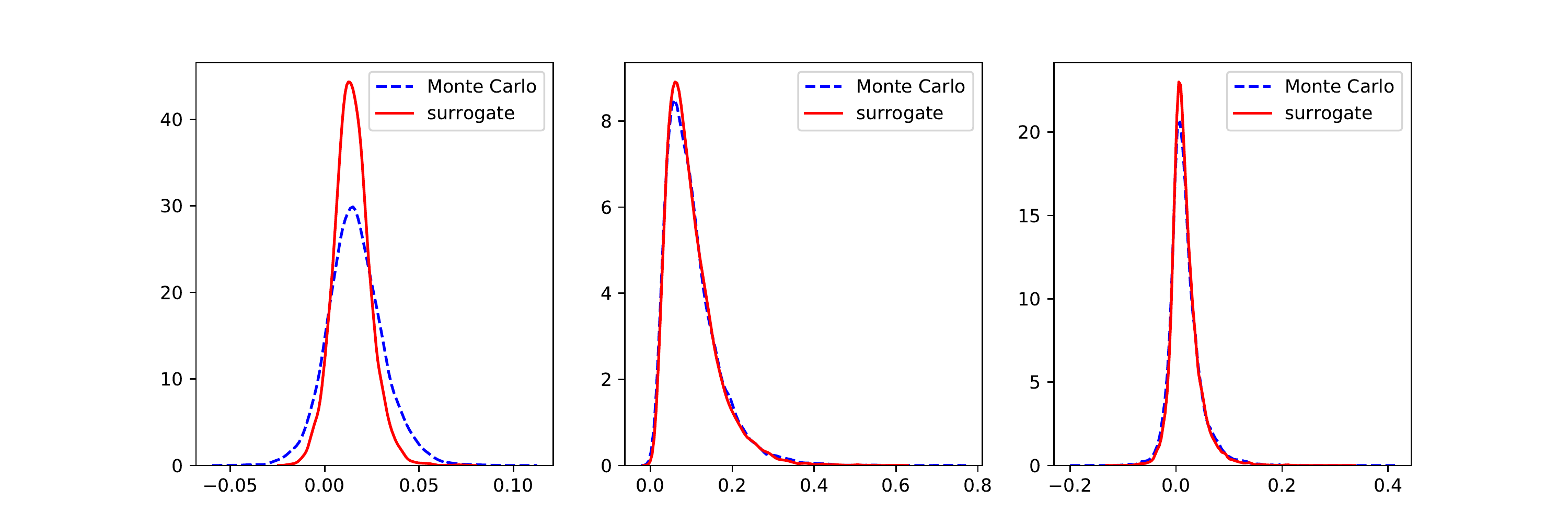}
		\caption{Location $(0.68, 0.05)$ $128$ training data}
		\label{fig:pdfs_kle4225_n128_loc_68_05} 
	\end{subfigure}
	
	\begin{subfigure}[b]{0.75\textwidth}
		\includegraphics[width=1\linewidth]{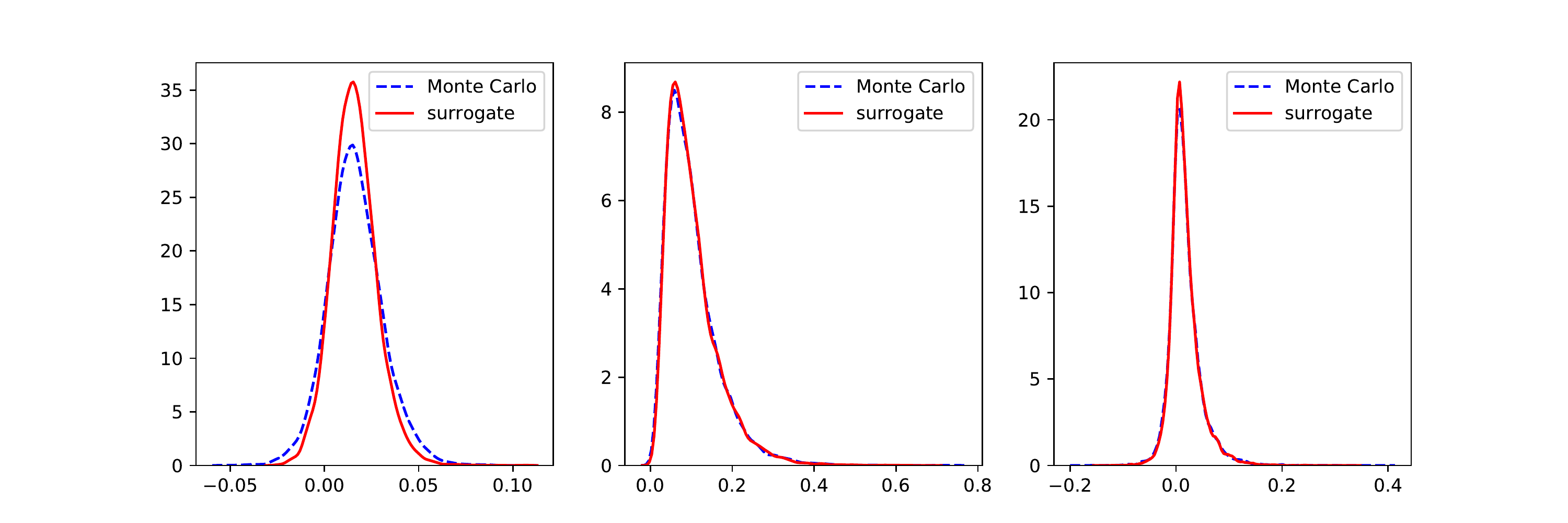}
		\caption{Location $(0.68, 0.05)$ $512$ training data}
		\label{fig:pdfs_kle4225_n256_loc_68_05}
	\end{subfigure}
	\caption{Distribution estimate for the pressure, and the two velocity components (from left to right) at location $(0.68, 0.05)$ 
		with Bayesian surrogate trained with KLE$4225$ dataset.}
	\label{fig:pdfs_kle4225_loc_68_05}
\end{figure}

We show the estimate of pressure $p$, velocity components $u_x$, $u_y$ at locations $(0.85, 0.88)$, and $(0.68, 0.05)$ on the unit square for $128$, and $512$ training data of KLE$4225$ in Fig.~\ref{fig:pdfs_kle4225_loc_85_88}, and~\ref{fig:pdfs_kle4225_loc_68_05}, respectively.
The PDF is obtained by kernel density estimation using the predictive mean. We can see that the density estimate is close to the Monte Carlo result even when the training dataset is small, and becomes closer as the training dataset increases. From Fig.~\ref{fig:pdfs_kle4225_loc_68_05}, we observe that the predictions for the velocity fields are better than the pressure field especially in locations away from the diagonal of the unit square domain, and this is in general the case for our current network architecture, where the three output fields are treated the same. 

\begin{figure}[hbtp]
	\centering
	\begin{subfigure}[b]{0.475\textwidth}
		\centering
		\includegraphics[width=\textwidth]{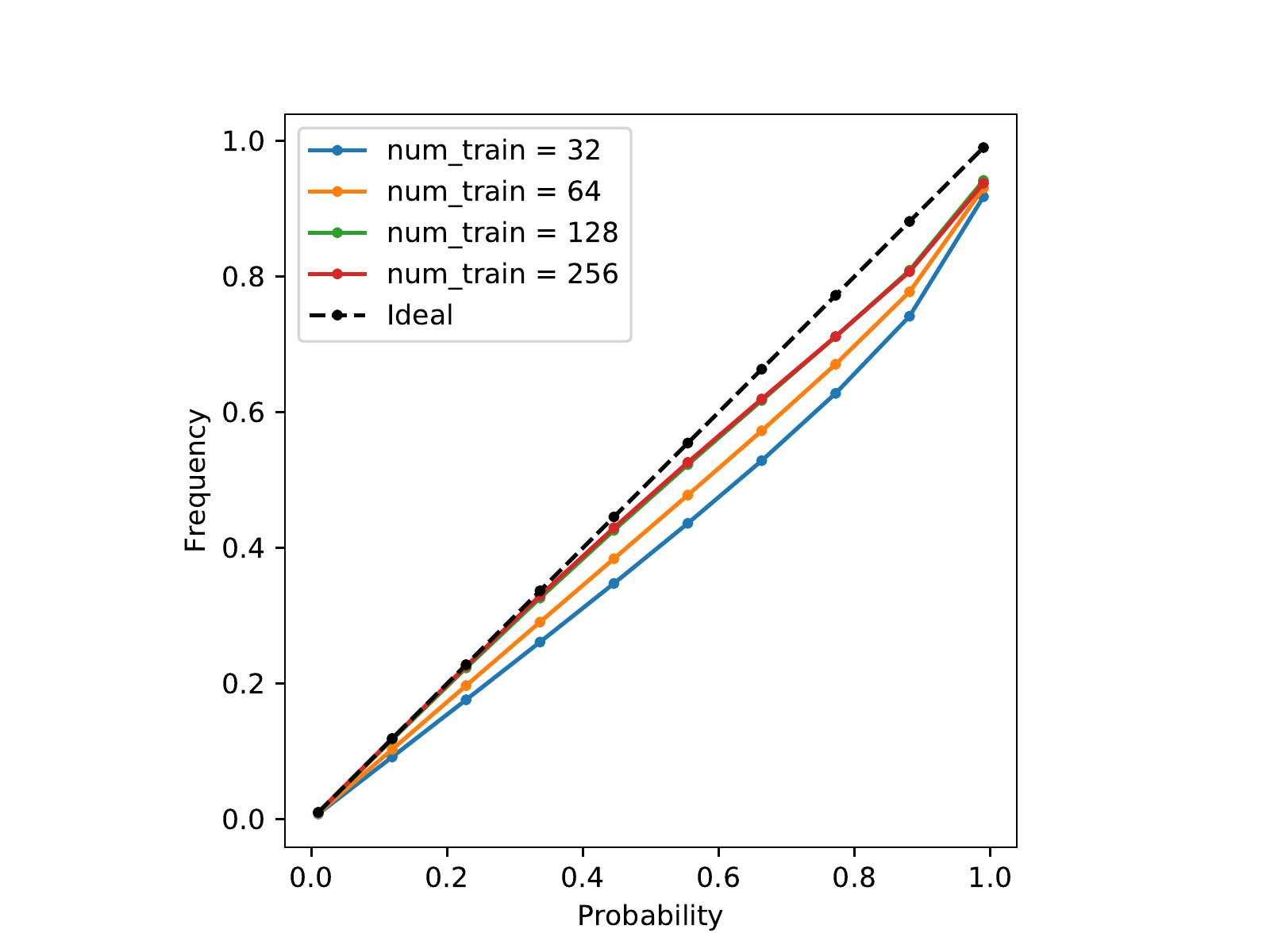}
		\caption{KLE$50$}
	\end{subfigure}
	~
	\begin{subfigure}[b]{0.475\textwidth}
		\centering
		\includegraphics[width=\textwidth]{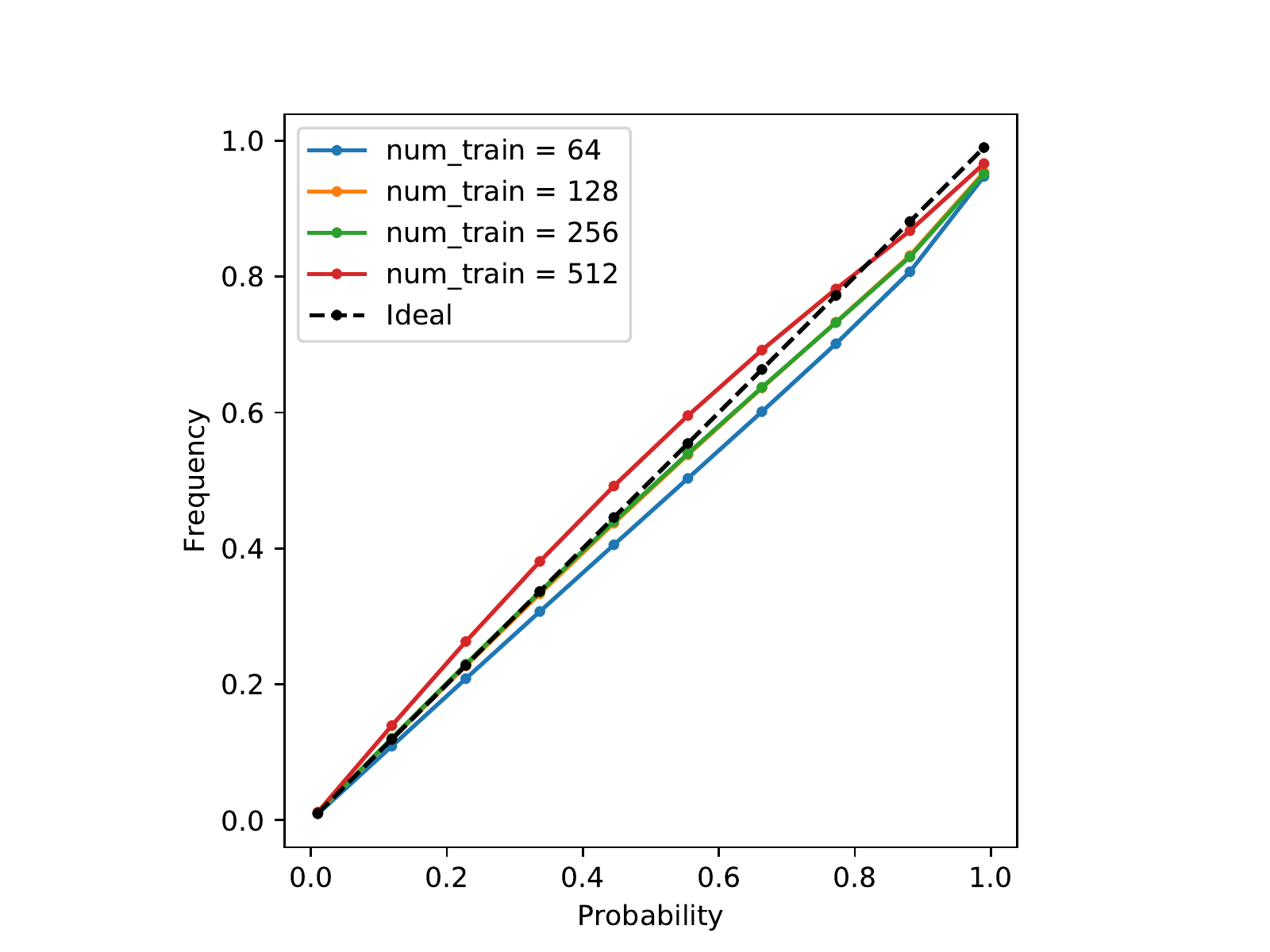}
		\caption{KLE$500$}
	\end{subfigure}
	
	\begin{subfigure}[b]{0.475\textwidth}
		\centering
		\includegraphics[width=\textwidth]{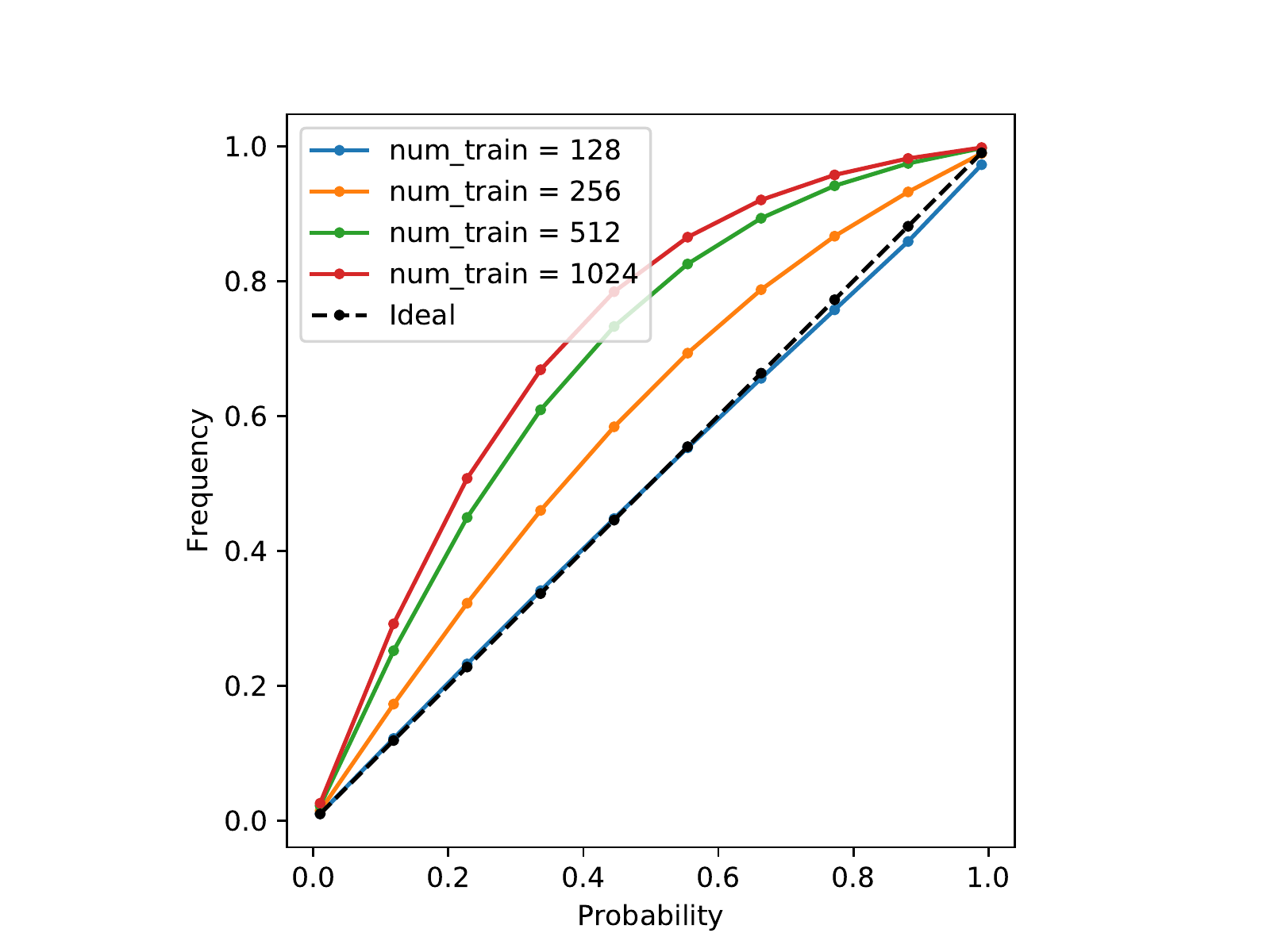}
		\caption{KLE$4225$}
	\end{subfigure}
	~
	\begin{subfigure}[b]{0.475\textwidth}
		\centering
		\includegraphics[width=\textwidth]{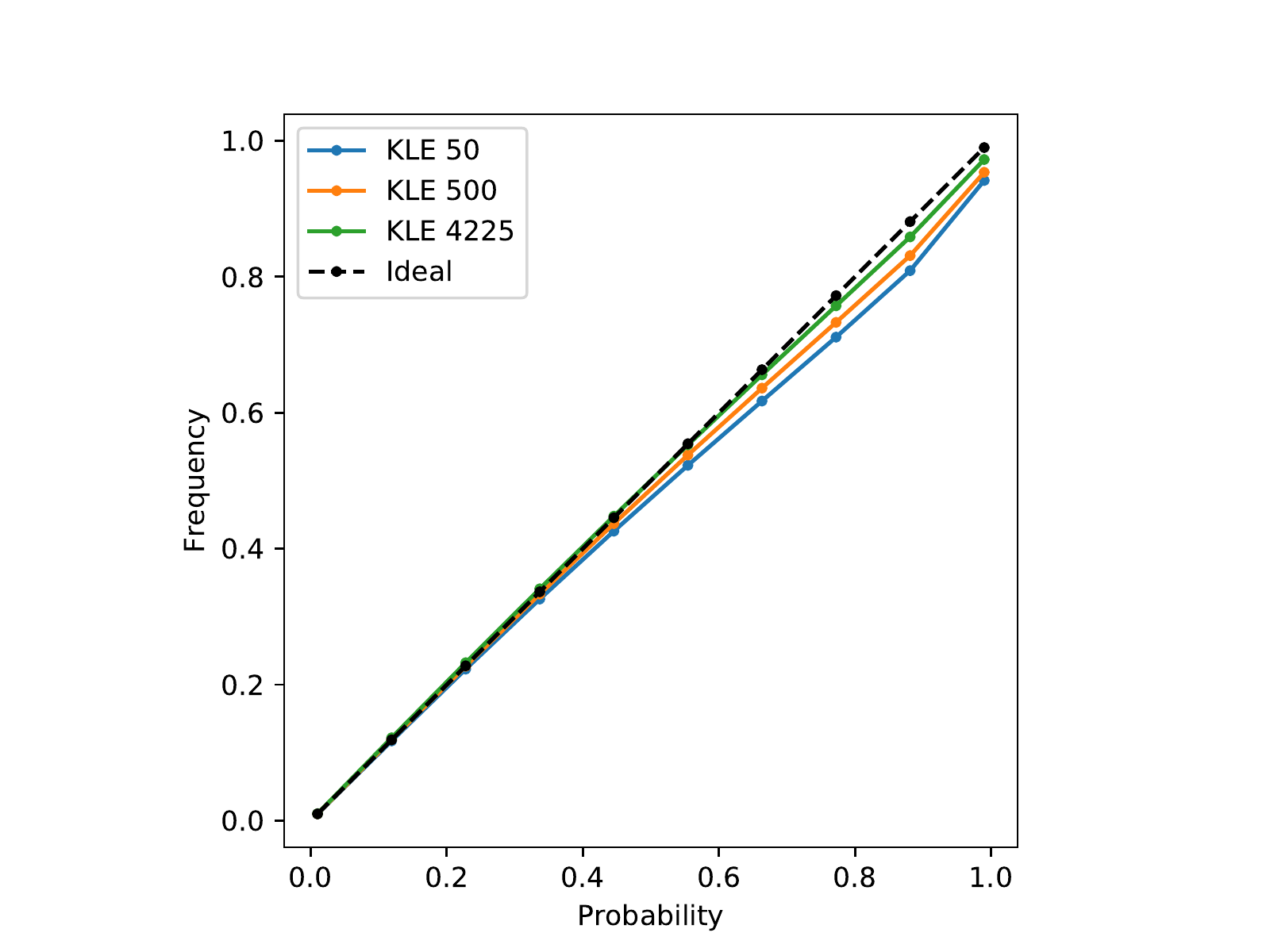}
		\caption{All cases with $128$ training data}
		\label{fig:reliability_n128}
	\end{subfigure}
	\caption{Reliability diagrams. Subfigures (a), (b), (c) show the reliability diagram for KLE$50$, KLE$500$, KLE$4225$, 
		respectively. (d) shows the  diagrams when the training dataset size is $128$ for all three datasets.}
	\label{fig:reliability}
\end{figure}

In order to access the quality of the computed uncertainty, we adopt the reliability diagram which expresses the discrepancy between the predictive probability and the frequency of falling in the predictive interval for the test data. The diagram is shown in Fig.~\ref{fig:reliability}.
Overall our models are well-calibrated since they are quite close to the ideal diagonal, especially the case when the training dataset size is $128$ as shown in Fig.~\ref{fig:reliability_n128}. In general the model turns to be over-confident (small predictive uncertainty) when the training data is small, and gradually becomes prudent (larger predictive uncertainty) when the training data increases. The main reason for this observation is that the predictive uncertainty is dominated by the variation seen in the training data, which is small when small data is observed. The initial learning rates and their scheduling scheme for network parameters $\bmw$ and noise precision $\beta$ may also play roles here since the uncertainty is determined by the optimum that that stochastic optimization obtained.

\begin{figure}[hbtp]
	\centering
	\includegraphics[width=0.5\textwidth]{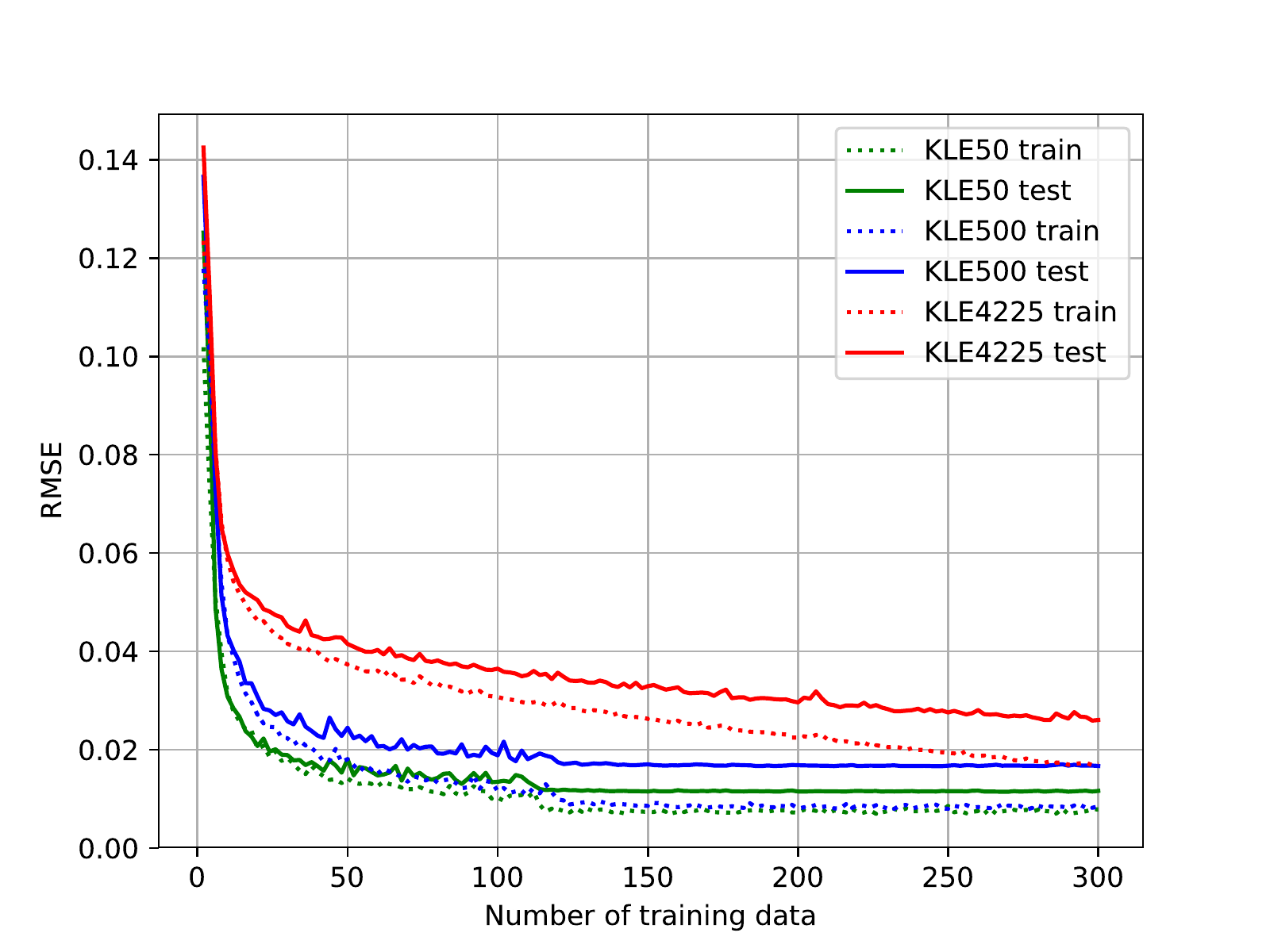}
	\caption{Test and training RMSE for training of SVGD using $128$ training data from KLE$50$, KLE$500$, and KLE$4225$.}
	\label{fig:rmse_training_n128}
\end{figure}

The training processes with different datasets are shown in Fig.~\ref{fig:rmse_training_n128}. We can see that the training and test RMSE converges around $50 \sim 75$ epochs of training for KLE$50$, and KLE$500$, but the convergence for KLE$4225$ seems to take longer time. The training dataset size is $256$ for all three sets.

\begin{figure}[hbtp]
	\centering
	\includegraphics[width=0.65\textwidth]{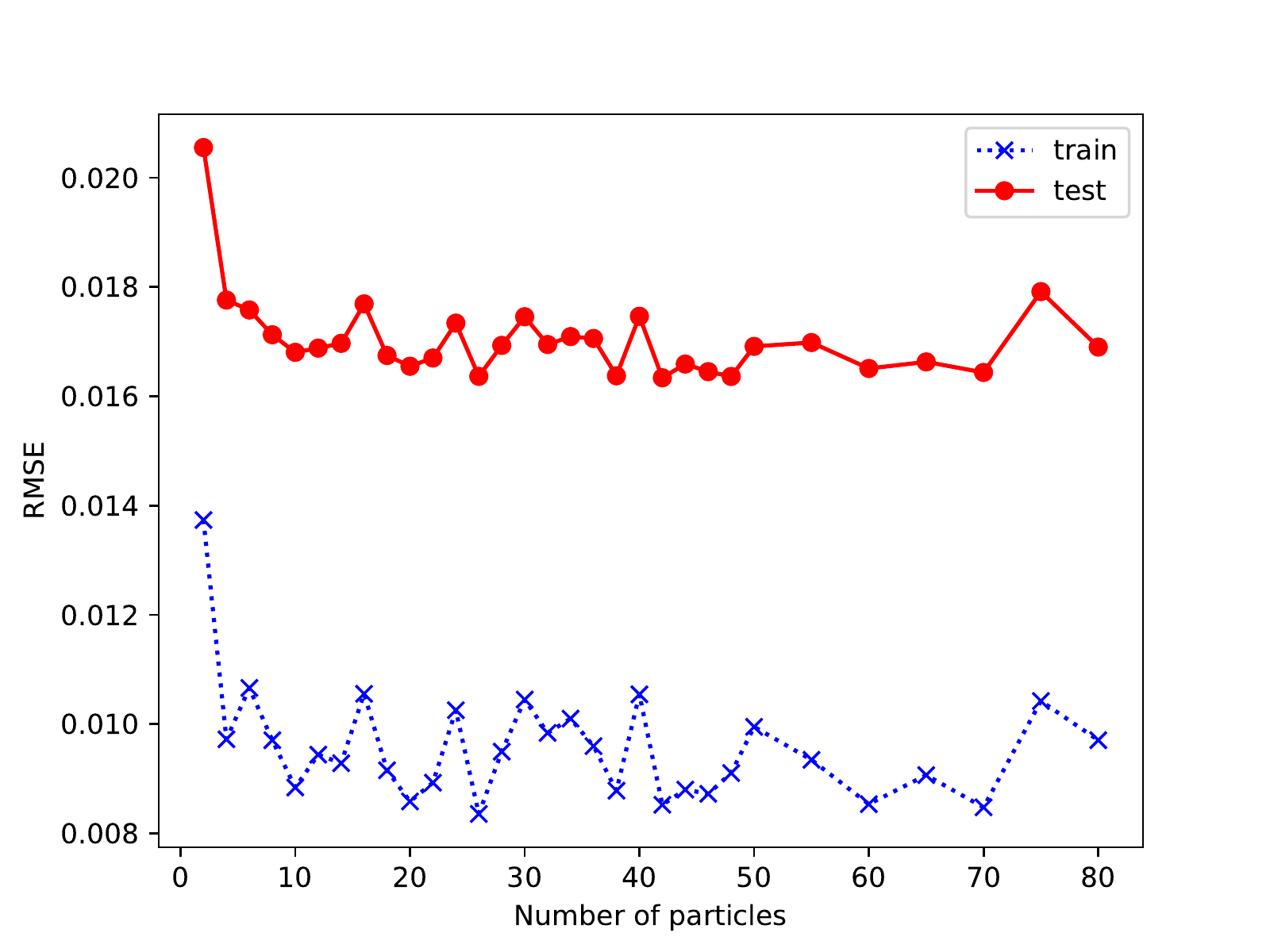}
	\caption{Test and training error using SVGD to train the Bayesian NN with $128$ training data of KLE$500$. As we increase the number of posterior samples for $\bmtheta$, i.e. the number of deterministic neural networks (and noise precision), both training and test errors drop and become steady. This plot empirically supports the reason we choose $20$ model instances/samples for SVGD.}
	\label{fig:mc_convergence}
\end{figure}

To empirically show that using $20$ samples of the Bayesian neural nets for the SVGD algorithm is sufficient in our problem, we show in Fig.~\ref{fig:mc_convergence} the convergence of the test and training RMSE  when we vary the number of samples.

\section{Conclusions}
\label{sec:conclusions}
In this work, we explore to use an alternative Bayesian surrogate, i.e. a Bayesian neural network, to predict the output fields of a systems governed by stochastic partial differential equations with high-dimensional stochastic input. The approach is built on the highly expressive deep convolutional encoder-decoder networks. The end-to-end image-to-image regression avoids the usual linear dimensionality reduction step, and achieves promising results in terms of predictive performance and uncertainty modeling, even with limited training data.

We show in experiments that one network (\texttt{DenseED-c16}) works well across problems with different intrinsic dimensionality. We believe the performance at the small dataset domain of deep neural networks is due to its unique generalization property~\cite{DBLP:journals/corr/ZhangBHRV16,2018arXiv180100173P} which effectively says the over-parameterized deep neural nets lack  over-fitting. 

The Bayesian model provides  uncertainty estimates for our  predictions thus accounting for epistemic uncertainty when training with small datasets. We show that Bayesian inference based on SVGD works well for training the Bayesian neural network, and presented results on uncertainty propagation tasks. The uncertainty of the trained model is well-calibrated by investigating the reliability diagrams.

There are three potential directions to improve the regression performance of non-Bayesian surrogate:
\begin{itemize}
	\item Add skip connections between the encoder and decoder path in each level of feature map size. The reason is that the input permiability field directly affects the output velocity field pixel-wise through the equation $\bmu(\bms) = -K(\bms) \nabla p(\bms)$. Concatenate the features extracted in the encoder path to the decoder path may help to recover the discontinuity of the output velocity fields which are caused by the discontinuity in the input.
	\item Conditional GANs loss. The idea is that we should use a stronger loss function such as an adversarial classifier to enforce the model output to obey the discontinuity in the data. Similar situation has been explored in jet maps in high-energy particle physics~\cite{de2017learning}. 
	\item Using group convolution in the last decoding layer to separate the influences between the pressure field and the velocity fields in the end. This is to address the observation that the prediction performance for the velocity fields is better than for the pressure field using our current network architecture \texttt{DenseED-c16}.
\end{itemize}

For the Bayesian neural network, the inference task is genuinely difficult since it is asked to find the posterior of millions of random variables based on hundreds of training data. Exploring other priors for the network weights, and using recent advances in variational inference for Bayesian neural networks are definitely valuable directions to pursue.

The image-to-image regression approach can be used to handle the prediction of systems with different source terms, by \textit{adding the source field as another channel in the input besides the material property field}. For implementation, one only needs to increase the number of input channels by $1$, and leave everything else unchanged.

Our current network does not utilize any information from physics, such as the governing equations or constraints between the three output fields. Incorporating physics information into  deep neural networks is a more principled ways to improve the model performance.

\section*{Acknowledgments}
This work was supported from the University of Notre Dame, the Center for Research Computing (CRC) and the Center for Informatics and Computational Science (CICS). Early developments were supported by the Computer Science and Mathematics Division of ORNL under the DARPA EQUiPS program. 
N.Z.\ thanks the Technische Universit\"at M\"unchen, Institute for Advanced Study for support through a Hans Fisher Senior Fellowship, 
funded by the German Excellence Initiative and the European Union Seventh Framework Programme under Grant Agreement No.~$291763$. The authors acknowledge Dr. Steven Atkinson from the CICS for generating and providing us the Darcy flow datasets. 


\appendix

\section{Experiments on Darcy Flow Problem}
\label{sec:DesignTests}
\noindent We optimize the network hyperparameters following the general architecture as shown in Fig.~\ref{fig:dense_ed}. The encoding layer in Fig.~\ref{fig:transition_layers}(a) down-samples its input feature maps by $2$ using convolution of stride $2$ as in the first convolution layer. The decoding layer in Fig.~\ref{fig:transition_layers}(b) up-samples the feature maps by $2$ using transposed convolution instead of pooling layers (as commonly seen in image classification networks) since the location information is critical for regression. The number of down-sampling and up-sampling layers are the same.
The datasets are described in Section~\ref{sec:dataset}. We aim to find one general network architecture that works well across different datasets from both Bayesian and non-Bayesian network models. \\

\noindent {\em Experiment 1 - Spatial dimension of feature maps at the coarsest scale:} This is determined by how many down-sampling layers are used. Shrinking the spatial dimension of feature maps can extract high-level or coarse information of the input permeability field, which is subsequently used to predict the output pressure and velocity fields. This design choice is related to a central concept in CNNs called \textit{receptive field}~\cite{luo2016understanding} of an unit within a certain layer, which is the region in the input image that affects this unit feedforward, or the region in the output image that affects this unit when back-propagating gradients. For dense prediction tasks such as our surrogate modeling problems, it is important for each pixel in the code feature maps to have the suitable receptive field in the input and output images.
We observe that for the dataset generated with fewer KLE terms, both the input and output velocity fields are smoother, or the output has stronger correlation across pixels, such as KLE$50$ in Fig.~\ref{fig:sample_kle50}. But for KLE$500$ or KLE$4225$ in Figs.~\ref{fig:sample_kle500} and~\ref{fig:sample_kle4225}, the fields vary more rapidly from pixel to pixel, but still there is weak long-range correlation between pixel values.
\\

\noindent In our experimental setup we have used no skip connections, the convolution kernels were fixed to be $k7s2p2$ for the first Conv layer, $k3s1p1$ in the layers within dense block,  $k1s1p0$ and $k3s2p1$ in the encoding and decoding layers, and $k5s2p2$ in the last ConvtT layer. No dropout was used and the growth rate of dense blocks was taken as $16$. 
\\

\noindent The loss function is taken as the regularized MSE in Eq.~(\ref{eq:loss_mse_regularized}), and the validation metric as  the $R^2$-score in Eq.~(\ref{eq:R2_score}). Adam optimizer is used for training $200$ epochs, with learning rate $0.001$ and a plateau scheduler on the test RMSE. Batch size is always smaller than the number of training data (e.g. $16, 32, 64$ for dataset with $32, 64, 128$ data, respectively). Weight decay is set to $0.0005$.

\begin{figure}[hbtp]
	\centering
	\begin{subfigure}[t]{0.6\textwidth}
		\includegraphics[width=0.95\linewidth]{dense_ed_c16_noskip.png}
		\caption{\texttt{DenseED-c16} with blocks $(3, 6, 3)$}
		\label{fig:dense_ed_c16_2} 
	\end{subfigure}
	~
	\begin{subfigure}[t]{0.38\textwidth}
		\includegraphics[width=0.85\linewidth]{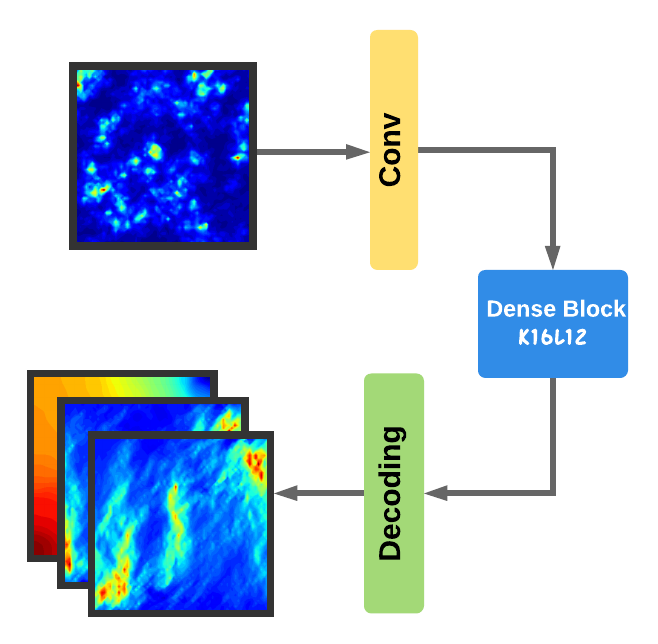}
		\caption{\texttt{DenseED-c32} with blocks $(12,)$}
		\label{fig:dense_ed_c32}
	\end{subfigure}
	\caption{Two configurations \texttt{DenseED-c16} and 
		\texttt{DenseED-c32} of \texttt{DenseED}.}
	\label{fig:dense_ed_c16_c32}
\end{figure}

We vary the {\em dense block configurations}, i.e. a list of integers specifying the number of layers within each dense block. This list contains odd number of integers because of the symmetry between encoding and decoding paths. For example, blocks $(3, 6, 3)$ specify the network architecture in Fig.~\ref{fig:dense_ed_c16_2}. $K16L6$ in the second dense block means there are $6$ layers within the dense blocks, and the growth rate for each layer is $16$. In this case, the code dimension (purple feature maps) is $16 \times 16$ since there are two down-sampling layers in the encoding path.

\begin{figure}[hbtp]
	\centering
	\includegraphics[width=0.55\textwidth]{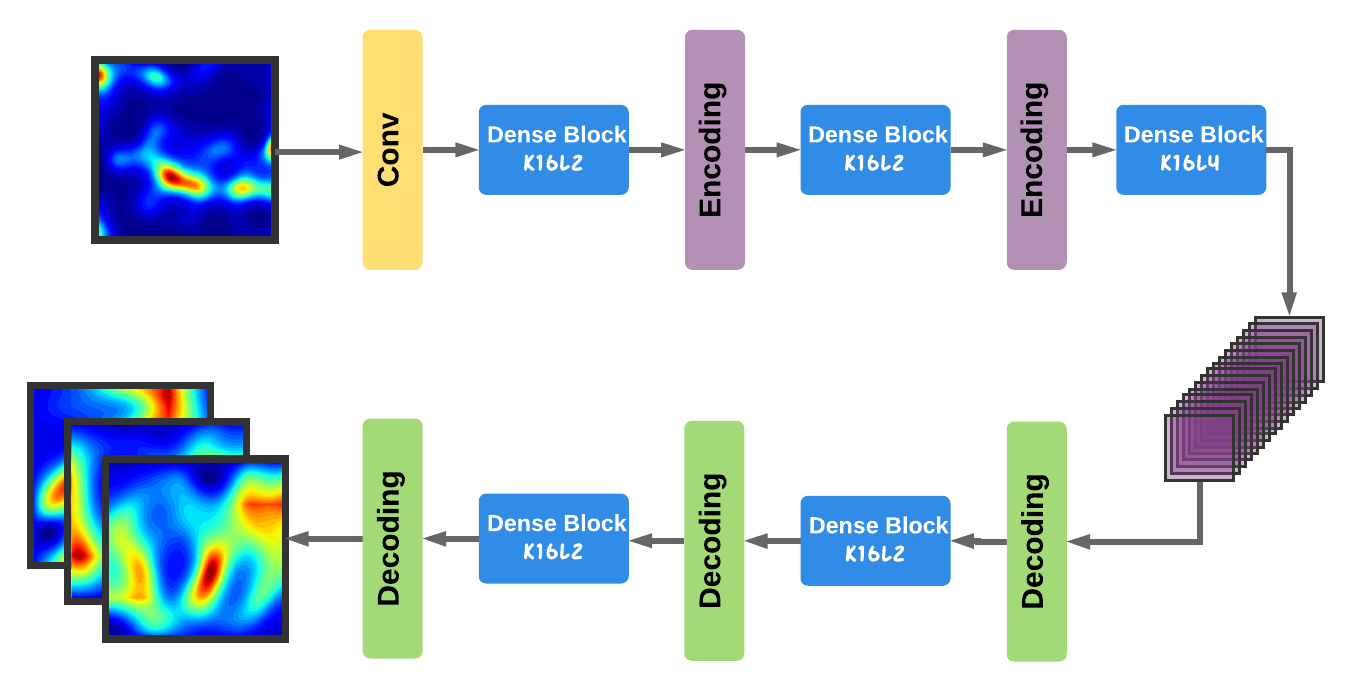}
	\caption{Configuration \texttt{DenseED-c8} with blocks $(2, 2, 4, 2, 2)$.}
	\label{fig:dense_ed_c8}
\end{figure}

The second block configuration is $(12,)$, i.e. only one dense block sits after the first Conv layer, as shown in Fig.~\ref{fig:dense_ed_c32} or $(2, 2, 4, 2, 2)$ as shown in Fig.~\ref{fig:dense_ed_c8}.

The validation $R^2$-scores of the above three networks \texttt{DenseED-c16}, \texttt{DenseED-c8}, \texttt{DenseED-c32} on test set ($500$ Monte Carlo samples from each KLE case) are summarized in Tables~\ref{table:r2_score_dense_ed_c16} ,~\ref{table:r2_score_dense_ed_c8},~\ref{table:r2_score_dense_ed_c32}.
From these tables, it can be seen that the $DenseED-c16$ network works well for all three datasets. The network \texttt{DenseED-c8} does not work well for the KLE$4225$ dataset, since its code dimension is too small. This enforces the output field to be too smooth, which is still favorable to the KLE$50$ dataset. We can also clearly note that the network \texttt{DenseED-c32} does not work well for KLE$50$, even though it performs well  for the KLE$4225$ dataset.
\begin{table}[hbtp]
	\centering
	\caption{Test $R^2$-score for \texttt{DenseED-c16}.}
	\begin{tabular}{ c c c c} 
		\hline
		& KLE$50$ & KLE$500$ & KLE$4225$ \\
		\hline
		$32$  & $0.718$ & $0.551$ & $0.280$ \\
		$64$  & $0.883$ & $0.817$ & $0.662$ \\
		$128$ & $0.947$ & $0.913$ & $0.829$ \\
		$256$ & $0.970$ & $0.954$ & $0.927$ \\
		$512$ & -     & $0.976$ & $0.963$ \\
		\hline
	\end{tabular}
	\label{table:r2_score_dense_ed_c16}
\end{table}

\begin{table}[hbtp]
	\centering
	\caption{Test $R^2$-score for \texttt{DenseED-c8}.}
	\begin{tabular}{ c c c c} 
		\hline
		& KLE$50$ & KLE$500$ & KLE$4225$ \\
		\hline
		$32$  & $0.661$ & $0.344$ & $0.052$ \\
		$64$  & $0.883$ & $0.687$ & $0.411$ \\
		$128$ & $0.931$ & $0.842$ & $0.607$ \\
		$256$ & $0.963$ & $0.934$ & $0.753$ \\
		$512$ & -     & $0.971$ & $0.829$ \\
		\hline
	\end{tabular}
	\label{table:r2_score_dense_ed_c8}
\end{table}

\begin{table}[hbtp]
	\centering
	\caption{Test $R^2$-score for \texttt{DenseED-c32}.}
	\begin{tabular}{ c c c c} 
		\hline
		& KLE$50$ & KLE$500$ & KLE$4225$ \\
		\hline
		$32$  & $0.563$ & $0.558$ & $0.413$ \\
		$64$  & $0.811$ & $0.807$ & $0.704$ \\
		$128$ & $0.893$ & $0.899$ & $0.863$ \\
		$256$ & $0.931$ & $0.937$ & $0.909$ \\
		$512$ & -     & $0.954$ & $0.939$ \\
		\hline
	\end{tabular}
	\label{table:r2_score_dense_ed_c32}
\end{table}
\noindent Predictions and learning curve for the selected network design \texttt{DenseED-c16} are presented in Section~\ref{sec:non_bayes_surrogate}.

\noindent {\em Experiment 2 -  Hyperparameter optimization with Hyperband:} The hyperparameters to select include the ones specifying the network architecture and  the training process. We separately optimize those two sets of hyperparameters with one of them fixed.
Hyperband~\cite{li2016hyperband} is a bandit random search algorithm which has several rounds of successive halving. There are two input for this algorithm, i.e. $R=243$, the maximum iterations for each hyperparameter configuration (here each iteration corresponding to one epoch of training) and $\eta=3$ which specifies to keep $1/3$ of the best configurations after running through all the candidate configurations.

The search space for network architecture with the search space specified in Table~\ref{table:hyperband_net_architecture} with the constraint that the number of parameters being less than $0.25$ million, gives the network architecture presented in Table~\ref{table:dense_ed_c16}.

\begin{table}[hbtp]
	\centering
	\caption{Network architecture hyperparameters and associated ranges for the \texttt{DenseED-c16} network.}
	\begin{tabular}{ c c c} 
		\hline
		Hyperparameters & Type & Values \\
		\hline
		Encoding dense block layers     & Q-uniform      &  $[1, 8]$ \\
		Bottom dense block layers       & Q-uniform      &  $[3, 8]$\\
		Decoding dense block layers     & Q-uniform      &  $[1, 8]$ \\
		Growth rate                     & Categorical   &  $16, 32, 48$\\
		Features after 1st Conv               & Categorical &  $32, 48$\\
		\hline
	\end{tabular}
	\label{table:hyperband_net_architecture}
\end{table}

\begin{table}[htbp]
	\centering
	\caption{Training Hyperparameters and associated ranges for the \texttt{DenseED-c16} network.}
	\begin{tabular}{ c c c} 
		\hline
		Hyperparameters & Type & Values \\
		\hline
		Initial learning rate   & Uniform &  [$10^{-4}$, $5 \times 10^{-3}$] \\
		Weight decay            & Uniform &  [$5 \times 10^{-5}$, $10^{-2}$]\\
		Batch size              & Categorical &  $16, 24$ \\
		Optimizer                & Categorical & Adam, RMSprop \\
		\hline
	\end{tabular}
	\label{table:hyperband_training_hyperparams}
\end{table}

The search for the training process hyperparameters as in Table~\ref{table:hyperband_training_hyperparams} gives approximately the following hyperparameters: initial learning rate $0.002$, weight decay $0.0005$, batch size $16$, and Adam optimizer.

Note that the hyperparameters found by Hyperband are only sub-optimal, but indicate the potential range. The actual hyperparameters used for training are presented in Section~\ref{sec:non_bayes_surrogate}.

\section{Generalization behavior}
\label{sec:Generalization}
\begin{figure}[htbp]
	\centering
	\includegraphics[width=0.7\textwidth]{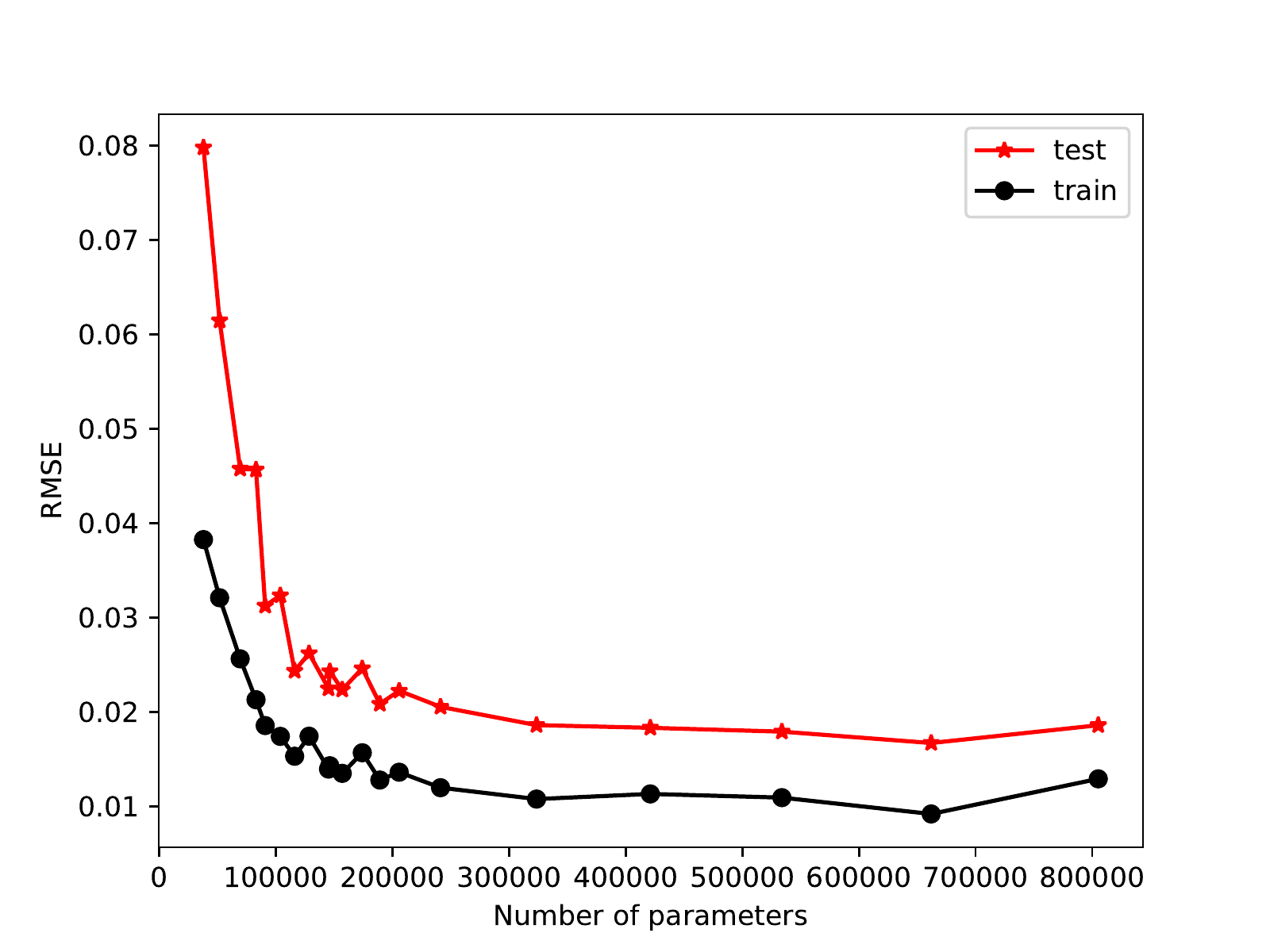}
	\caption{Non-overfitting of over-parameterized neural network \texttt{DenseED-c16}.}
	\label{fig:overparames_non_overfitting}
\end{figure}

Several numerical experiments were conducted to show the generalization behavior of the network \texttt{DenseED-c16}.
The network was trained with different block configurations while keeping the number of training data to be $256$. The number of parameters ranged from $37,892$ to $805,204$. The training and test RMSE are shown in Fig.~\ref{fig:overparames_non_overfitting}. Clearly, the model is over-parameterized but still shows no overfitting behavior, i.e. the training error does not become smaller and the test (generalization) error does not go higher as we increase the number of parameters. When the number of parameters is less than $200,000$ (still in the over-parameterized regime), there is plenty of room to improve in both training loss and test loss. We configure the number of parameters of the baseline network \texttt{DenseED-c16} as in Fig.~\ref{fig:dense_ed_c16} to have $241,164$ parameters, which is favorable according to the generalization curve in Fig.~\ref{fig:overparames_non_overfitting}.

\bibliographystyle{elsarticle-num}
\bibliography{refs}

\end{document}